%% file: main.tex
\documentclass[10pt,letterpaper]{article}
\input{macros}

\title{Composition of nested embeddings\\ with an application to outlier removal}

\usepackage{authblk}
\newif\ifanonymous

\ifanonymous
\author{Anonymous}
\else 
\newcommand*\samethanks[1][\value{footnote}]{\footnotemark[#1]}
\author{\begin{tabular}{@{}c@{}}
    Shuchi Chawla\thanks{Department of Computer Science, University of Texas at Austin.} \\
    \normalsize \texttt{shuchi@cs.utexas.edu}
  \end{tabular}
\and \begin{tabular}{@{}c@{}}
    Kristin Sheridan\samethanks \\
    \normalsize \texttt{kristin@cs.utexas.edu}
    \end{tabular}
}
\fi

\begin{document}
\date{}
\maketitle

\input{maincontent/abstract}
\newpage
%% \begin{keyword}



%% \end{keyword}

% \end{frontmatter}

% \linenumbers

\section{Introduction}

\input{maincontent/intro}

\section{Definitions and main results}

\input{maincontent/prelims}

\section{SDP relaxation and approximation}

\input{maincontent/sdp}

\section{Lipschitz extensions \label{sec:nested}}

% \kristin{what was the original paper for the randomized tree embeddings? should we mention it as an inspiration and/or point out how we've changed things? }

\input{maincontent/nested_embed}

\section{Nested compositions into $\ell_1$}

\label{sec:full_nested}

\input{maincontent/full_nested}

\section{Conclusion}

\input{maincontent/conclusion}

\section*{Acknowledgements}

We are grateful to Ola Svensson for comments on a previous version of this paper that allowed us to simplify Algorithm~\ref{alg:extend} and improve our approximation factor for the number of outliers from $O(\log^4 k)$ to $O(\log^2 k)$. 

This work was supported in part by NSF awards CCF-2217069 and CCF-2225259.



\bibliographystyle{plain} 
\bibliography{main.bib}


%% \begin{thebibliography}{00}


%% \bibitem{}

%% \end{thebibliography}

\appendix

\section{Full constants for Sections \ref{sec:nested} and \ref{sec:full_nested} \label{sec:full_constants}}

\input{maincontent/full_constants}

\section{Hardness of finding outlier sets}
\input{maincontent/hardness}

\end{document}

%% file: macros.tex
\usepackage[utf8]{inputenc}
\usepackage{fullpage,times}
\usepackage[numbers]{natbib}
\usepackage{enumerate}
\usepackage[dvipsnames]{xcolor}

\newcommand{\change}[1]{{\color{Mahogany} #1}}

\usepackage{amsmath,amsfonts,amsthm,amssymb,multirow,hyperref}
\usepackage{cancel} 
\usepackage{mathtools}
\usepackage[normalem]{ulem}

\newcommand{\reals}{\mathbb{R}}
\newcommand{\perm}{\pi}
\newcommand{\ZZ}{\mathbb{Z}}
\newcommand{\st}{^*}
\newcommand{\decider}{b}
\newcommand{\cA}{\mathcal{A}}

\newcommand{\ind}{\texttt{index}}
\newcommand{\factor}{382H_k} %{(1,191H_k)}

\newcommand{\lpspace}{\ell_{p}}
\newcommand{\lspace}[1]{\ell_{#1}}

\newcommand{\origdist}{c_X}
\newcommand{\subdist}{c_S}
\newcommand{\from}{\leftarrow}
\newcommand{\fto}{\rightarrow}
\newcommand{\neigh}{\gamma}
\newcommand{\comparison}{\kappa}
\newcommand{\comparisontwo}{2}

\newcommand{\shrink}{\ensuremath{3^{-1+\frac{1}{p}}}}

\newcommand{\oneineach}{[(\tau+5)\subdist+(2\tau+5)\origdist] \cdot}
\newcommand{\closexy}{\left[ \left[ \frac{\tau+3}{(\kappa-1)\tau}\cdot H_k\cdot (\kappa(\tau+3)+(\kappa+1)(\tau+5)) \right] \subdist + \left[1+\frac{\tau+3}{(\kappa-1)\tau}\cdot H_k\cdot (2\tau+5)(2\kappa+1) \right]\origdist \right]}
\newcommand{\rangwidth}{\tau}
\newcommand{\randrang}{[2,\tau+2]}

\newcommand{\bothinstwo}{ \subdist}
\newcommand{\distp}[2]{\ensuremath{||#1-#2||_p}}
\newcommand{\distl}[3]{\ensuremath{||#2-#3||_{#1}}}
\newcommand{\bothinkitwo}{ \origdist}
\newcommand{\oneineachtwo}{ [7\cdot \subdist+9\cdot \origdist] \cdot}
\newcommand{\farxytwo}{ [31\cdot \subdist + 45\cdot  \origdist ]}
\newcommand{\closexytwo}{ \left(\frac{155}{2}\cdot H_k\cdot \subdist + (\frac{225}{2}\cdot H_k+1)\cdot \origdist\right)}
\newcommand{\concat}{|}

\usepackage{lineno}
\usepackage{algorithm}
\usepackage[noend]{algpseudocode}

\usepackage{parskip} %% used to replace indents with space between paragraphs

\newcommand{\kristin}[1]{\color{blue}[{\small {\bf Kristin:} { {#1}}}]\color{black}}

\newcommand{\hide}[1]{}

\usepackage{braket,amsfonts}

\usepackage{array,subcaption}

\usepackage{pgfplots}

\newcounter{names}
\counterwithin{names}{section}

\newtheorem{thm}[names]{Theorem}
\newtheorem{cor}[names]{Corollary}
\newtheorem{lem}[names]{Lemma}
\newtheorem{definition}[names]{Definition}

\usepackage{graphicx,epstopdf}

\usepackage{amsopn}

\DeclareMathOperator{\expect}{E}

\patchcmd\newpage{\vfil}{}{}{}
\newcommand{\remove}[1]{}

\newenvironment{proofof}[1]{\vspace{0.1in}\noindent{\em Proof of #1.}}{\qed}

%% file: maincontent/abstract.tex
\begin{abstract}
    We study the design of embeddings into Euclidean space with outliers. Given a metric space $(X,d)$ and an integer $k$, the goal is to embed all but $k$ points in $X$ (called the ``outliers") into $\ell_2$ with the smallest possible distortion $c$. Finding the optimal distortion $c$ for a given outlier set size $k$, or alternately the smallest $k$ for a given target distortion $c$ are both NP-hard problems. In fact, it is UGC-hard to approximate $k$ to within a factor smaller than $2$ even when the metric sans outliers is isometrically embeddable into $\ell_2$. We consider bi-criteria approximations. Our main result is a polynomial time algorithm that approximates the outlier set size to within an $O(\log^2 k)$ factor and the distortion to within a constant factor. 

    The main technical component in our result is an approach for constructing Lipschitz extensions of embeddings into Banach spaces (such as $\ell_p$ spaces). We consider a stronger version of Lipschitz extension that we call a \textit{nested composition of embeddings}: given a low distortion embedding of a subset $S$ of the metric space $X$, our goal is to extend this embedding to all of $X$ such that the distortion over $S$ is preserved, whereas the distortion over the remaining pairs of points in $X$ is bounded by a function of the size of $X\setminus S$. Prior work on Lipschitz extension considers settings where the size of $X$ is potentially much larger than that of $S$ and the expansion bounds depend on $|S|$. In our setting, the set $S$ is nearly all of $X$ and the remaining set $X\setminus S$, a.k.a. the outliers, is small. We achieve an expansion bound that is logarithmic in $|X\setminus S|$.
    
    %This is a composition of two given embeddings from subsets of $X$ into $\ell_2$ which inherits the distortions of each to within small multiplicative factors. Specifically, given a low $c_S$ distortion embedding from $S\subset X$ into $\ell_2$ and a high(er) $c_X$ distortion embedding from the entire set $X$ into $\ell_2$, we construct a single embedding that achieves the same  distortion $c_S$ over pairs of points in $S$ and an expansion of at most $O(\log k)\cdot c_X$ over the remaining pairs of points, where $k=|X\setminus S|$. Our composition theorem extends to embeddings into arbitrary $\ell_p$ metrics for $p\ge 1$, and may be of independent interest. 
    % While unions of embeddings over disjoint sets have been studied previously, to our knowledge, this is the first work to consider compositions of {\em nested} embeddings.

    % \kristin{I made some changes to the second paragraph - lmk what you think}

\end{abstract}

%% file: maincontent/intro.tex
Low distortion metric embeddings are an important algorithmic tool with a myriad of applications. The goal is to transform a dataset that lies in an unwieldy metric space into one lying in a nicer space, enabling clean and fast algorithms. Embeddings play an important role in the design of approximation algorithms, in finding good low dimensional representations for data in machine learning and data science, in data visualization, in the design of fast algorithms, and more. (See, e.g., \cite{indyk2004}.)

One of the most enduring concepts from the vast literature on embeddings is that of distortion, which is defined to be the maximum ratio over all pairs of points in the metric by which the distance between the points is expanded or contracted by the embedding. Since this is a worst case notion, it can be particularly sensitive to errors or noise or even intentional corruption of the underlying data. Indeed adversarial or random data corruption is a frequent problem in data science contexts. 

Motivated by these applications, Sidiropoulos et al.~\cite{sidiropolous17} introduced the notion of embeddings with outliers. Given a metric space $(X, d_X)$ of finite size and a target space $(Y, d_Y)$ our goal is to find a low-distortion embedding into $Y$ of all but a few points in $X$; these points are called the outliers. Formally, we say that the space $(X, d_X)$ has a $(k, c)${\em -outlier embedding} into space $(Y,d_Y)$ if there exists an outlier set $K\subset X$ of size at most $k$ and a map $\alpha$ from $X\setminus K$ to $Y$ with distortion at most $c$. \cite{sidiropolous17} showed that for many host spaces $(Y, d_Y)$ of interest, computing the optimal outlier set size $k$ for a given target distortion $c$, or the optimal distortion $c$ for a given target outlier set size $k$, is NP-hard. 

%3. Results and comparison to previous work.

In this work, we study the design of approximately optimal outlier embeddings into the Euclidean metric. Given a metric $(X,d_X)$ that admits a $(k,c)$-outlier embedding into $\ell_2$, we provide a polynomial time algorithm that constructs an $(O(k\operatorname{polylog} k), O(c))$-outlier embedding into $\ell_2$.  In other words, our algorithm removes $O(k\operatorname{polylog} k)$ outliers and returns an embedding of the remaining metric into $\ell_2$ that has distortion only a small constant factor worse than the desired one.  In fact, our algorithm allows for a tradeoff between the outlier set size and the distortion obtained -- allowing us to obtain a distortion of $(1+\epsilon) c$ for any $\epsilon>0$ at the cost of blowing up the outlier set size by an additional multiplicative factor of at most {$1/\epsilon$}. %\kristinEDIT{}{$^2$}. \kristin{with the update we made, I think it might need to be $3+\epsilon$, unless we scale the embedding down in which case it may not remain expanding but we would be able to get this better distortion }

To our knowledge, the only approximations to multiplicative distortion for outlier embeddings known prior to our work considered the special case of embedding unweighted graphs into a line. For this setting, Chubarian and Sidiropoulos~\cite{Chubarian20} developed an algorithm that constructs an $(O(c^6k\log^{5/2}(n)), O(c^{13}))$-outlier embedding when the input metric is an unweighted graph metric. Here $n$ is the size of the given metric $X$. While our result is incomparable to theirs (as it does not limit the dimension of the embedding), we emphasize that our approximation factors do not depend on the size of the metric $X$ and has a vastly improved dependence on the distortion $c$, as we discuss below.

%4. Primary technique: composition of two embeddings. Relate to previous work on union of embeddings. How does our model and results differ from what was peviously known.

\paragraph{Lipschitz extension and composition of nested embeddings.} 
The main technical tool in our work is a stronger version of Lipschitz extensions that we call a composition of nested embeddings. Formally, consider a metric space $(X,d_X)$ and a subspace $S\subset X$. Let $\alpha_S:S\rightarrow Y$ be an embedding from $S$ into $Y$ with distortion $c_S$. A Lipschitz extension is an embedding $\alpha$ of the entire set $X$ into $Y$ such that $\alpha(s)=\alpha_S(s)$ for all $s\in S$ and the expansion on any pair of points in $X$ is not too much more than $c_S$. Observe that we only require the {\em expansion} on points in $X$ to be bounded -- distances are allowed to shrink arbitrarily. Results for Lipschitz extensions typically achieve expansion bounds that depend on the size of $S$ and are independent of the size of $X$. In particular, $X$ can be arbitrarily larger than $S$, even unbounded.

In our setting, we are interested in the case where $S$ (i.e. the ``good" set) is almost all of $X$ and $X\setminus S$ is much smaller. We desire an extension where the expansion bounds depend on the size of $|X\setminus S|$. We call such a Lipschitz extension a {\em weak} composition of nested embeddings; the terminology is explained below. In Section \ref{sec:nested} we design weak nested compositions for embeddings into $\ell_p$ spaces (or more generally Banach spaces) where expansion on any pair of points is at most $O(H_k)$ times $c_S$.

Weak nested compositions into the Euclidean metric allow us to identify outliers with the help of a fractional relaxation. We express the problem of finding an outlier embedding into the Euclidean metric as a semi-definite program where all but $k$ fractionally chosen points are required to have small distortion between them. (See Section~\ref{sec:sdp}.) The distortion allowed for any pair of points depends on the fractional extent to which either is chosen as an outlier. The existence of a  weak $O(H_k)$-nested composition implies that we can find a feasible fractional solution where the outliers suffer only a small factor larger distortion than the non-outliers. This enables a rounding scheme that approximates the number of outliers to within an $O(H_k^2)$ factor and preserves the distortion over non-outliers to within a constant factor. 

%In fact, while we will define weak nested embeddings as a more general version of this problem (allowing dependence on the distortion of $\alpha_S$ and on the distortion of a secondary embedding $\alpha_X$ of the entire set into the target space and requiring bounds on the distortion of points in $S$ but not restricting the new embedding to specific points for elements in $S$), we will  provide a Lipschitz extension  in Section \ref{sec:nested} for embeddings into $\ell_p$ spaces (or more generally Banach spaces) where expansion on any pair of points is at most $O(H_k)$ times the distortion of the original embedding. 

\paragraph{Strong nested composition.} 
% The main technical tool in our work is what we call a composition of nested embeddings, which may be of independent interest. 
Although our application to outlier embeddings requires only bounding the {\em expansion} of the Lipschitz extension from $S$ to $X\setminus S$, more generally we ask whether we can construct an extension which has small {\em distortion} over all of $X$, that is, it doesn't expand or contract by much. We call such an extension a {\em strong nested composition}, and we believe this concept is of independent interest. Observe that the distortion of such an embedding cannot simply be a function of $S$ alone, as $X\setminus S$ may not admit any good embedding at all. 

%We also discuss a new concept we call nested compositions of embeddings. 
Formally, we are given two {\em nested} embeddings $\alpha_X: X\rightarrow Y$ and $\alpha_S:S\rightarrow Y$  with distortions $c_X$ and $c_S$ respectively. In general we would expect that the distortion of $\alpha_S$ is smaller than that of $\alpha_X$. In fact, the distortion of $\alpha_X$ even restricted to just the subset $S$ may be larger than $c_S$. Our goal is to find a new embedding from $X$ into $Y$ such that the distortion of this embedding over the good set $S$ is the same as before -- $c_S$ -- while the distortion over all of $X$ is comparable to $c_X$. In particular, we want a combination of the two embeddings that inherits their respective distortions, with small multiplicative worsening, over the corresponding sets of points. In Section~\ref{sec:full_nested} we show that strong nested compositions exist for the $\ell_1$ metric, and leave open the question of extending this result to other $\ell_p$ metrics.

% looks good
% {We show that for embeddings into the $\ell_p$ metric for $p\ge 1$, this is indeed achievable with a small caveat. %\kristinEDIT{}{in expectation for each pair} 
% In particular, we construct a new embedding based on $\alpha_S$ and $\alpha_X$ that achieves a distortion of $c_S$ over the set $S$, while at the same time expanding distances over the remaining pairs of points in $X$ by a factor of at most $c_X$ times $O(\log (|X\setminus S|))$. The caveat is that some of the latter distances may shrink. We call such a composition a {\em weak nested composition} (see Definition~\ref{def:nested-det}).\footnote{When the host metric is $\ell_1$, we also achieve expansion, obtaining a net distortion of at most $O(\log(|X\setminus S|))c_X$ over all pairs of points.}}

%\kristinEDIT{does not depend on the sizes of $X$ or $S$}{depends only logarithmically on the size of $X\setminus S$}. 

The concept of nested composition is similar in some ways to unions of embeddings studied previously in \cite{makarychev2016} and \cite{ostrovskii2022}. The goal in these works is to combine two embeddings over {\em disjoint} subsets $A$ and $B$ of a metric space into a single embedding over their union such that the distortion over the union is comparable to the distortions $c_A$ and $c_B$ of the given embeddings. When the host metric is $\ell_2$ or $\ell_\infty$, these works construct unions whose distortion is some constant factor times the {\em product}, $c_A\cdot c_B$, of the distortions of the given embeddings. Additionally, both the sets inherit the same distortion bound -- if, for example, $c_A\ll c_B$, it is not guaranteed that points in $A$ will retain similarly lower distortion in the composition. Extending these results to $\ell_p$ metrics for $p\not\in\{2,\infty\}$ is open. By contrast, our composition guarantees {\em the same} distortion $c_S$ as the given embedding on the ``inner" set of points $S$, and a bound on the expansion over the remaining points that is linear in $c_X$ (although the latter does not imply a bound the distortion over $X$ as the composed embedding may contract some pairs of points).

Our results are also reminiscent of local versus global guarantees for metric embeddings. In particular, Arora et al.~\cite{arora2012} ask: suppose that {\em every} subset of metric space $X$ of size at most $k$ admits a low distortion embedding into $\ell_1$, does $X$ also admit a low distortion embedding into $\ell_1$? Charikar et al.~\cite{charikar2010} show that this is indeed possible but that the distortion blows up by a factor of $\Theta(\log n/k)$. Our setting is slightly different in that we not only need every small set of size $k$ to be embeddable with low distortion, but we also need a good embedding for {\em one} set of size $n-k+1$.

%unions of embeddings (I still need to look more closely at this especially the twisted union ones because I'm not quite sure I understand what that means) and local/global tradeoffs \cite{abraham2009},\cite{arora2012},\cite{charikar2010},\cite{ostrovskii2022}, \cite{makarychev2016}

\paragraph{\bf A tradeoff between outlier set size and distortion.} Armed with a weak nested composition into $\ell_2$ space, we develop an SDP-rounding algorithm for obtaining bicriteria approximations for outlier embeddings into $\ell_2$. For a metric $(X,d_X)$ that admits a $(k,c)$-outlier embedding into $\ell_2$, and any given target distortion $c'=\gamma c$, {$\gamma\geq 1$}, our algorithm constructs a $(k',c')$-outlier embedding with at most $k'=O(\frac{\log^2k}{\gamma^2-1}\cdot k)$ outliers. 
% Here $\beta$ is the multiplicative factor incurred in the distortion of the nested composition we construct, and $\zeta$ is the worst-case distortion of embedding any size $k$ subset of $X$ into $\ell_2$. Both $\beta$ and $\zeta$ are $O(\log k)$, implying a polylogarithmic approximation for the number of outliers. Note further that our approximation factor depends inversely on $\zeta/c$ -- as the distortion of the non-outliers becomes closer and closer to the overall worst case distortion for $X$, it becomes easier to approximate the outlier set.\footnote{In contrast, \cite{Chubarian20}'s approximation factor for $k$ for embeddings into a line has a polynomial dependence on $c$.} 
Additionally, we achieve a tradeoff between the outlier set size $k'$ and the target distortion $c'$. Setting $c'=(1+\epsilon)c$ for a small $\epsilon>0$, for example, increases the outlier set size by an {$O(1/\epsilon)$} factor.

%\kristin{Should we include the $1/c^2$ term here? I noticed that the other paper had a dependence on $c^6$ they mentioned in their big O term. I need to think more on why our term is inversely proportional in poly $c$ and theirs is proportional in poly $c$. It seems to me that if you allow more slack in the distortion, your approximation should get better (eventually getting precise if $c$ gets past $O(\log n)$ since $k=0$ in that case), so the direction of our dependence doesn't seem unintuitive to me }

\paragraph{\bf Hardness of approximation.} Finally, we note that Sidiropolous \textit{et al.} \cite{sidiropolous17} showed that it is NP-hard to determine the size of the smallest outlier set such that the remaining metric is isometrically embeddable into $\ell_2^d$ for any fixed dimension $d>1$. Designing an outlier embedding into $\ell_2$ with arbitrary dimension is potentially an easier problem. Since we do not limit the dimension of the outlier embeddings we construct, we revisit and strengthen \cite{sidiropolous17}'s result along these lines. We show that NP-hardness continues to hold even for embeddings into $\ell_2$ without a specified dimension bound, and also when the given metric is the shortest path metric of an unweighted undirected graph. Our construction of a hard instance is arguably simpler than that of Sidiropolous \textit{et al.}, and is readily seen to extend to $\ell_p$ metrics for $p>1$. We present a separate, more involved, construction for $p=1$. As with Sidiropolous \textit{et al.}'s results we show that, under the unique games conjecture, it is also hard to obtain a $2-\epsilon$ approximation for the minimum outlier set size, for any $\epsilon>0$. 

\subsection*{Further related work}

\paragraph{Approximations for distortion.}
Much of the work on low distortion embeddings focuses on providing uniform bounds for embedding any given finite metric into a structured space.
Indyk and Matousek \cite{indyk2004} give an excellent overview of many of these results. Bourgain's Theorem \cite{bourgain1985,linial1994} shows that all finite metric spaces of size $n$ have an $O(\log n)$-distortion embedding into $\ell_p$-space for $p\geq 1$ and that a randomized version of such an embedding can be computed quickly. A derandomized such embedding can be computed in polynomial time with $\Theta(n^2)$ dimensions.

More closely related to our work, a number of papers study the objective of approximating the instance-specific minimum distortion for embedding into various host metrics. This includes, e.g., embeddings into constant dimensional Euclidean space \cite{badoiu2006,deberg2010,edmonds2010,matousek2008,sidiropulos2019}, the line  \cite{matousek2008,fellows2013,badoiu2005,nayyeri2015}, trees \cite{chepoi2012,badoiu2007}, and ultrametrics \cite{alon2008}. These works tend to use combinatorial arguments as they focus on low dimensional embeddings, whereas ours relies on an SDP formulation of the problem. 
% \kristin{maybe add some comments about hardness for low dimensional target spaces?}

\paragraph{Lipschitz extension.} Much of the prior work on Lipschitz extensions has largely focused on extending embeddings of small subsets of a metric $(X,d)$ into a target space. The Johnson-Lindenstrauss extension theorem shows that for any metric $(X,d)$ and Lipschitz embedding $\alpha_S:S\subseteq X\rightarrow \ell_2$ with $|S|=n$, there exists an embedding $\alpha:X\rightarrow \ell_2$ with Lipschitz constant at most $O(\sqrt{\log n})$ times the Lipschitz constant of the original embedding \cite{Johnson1984}. (Note that contraction may be arbitrary in this embedding, and $X$ does not have to be a finite space.) More recent work by Naor and Rabani \cite{naor2017} shows that in general, some Lipschitz extensions into Banach spaces require $\Omega(\sqrt{\log n})$ increase in the Lipschitz factor. Other work on Lipschitz extensions has largely focused on extending embeddings of subsets of $X$ in which all point-wise distances in $S$ are at least an $\epsilon$ fraction of the subspace's diameter (i.e. the distances between points in $S$ aren't too different from each other). 

In this paper, we focus on extending embeddings of subsets $S\subseteq X$ of relatively large size  compared to $X$ into $\ell_p$ spaces. In particular, we show that if $|X|=n$ and $|X\setminus S|=k$, then there exists a Lipschitz extension of any embedding of $S$ into $\ell_p$ that increases the Lipschitz constant by at most a factor of $O(\log k)$. Because we are considering outlier sets $X\setminus S$ that we expect to be small, this is generally going to give us a better approximation factor for our purposes. 
%However, using the results of Johnson and Lindenstrauss, we can actually adjust out outlier embedding results to give an embedding into $\ell_2$ with at most $O\left(\frac{\min(\log^2k,\log(n-k)}{\epsilon }\right)\cdot k$ outliers and at most $(1+\epsilon)c$ distortion, where $k$ is the optimal number of outliers for target distortion $c$. Because we expect $k$ to be small relative to $n$, we focus on the first part of this bound. \shuchi{This seems repetitive.}

% In particular, given a metric space $(X,d_X)$ and Banach space $(Y,||\cdot ||_Y)$, the optimal (non-instance specific) Lipschitz extension factor for a Lipschitz embedding $\alpha_S:S\subseteq X\rightarrow Y$ with $|S|=n$ is $\Theta(\sqrt{\log n})$ \cite{naor2017}. 

\paragraph{Outlier embeddings.}
The notion of outlier embeddings was first introduced by Sidiropolous \textit{et al.} in \cite{sidiropolous17}. In that paper, they showed that it is NP-hard to find the size of a minimum outlier set for embedding a metric into ultrametrics, tree metrics, or $\ell_2^d$ for constant $d$. Under the Unique Games Conjecture, it is also NP-hard to approximate these values to a factor better than $2$. On the algorithmic side, they gave polynomial time algorithms to $3,4,$ or $2$-approximate minimum outlier set size for isometric embeddings into ultrametrics, tree metrics, or $\ell_2^d$ for fixed constant $d$, respectively. The algorithm for $\ell_2^d$ embeddings is exponential in $d$, so $d$ cannot grow with the size of the input while remaining efficient. 

Sidiropolous \textit{et al.} also gave bi-criteria approximations for $\ell_\infty$ (i.e. additive) distortion. In particular, they give a polynomial time algorithm to find embeddings with at most $2k$ outliers and $O(\sqrt \delta)$ $\ell_\infty$-distortion when there exists an embedding of the metric with  at most $k$ outliers that has $\ell_\infty$-distortion at most $\delta$. The algorithm is polynomial in $k$, $\delta$, and $n$ but exponential in $d$ which is taken to be a constant.

Chubarian \textit{et al.} \cite{Chubarian20} expanded on the results of Sidiropolous \textit{et al.} by giving the first bicriteria approximation for minimum outlier sets with multiplicative distortion. In particular, they showed that given an unweighted graph metric and tuple $(k,c)$, there is a polynomial time algorithm that either correctly decides that there does not exist an embedding of the metric into the real line with at most $k$ outliers and at most $c$ distortion, or outputs an $(O(c^6k\log^{5/2}n),O(c^{13}))$-outlier embedding into the real line.

\paragraph{Embeddings with slack.}
A different notion of distortion that is robust to noise in the data was introduced by Abraham {\em et al.}~\cite{abraham2005}. In this {\em embeddings with slack} model, a budget of slack is applied to pairs of vertices in the metric space (as opposed to individual vertices, as in our model). An embedding of a metric space $(X,d_X)$ into another space $(Y,d_Y)$ has distortion $c$ with $\epsilon$-slack if all but an $\epsilon$ fraction of the distances are distorted by at most $c$. Abraham {\em et al.}~\cite{abraham2005} showed that there exists a polynomial time algorithm that finds an $O(\log \frac{1}{\epsilon})^{1/p}$-distortion embedding with $\epsilon$ slack for embeddings into $\ell_p$ for $p\geq 1,\epsilon>0$. Chan \textit{et al.} \cite{chan2006} showed that there is a polynomial time algorithm for embedding a metric $(V,d)$ of $n$ points into a spanner graph of at most $O(n)$ edges with $\epsilon$-slack and $O(\log \frac{1}{\epsilon})$ distortion. Lammersen \textit{et al.} \cite{lammersen2009} extended results in this topic to the streaming setting by giving an algorithm using poly-logarithmic space that computes embeddings with slack into finite metrics. From an algorithmic viewpoint, defining outliers in terms of edges versus nodes leads to very different optimization problems. Furthermore, an important difference between our work and these previous works on embeddings with slack is that we are interested in instance-specific approximations, whereas these latter works aim to find uniform bounds on distortion with slack that hold for all input metric spaces.

\hide{
\paragraph{Lipschitz extensions}
In proving correctness of our algorithm for approximating outlier sets, we show discuss a new concept we call nested embeddings. In a nested embedding we want to take an embedding $\alpha_X$ on some finite metric space $(X,d)$ and a lower distortion embedding $\alpha_S$ on $S\subseteq X$ and create a new embedding that still has the distortion of $\alpha_S$ on pairs of points in $S$ but does not have distortion much more than that of $\alpha_X$ on any pair of points. If we relax our guarantees to only limit the amount of expansion on pairs of points with at least one in $X\setminus S$ but allow arbitrary contraction, we get a concept very similar to Lipschitz extensions. In particular, a Lipschitz extension of a Lipschitz embedding $\alpha_S$ for $S\subseteq X$  is an embedding $\alpha$ of $X$ such that $\alpha(s)=\alpha_S(s)$ for all $s\in S$ but 
}

%% file: maincontent/prelims.tex
We begin by defining terms used in this paper and discussing our main results.

\subsection*{Outlier embeddings and distortion}

\begin{definition}\label{def:metricspace}
A \emph{metric space} is a pair $(X,d_X)$ such that $X$ is a set of elements we call \emph{points} or \emph{nodes} and $d_X:X\times X\rightarrow \reals_{\geq 0}$ is a function that has the following properties:

\begin{enumerate}
    \item For all $x,y\in X$, $d_X(x,y)=0$ if and only if $x=y$ 
    \item For all $x,y\in X$, $d_X(x,y)=d_X(y,x)$ 
    \item For all $x,y,z\in X$, $d_X(x,z)\leq d_X(x,y)+d_X(y,z)$ 
\end{enumerate}
\end{definition}

%We are particularly interested in maps between metric spaces, which we call embeddings and their corresponding distortion. 

In this paper we will focus on expanding embeddings from a given finite metric into $\ell_p$. 

\begin{definition}\label{def:distortion}
    An {expanding embedding} $\alpha:X\rightarrow Y$ of a metric space $(X,d_X)$ into another metric space $(Y,d_Y)$ has \emph{distortion} $c\geq 1$ if %there exists a real number $c\st$ such that 
    for all $u,v\in X$:
    \begin{align*}
        d_X(u,v) \leq d_Y(\alpha(x),\alpha(y)) \leq c \cdot d_X(u,v)
    \end{align*}
\end{definition}

%Further, the multiplicative, or bi-Lipschitz distortion of a more general embedding is defined as follows.

%\begin{definition}\label{def:distortion}
%    An { embedding} $\alpha:X\rightarrow Y$ of a metric space $(X,d_X)$ into another metric space $(Y,d_Y)$ has \emph{distortion} $c\geq 1$ if there exists a constant $a$ such that 
    %there exists a real number $c\st$ such that 
%    for all $u,v\in X$:
%    \begin{align*}
%        d_X(u,v) \leq a\cdot d_Y(\alpha(x),\alpha(y)) \leq c \cdot d_X(u,v)
%    \end{align*}
%\end{definition}

%An embedding will be called \emph{expanding} with distortion $c$ if the above holds for $c\st=1$. 

% Note that existence of any $c$-distortion embedding for $(X,d_X)$ into $(Y,d_Y)$ when $(Y,d_Y)$ is an $\ell_p$ space implies existence of a $c$-distortion embedding $\alpha'$ such that $d_X(u,v)\leq d_Y(\alpha'(u),\alpha'(v))$ for all $u,v\in X$ (by multiplying all indices by $1/c\st$). Since we are 

%Additionally, we may be interested in embeddings of subsets of nodes of a metric space, or outlier embeddings as defined by 
Following \cite{Chubarian20} and \cite{sidiropolous17}, we consider so-called {\em outlier embeddings} that embed all but a small set of outliers from the given metric into the host space. 

\begin{definition}\label{def:kcembed}
    An embedding $\alpha:X\rightarrow Y$ of a metric space $(X,d_X)$ into another metric space $(Y,d_Y)$ is a \emph{$(k,c)$-outlier embedding} if there exists $K\subseteq X$ such that $|K|\leq k$ and $\alpha|_{X\setminus K}$ (the restriction of $\alpha$ to the domain $X\setminus K$) is an embedding of $(X\setminus K,d|_{X\setminus K})$ with distortion at most $c$. 
\end{definition}

\subsection*{Nested compositions and Lipschitz extensions}

A main component of our approach is showing that a low-distortion embedding of a subset of the given metric space into some $\ell_p$ space can be extended into an embedding of the entire metric with small expansion. 

\begin{definition}\label{def:lip-ext}
    Let $(X,d_X)$ and $(Y,d_Y)$ be two metric spaces and $\alpha_S:S\subseteq X\rightarrow Y$ be an embedding with Lipschitz constant at most $L$. Then $\alpha:X\rightarrow Y$ is a \emph{Lipschitz extension} of $\alpha_S$ with \emph{extension factor} $g(|S|,|X|)$ if for all $x\in S$, $\alpha(x)=\alpha_S(x)$ and for all $x,y\in X$, 
    \begin{align}
        d_Y(\alpha(x),\alpha(y)) &\leq g(|S|, |X|)\cdot L\cdot d_X(x,y). \label{eq:lip-ext}
    \end{align}
\end{definition}

We introduce a new variant of Lipschitz extension called nested composition, which focuses on parameters of interest for us, and aims to preserve both the expansion and contraction of the embedding. In particular, we start with {\em expanding} nested embeddings of $S$ and $X$ into $Y$ with distortions $c_S$ and $c_X$ respectively. Our goal is to produce a single expanding embedding that preserves the smaller distortion $c_S$ over pairs of points in $S$, and bounds the distortion over $X$ by $c_X$ times some function $g$ of the size of $X\setminus S$. Importantly, the factor $g$ depends only on the size of $X\setminus S$, and not on the size of $S$, that may be much larger. The weak variation of this notion is similar to Lipschitz extension in that it will still allow arbitrary contraction over $X$, but it will no longer require that the exact points of the original embedding be preserved.

\begin{definition}[Composition of nested embeddings]\label{def:nested-det}
    Let $(X,d_X)$ and $(Y,d_Y)$ be two metric spaces and $g:[0,\infty)^2\times \mathbb{N}\rightarrow [1,\infty)$.
    %, let $S\subseteq X$, and $\beta\ge 1$.
    %let $f:[1,\infty)\rightarrow [1,\infty)$ 
    %and $g:[1,\infty)\rightarrow [1,\infty)$. 
    A weak $g$-\emph{nested composition} is an algorithm that, given a set $S\subseteq X$ with $k:=|X\setminus S|$, and two expanding embeddings, $\alpha_S:S\rightarrow Y$ with distortion $c_S$ and $\alpha_X:X\rightarrow Y$ with distortion $c_X\geq c_S$, returns an embedding $\alpha:X\rightarrow Y$ such that, 
    %there exists constants $a,a'$ such that, 
    \begin{align}
        \text{for all } u,v\in S,\;\; d_X(u,v)\leq d_Y(\alpha(u),\alpha(v)) \leq c_S\cdot d_X(u,v), \label{eq:nested-constraint1}
    \end{align}
    %and for all $u,v\in X$,
    \begin{align}
        \text{and, for all } u,v\in X,\;\;\;\;\;\;\;\;\;\;\;\;\;\;\;\;\;\;\;\;  d_Y(\alpha(u),\alpha(v))\leq g(c_S,c_X,k) \cdot d_X(u,v). \label{eq:nested-constraint2}
    \end{align}
%    When $g$ is the linear functions $f(x)=\beta x$ and $g(x)=\beta x$ for some constant $\beta\ge 1$, we call $\alpha$ a $\beta$-nested composition. 
We say that it is a nested composition if the embedding $\alpha$ is additionally an expanding embedding. That is, %the inequality $d_X(u,v) \leq  d_Y(\alpha(u),\alpha(v))$ holds for all $u,v\in X$. In other words, we have
    \begin{align}
        \text{for all } u,v\in S,\;\; d_X(u,v)\leq d_Y(\alpha(u),\alpha(v)) \leq c_S\cdot d_X(u,v), \label{eq:nested-constraint1}
    \end{align}
    %and for all $u,v\in X$,
    \begin{align}
        \text{and, for all } u,v\in X,\;\; d_X(u,v) \leq  d_Y(\alpha(u),\alpha(v))\leq g(c_S,c_X,k) \cdot d_X(u,v). \label{eq:nested-constraint2}
    \end{align}
\end{definition}

% \shuchi{Let's discuss how to parameterize $g$.}

We use randomness in our construction of nested compositions. For a randomized Lipschitz extension, we require that $\alpha(x)=\alpha_S(x)$ for all $x\in S$ for any possible $\alpha$ in the distribution, and that expansion is bounded in expectation over the randomness in the construction. For a randomized nested composition, 
all contraction bounds should be satisfied with probability $1$ and expansion bounds in expectation.

%but instead of Equation \ref{eq:lip-ext}, we will require
%\begin{align*}
%    \mathop{E}_\alpha[d_Y(\alpha(x),\alpha(y))] &\leq f(|X\setminus S|)\cdot L\cdot d_X(x,y),
%\end{align*}
%where the expectation is taken over the distribution of possible embeddings $\alpha$. 

In our work, we will show that there is an efficient algorithm that finds a randomized Lipschitz extension (or weak nested composition) with extension factor $O(\log |X\setminus S|)$ where $(Y,d_Y)$ is \textit{any} target metric space. However, the application to outlier embeddings requires the existence of a single deterministic extension/composition, so our deterministic extension requires that the target metric be a Banach space (such as an $\ell_p$ space).

\hide{
\kristin{I'm not sure if we need the below paragraph and the previous one? if we do keep them, we should probably adjust them to clarify that the distribution needs to be finite. I tried to address this issue in the discussion on Lipschitz extensions above 

original: For nested compositions where the host space $Y$ is the $\ell_p$ metric, we will consider randomized embeddings. For a randomized embedding $\alpha$ into $\ell_p$ space, the distortion is measured relative to the expectation of the $p$th power of the $\ell_p$ norm. In particular, we will require the following version of the constraint \eqref{eq:nested-constraint2} above:
    \begin{align*}
    \expect\left[\distp{\alpha(x)}{\alpha(y)}^p\right]^{1/p} \leq g(k) c_X\cdot d_X(u,v).
    \end{align*}
With this definition, it is easy to see that the existence of a randomized (weak) nested embedding into $\ell_p$ implies the existence of a deterministic (weak) nested embedding into $\ell_p$ with the same parameters. In particular, a scaled concatenation (concatenation of vectors with scalars applied to them) of the embeddings in the support of the claimed distribution provides such an embedding.}
}

\subsection*{Main results}

\paragraph{\bf Nested compositions.} Our main technical result shows that we can efficiently construct randomized weak $O(\log k)$-nested compositions when the host space $Y$ is an $\ell_p$ metric. For $p=2$, the case of interest for us, we can even efficiently construct a {\em deterministic} weak $O(\log k)$-nested composition into $\ell_2$ space by solving an appropriate semi-definite program. Finally, for $p=1$, our construction ensures expansion and satisfies both inequalities in constraint~\eqref{eq:nested-constraint2}. We leave open the question of constructing (strong) nested compositions satisfying expansion for $p>1$.

%that we can efficiently construct $O(1)$-weak nested compositions when the host space $Y$ is an $\ell_p$ metric.
\begin{thm}\label{thm:extension}
Let $(X,d_X)$ be any finite metric and $(Y,d_Y)$ be any Banach space. Let $\alpha_S:X\rightarrow Y$ be any Lipschitz embedding of $S\subseteq X$ into $Y$ with $k:=|X\setminus S|$. Then there exists a $125H_k$-Lipschitz extension (and thus a weak $125H_kc_S$-nested composition) from $X$ into $Y$, where $H_k$ is the $k$th Harmonic number.
\end{thm}

\begin{thm}\label{thm:nested-comp}
Let $(X,d_X)$ be any finite metric. Then there exists 
% a $\factor$-weak nested composition from $X$ into the $\ell_p$ metric, where $H_k$ is the $k$th Harmonic number.
% %set of embeddings $\set{\alpha_T:T\rightarrow \lspace{p}}_{T\subseteq X, |T|=t}$ with $t\geq |X|-|S|+1$,  there exists a $\factor$-nested composition of embeddings $\alpha_S$ and  $\set{\alpha_T}_{T\subseteq X, |T|=t}$.
%     Further, for $p=1$, there exists a
a $\factor c_X$-nested composition from $X$ into $\ell_1$. 
    % \kristin{should we define $k$ here/metion something about the smaller embedding?}
\end{thm}
% \shuchi{Should that be $\factor$ times $c_X$?}

%\kristin{As I'm working on the sdp section I'm running into the issue of needing to reference the algorithm in the next section of needing to assume existence where this bound holds for everyone, so I was thinking it would be nice to be able to reference the following corollary below - do you think this an appropriate place for it? I added something similar to follow the next theorem} 

%The following corollary asserts that in fact under the conditions of Theorem~\ref{thm:nested-comp}, a {\em deterministic} nested composition with the bounded expansion exists.
% desired distortion bounds exists. 
%In particular, a scaled concatenation of the embeddings in the support of the distribution produced by the above theorem provides such an embedding. 

%\begin{cor}\label{cor:exists-embedding-weak}
%    Let $(X,d_X)$ be any finite metric with $S\subseteq X$ and let $p\ge 1$. Then, given any expanding embeddings $\alpha_S:S\rightarrow \lspace{p}$ and $\alpha_X:X\rightarrow \lspace{p}$, there exists a {\em weak} $\factor$-nested composition of $\alpha_S$ and $\alpha_X$ where $k=|X\setminus S|+1$. 
    
%    Further, if $p=1$, there exists a {\em strong}  $\factor$-nested composition of $\alpha_S$ and $\alpha_X$ where $k=|X\setminus S|+1$.
%\end{cor}

For metric spaces $(X, d_X)$ where the distortion of the embedding into $\ell_1$ depends on the size of the subset being embedded, we can in fact obtain a stronger guarantee -- the distortion of the composition depends only on the size of the outlier set $k=|X\setminus S|$ and not on the size of the entire space $X$. In particular, we can replace the quantity $c_X$ in constraint~\eqref{eq:nested-constraint2} by the worst case distortion from embedding any subset of size $k+1$ into the host metric.

\begin{thm}\label{thm:nested-comp-strong}
Let $(X,d_X)$ be any finite metric. Suppose that for $k\in\ZZ^+$ every subset of $X$ of size $k+1$ can be embedded into $\ell_1$ with distortion $\zeta_k$. Then, there exists a $\factor \zeta_k$-nested composition from $X$ into $\ell_1$.
%a weak nested composition that given a subset $S\subseteq X$ with $|X\setminus S|\le k$, and an expanding embedding $\alpha_S$ from $S$ into $\ell_1$ with distortion $c_S$, returns an embedding $\alpha:X\rightarrow \ell_1$ such that, 
    %there exists constants $a,a'$ such that, 
%    \begin{align*}
%        \text{for all } u,v\in S,\;\; & & d_X(u,v)\leq \distl{1}{\alpha(x)}{\alpha(y)} & \leq c_S\cdot d_X(u,v), \\
%        \text{and, for all } u,v\in X,\;\; & & d_X(u,v)\leq \distl{1}{\alpha(x)}{\alpha(y)} & \leq \factor \zeta_k\cdot d_X(u,v).    \end{align*}
    %\kristin{Is the  second one supposed to be 191?}
%achieves a distortion of $ c_S$ over pairs of points in $S$ and an expected distortion of at most $381 H_{k+1}\cdot  \zeta_k$ over the remaining pairs of points in $X$.
\end{thm}

\paragraph{\bf Outlier embeddings.} With these results in hand, we obtain the following bicriteria approximation for outlier embeddings from finite metrics into $\lspace{2}$.

%First, we will apply the Lipschitz extension results from Theorem \ref{thm:extension} to show the following theorem.

\begin{thm}\label{thm:overall-approx}
Let $(X,d_X)$ be a metric space that admits a $(k,c)$-outlier embedding.  Then there exists a polynomial time algorithm $\cA$ that, for any $\gamma>1$, finds a subset $K\subseteq X$ and an embedding $\alpha:X\setminus K\rightarrow \lspace{2}$ such that $\alpha$ has distortion at most $\gamma c$, and
    \[|K|\le 2\frac{(125\cdot H_k)^2+\gamma^2}{\gamma^2-1} k\]
    %\[|K|\le \frac{(\finfact \cdot \zeta)^2}{(\finfact \cdot \zeta)^2-((\desdist)^2-c^2)} \cdot  \frac{(\finfact\cdot \zeta)^2}{(\desdist)^2-c^2}\cdot k.\]

    Choosing $\gamma=1+\epsilon$ for $\epsilon\in (0,1]$, in particular, provides an 
    %$(O(kH_k^2\cdot\zeta^2), 2c)$-outlier
    {$\left(O(\frac{\log^2k}{\epsilon }k), (1+\epsilon)c\right)$}-outlier embedding from $X$ into $\lspace{2}$. 
% \kristin{do we need to drop the $c^2$ term in the denominator? it seems like we may have been missing this part from before? the $c^2$ term maybe should have only been there when we were offsetting with the $g(k)$ think - the $c$ probably should have appeared in the $g(k)$ term?}
\end{thm}

\paragraph{\bf Hardness of approximation.} Finally, we provide a strengthening of Sidiropolous \textit{et al.} \cite{sidiropolous17}'s hardness result for outlier embeddings, showing that it is NP-hard to determine the size of the smallest outlier set such that the remaining metric is isometrically embeddable into $\ell_2$ even when the dimension of the embedding is unrestricted. As with Sidiropolous \textit{et al.}'s results, under the unique games conjecture, it is also hard to obtain a $2-\epsilon$ approximation for the minimum outlier set size, for any $\epsilon>0$. Furthermore, our construction achieves two other properties that \cite{sidiropolous17}'s doesn't: (1) Our hardness results apply also to shortest path metrics over unweighted undirected graphs. (2) We show that the hardness result holds for embedding into the $\ell_p$ metric for any $p\ge 1$.

\begin{thm}\label{thm:np-hard}
    Let $(X,d)$ be the distance metric for an unweighted undirected graph $G=(V,E)$. then, given $(X,d,k)$ it is NP-hard to decide if there exists a subset $K\subseteq X$ with $|K|=k$ such that $(X\setminus K,d|_{X\setminus K})$ is isometrically embeddable into $\lspace{p}$ for any finite integer $p\geq 1$. 
    
    Under the unique games conjecture, it is NP-hard to find a $2-\epsilon$ approximation for the minimum such $k$, for any $\epsilon>0$.
\end{thm}

% \shuchi{\\
% 1. Distortion

% 2. $(k,\alpha)$ outlier embedding

% 3. Composition of nested embeddings

% 4. Statements of the two main theorems: (1) Guarantee for composition of nested embeddings; (2) Bi-criteria approx for outlier embeddings.
% }

%% file: maincontent/sdp.tex
\label{sec:sdp}
In this section we will prove Theorem \ref{thm:overall-approx}.
%and Corollary \ref{cor:overall-approx} assuming that Theorems~\ref{thm:nested-comp} and \ref{thm:nested-comp-strong} hold.
%Corollaries \ref{cor:exists-embedding-weak} and  \ref{cor:exists-embedding} hold. 
% \kristin{to do: modify to swap $c^2$ for $9c^2$ }
We begin with a semi-definite programming formulation for constructing an outlier embedding into $\ell_2$. In the absence of outliers, the optimal embedding of any finite metric into $\ell_2$ can be found using an SDP. In particular, for a given such metric $(X,d)$, let $\vec{v_x}$ for $x\in X$ denote the mapping of $x$ into $\ell_2$. Then, the constraint $d^2(x,y) \leq ||\vec{v_x}-\vec{v_y}||^2\leq c^2\cdot d^2(x,y)$ ensures that the distance between points $x$ and $y$ is distorted by a factor of at most $c$. The challenge is to incorporate outliers into this formulation.

% \shuchi{Removed the shrink factor above, as here we're talking about no outliers.}

%There exists a standard method to solve for the minimum distortion needed to embed a finite metric into $\lspace{2}$, using the concept of semi-definite programming (SDP) [cite]. We will work from a variation of the program used in this technique. 
% \shuchi{Citation and explanation not needed as this is straightforward.}

Consider the finite metric space $(X,d)$, and suppose that there exists a $(k,c)$-outlier embedding from $(X,d)$ into Euclidean space for some integer $k>0$ and real number $c\geq 1$. We use the vector $\vec{v_x}$ for $x\in X$ to denote the mapping of $x$ into $\ell_2$, and $\delta_x\in [0,1]$ as an indicator for whether $x$ is an outlier. %For brevity, we write $\delta_x:=\langle \vec{\delta_x},\vec{\delta_x}\rangle$. 
We then construct the following SDP:

\begin{align}
\min_{\delta, \vec{v}} \quad & \sum_{x\in X}\delta_x & & \label{eqn:sdp1}\tag{Outlier SDP} \\ 
%\begin{aligned}
\textup{s.t.}\quad & \forall x,y\in X: & & (1-\delta_x-\delta_y)\cdot  d^2(x,y) \leq ||\vec{v_x}-\vec{v_y}||^2\leq ( c^2+(\delta_x+\delta_y) f(k))\cdot d^2(x,y), \label{eqn:sdp2} \\
\notag & \forall x\in X: & & \delta_x \in [0,1].
%\end{aligned}
\end{align}

Here $f(k)$ is a function to be determined. We claim that for an appropriate choice of $f$, this SDP is a relaxation for the problem of minimizing the outlier set size such that all non-outlier elements in $X$ can be embed into $\ell_2$ with distortion $c$. In particular, given a $(k,c)$-outlier embedding from $(X,d)$ into $\ell_2$, we can find a feasible solution for the SDP with value at most $k$. For Theorem \ref{thm:overall-approx}, it will be sufficient to set $f(k)=(125\cdot c\cdot H_k)^2$. 
% For {Theorem \ref{thm:l2embed}}, it will be sufficient to set $f(k):=(g(k)\zeta)^2$ where $\zeta$ is the distortion of an outlier-free embedding of $X$ into $\ell_2$ and $g$ is the function corresponding to the weak nested composition guaranteed in the statement of {Theorem \ref{thm:l2embed}}. 

\begin{lem} \label{lem:sdp-sol-cx}
Let $(X,d)$ be a finite metric space with expanding embedding $\alpha:S\rightarrow \ell_2$ of distortion at most $c$ for $S\subseteq X$. Then if there exists a $(g(k)/c)$-Lipschitz extension (or a weak $g(k)$-nested embedding) of $\alpha$ with $k=|X\setminus S|$, \eqref{eqn:sdp1} with $f(k):=g(k)^2$ has a feasible solution with value equal to $k$.
% .eqref{eqn:sdp1} with $f(k):=(125c\cdot H_k)^2$ has a feasible solution with value equal to $k$.
% \begin{itemize}
%     \item \eqref{eqn:sdp1} with $f(k):=(125c\cdot H_k)^2$ has a feasible solution with value equal to $k$
%     \item If $(X,d)$ admits a $(0,\zeta)$-outlier embedding as well as a $(k, c)$-outlier embedding into $\ell_2$ for $c\le\zeta$. If there exists a weak $g$-nested composition from $(X,d)$ into $\ell_2$, then \eqref{eqn:sdp1} with {$f(k):=(g(k)\zeta)^2$} has a feasible solution with value equal to $k$. 
% \end{itemize}
\end{lem}

% \shuchi{What are the two cases of the lemma? Also, why not state the general version that holds for any given weak nested composition?}

\begin{proof}
Let  $K=X\setminus S$. We will  construct a feasible solution for \eqref{eqn:sdp1}. Set $\delta_x$ to $1$ if $x\in K$ and $0$ otherwise. Set $\vec{v_x}$ to $\alpha(x)$. Clearly $\sum_{x\in X}\delta_x=k$. We show that this setting of the variables satisfies the given constraints. 

Consider a constraint corresponding to $x,y\in X\setminus K$. Then we have that $(1-\delta_x-\delta_y)\cdot d(x,y)^2 = d(x,y)^2\leq ||\alpha(x)-\alpha(y)||_2^2\leq c^2\cdot d(x,y)^2 = (c^2+(\delta_x+\delta_y)f(k))\cdot d(x,y)^2$ by the facts we have already asserted about $\alpha$. 

Next, consider a constraint corresponding to $x\in X,y\in K$. In this case, $\delta_y=1$ and $(1-\delta_x-\delta_y)\le 0$, so the first inequality in \eqref{eqn:sdp2} is satisfied. On the other hand, by the definition of $f$ and $\alpha$, $||\alpha(x)-\alpha(y)||_2^2\leq f(k)\cdot (d(x,y))^2\leq (c^2+(\delta_x+\delta_y)f(k))\cdot (d(x,y))^2$, which gives us the second inequality. Thus, the solution $(\delta,\vec{v})$ is a feasible solution with value $k$.
%We get by definition of $\beta$ that $(1-\delta_x-\delta_y)\cdot d(x,y)^2\leq ||\beta(x)-\beta(y)||_2^2\leq f(c,k)\cdot (d(x,y))^2$. However, note that we must also have $\delta_y$ set to $1$ and distances are always non-negative, so this satisfies the corresponding constraint as well. Thus, we have a valid solution to the SDP and an upper bound on its minimum value.
\end{proof}

Observe that $f$ is a function of $k$, and so in order to set up and solve the SDP, we require knowing the value of the parameter $k$. We can get around this by setting $k=1, 2, \cdots$, and so on until we find the smallest value of $k$ for which the SDP with parameter $f(k)$ has a feasible solution of value at most $k$. Rounding this solution then gives the desired theorem.

We are now ready to prove Theorem~\ref{thm:overall-approx}.

\begin{proof}[Proof of Theorem \ref{thm:overall-approx}]
Suppose that $(X,d)$  a $(k,c)$-outlier embedding into $\ell_2$.
By Lemma \ref{lem:sdp-sol-cx}, there exists a solution to the SDP \ref{eqn:sdp1} with value at most $k$, which we can find efficiently by solving the SDP. Let $\{(\vec{v_x}, \delta_x)\}_{x\in X}$ denote such a solution. Let $\Delta$ be a parameter to be defined. Define $\alpha(x)\mapsto \frac{1}{\sqrt{1-2\Delta}}\vec{v_x}$, and $K \mapsto \{x: \delta_x\ge \Delta\}$. We claim that for an appropriate choice of $\Delta$, the solution $(K,\alpha)$ satisfies the requirements of the theorem.  

In particular, we note that $|K|\le (\sum_{x\in X}\delta_x)/\Delta \le k/\Delta$. To bound the distortion of $\alpha$ restricted to $X\setminus K$ by $\gamma c$, let us consider some pair of points $x,y\in X\setminus K$, and recall that we have $\delta_x, \delta_y<\Delta$. Then, substituting $\vec{v_x} = \sqrt{1-2\Delta}\alpha(x)$ in \eqref{eqn:sdp2} gives us:
\[(1-2\Delta)\cdot d^2(x,y) \leq (1-2\Delta) ||\alpha(x)-\alpha(y)||_2^2\leq (c^2+2\Delta f(k))\cdot d(x,y)^2
\]
The first inequality implies expansion. The second provides an upper bound on the distortion of: 
\[  \frac{c^2+2\Delta f(k)}{1-2\Delta}\]
Setting this quantity equal to $\gamma^2 c^2$ and solving for $\Delta$ gives us $\Delta = \frac{c^2(\gamma^2-1)}{2f(k)+2c^2\gamma^2}$ and $|K|\le \frac{2f(k)+2c^2\gamma^2}{c^2(\gamma^2-1)}k$. 
\end{proof}

% The proof of Theorem \ref{thm:l2embed} is identical to that of Theorem \ref{thm:overall-approx}, but we apply case 2 of Lemma \ref{lem:sdp-sol-cx} instead of case 1.

%% file: maincontent/nested_embed.tex
% In the previous subsection, we assumed that $k$ is an upper bound on the value of the given SDP. 
%  In this seciton, we show that this is the case, in particular showing that if $K$ is a minimum outlier set for embedding a metric $(X,d)$ into $\ell_p$ space, then there exists an embedding $\alpha$ such that $d_{\ell_p}(\alpha(u),\alpha(v))=d(u,v)$ for all $u,v\in X-K$, and for any $u\in V,v\in K$, we have distortion at most $c\log n$ for some constant $c$. In other words, $ d_{\ell_p}(\alpha(u),\alpha(v))\leq (c\log n )\cdot d(u,v)$. 

In this section, we will show existence of Lipschitz extensions of embeddings into $\ell_p,p\geq 1$ space for arbitrary $\ell_p$. To do this, we first give a randomized Lipschitz extension of an embedding $\alpha_S:S\subseteq X\rightarrow Y$ for $(X,d)$ and $(Y,d_Y)$ being metric spaces with $|X\setminus S|$ finite.\footnote{Note that line 4 of the randomized algorithm given here is inspired by the algorithm given by \cite{trees} for randomized embeddings of metrics into trees. In this case however, we want to group together nodes that are close to each other relative to the ``good" set of nodes, as this will allow us to place nodes at the same spot as good nodes that they are relatively close to. } In the case that $(Y,d_Y)$ is an $\ell_p$ metric, we show that because the output of the algorithm is over a distribution of finite support, averaging over the distribution will result in an embedding in which no distance is stretched too much. In this step, we fundamentally use the fact that the embedding is a Lipschitz extension and not just a weak nested embedding, as the averaging leaves the embeddings of nodes in $S$ the same, whereas averaging in general may cause contraction compared to the expectation itself.
% In this section, we will present an algorithm for composing nested embeddings into $\ell_p$ spaces, thereby proving Theorems \ref{thm:nested-comp} and \ref{thm:nested-comp-strong}. 
Our algorithm for the randomized Lipschitz extension, Algorithm \ref{alg:extend}, is formally specified below.

% \shuchi{Don't we need $p\ge 1$?}

\input{maincontent/alg1}

Let $c_S$ denote the distortion of $\alpha_S$. We now show that if $\alpha$ is the output of Algorithm \ref{alg:extend} on input $((X,d),S, \alpha_S,\rangwidth)$ with $\rangwidth=2$, the distortion between elements in $S$ is at most $c_S$ and all other distortion is at most $125\subdist\cdot H_k$  in expectation. 
% Our argument is broken into two parts: Lemma~\ref{lem:expanding} shows that the embedding $\alpha$ has low contraction; Lemma~\ref{lem:expansion-weak} bounds the amount by which every distance expands. 

The following lemma states our expansion bounds; we  prove it in the following subsection. Throughout these arguments we assume that $\rangwidth=2$, although it is possible to obtain slightly better distortion bounds by choosing a value for $\rangwidth$ carefully. We present general versions of the lemmas, exhibiting the dependence of the bounds on $\rangwidth$ in Appendix \ref{sec:full_constants}. Section \ref{sec:improvements} gives a deterministic construction of a Lipschitz extension and thus proves Theorem \ref{thm:extension}.
% Theorem~\ref{thm:nested-comp} follows in a straightforward manner from these lemmas; Section~\ref{sec:improvements} contains a proof of this theorem as well as a proof of the strengthened composition result -- Theorem~\ref{thm:nested-comp-strong}.

%Finally, in Section~\ref{sec:improvements}, we prove the strengthened versions of our composition results: Theorem~\ref{thm:nested-comp-strong} and Corollary~\ref{cor:exists-embedding}.

% \begin{lem}\label{lem:expanding}
% Let $\alpha\from$ Algorithm \ref{alg:extend}$((X,d),S,p,\alpha_S,\alpha_X,\rangwidth)$ with $\rangwidth=2$. Then for all $x,y\in X$, we have $\shrink\cdot \distp{\alpha(x)}{\alpha(y)}\geq d(x,y)$. 
% Furthermore, for $x,y\in S$, we have $\distp{\alpha(x)}{\alpha(y)}\geq d(x,y)$. 
% \end{lem}

\begin{lem}\label{lem:expansion-weak}
Let $\alpha\from$ Algorithm \ref{alg:extend}$((X,d),S,p,\alpha_S,\alpha_X,\rangwidth)$ with $\rangwidth=2$. Then we have the following bounds on the expansion for each pair $x,y\in X$:
\begin{enumerate}[(a)]
    \item \label{lem:not-outlier-weak} If $x,y\in S$, then $d_Y(\alpha(x),\alpha(y))\leq  c_S\cdot d(x,y)$.
    % \item \label{lem:sameset} If $x,y\in K_i$, then  $d_Y(\alpha(x),\alpha(y)\leq c_S\cdot d(x,y)$.
    \item \label{lem:ui-weak} If $x\in S,y\in X\setminus S$, then $d_Y(\alpha(x),\alpha(y))\leq 10\subdist \cdot d(x,y)$
    \item \label{lem:farxy-weak} If $x,y\in X\setminus S$
    { and $d(x,\gamma(x))\leq 2\cdot d(x,y)$ for $\gamma$ as defined in line 2 of the algorithm}, then $d_Y(\alpha(x),\alpha(y))\leq 50\subdist \cdot d(x,y)$.
    % \kristin{don't we need a condition on their relative distance for this one? otherwise it looks odd to have d and e. Also should we say expected expansion at the top?}
    \item \label{lem:closexy-weak} If $x,y\in X\setminus S$ { and $d(x,\neigh(x)),d(y,\neigh(y))> 2\cdot d(x,y)$ for $\neigh$ as defined in line 2 of the algorithm}, then $E_\alpha[d_Y(\alpha(x),\alpha(y))]\leq 125\subdist \cdot d(x,y)$.
\end{enumerate}
\end{lem}

% \shuchi{Should we remove case b? It is captured by case e.}

%The summary of the results of Lemmas \ref{lem:not-outlier-weak}, \ref{lem:sameset}, \ref{lem:ui-weak}, \ref{lem:farxy-weak}, and \ref{lem:closexy-weak}  proving this result appears in Table \ref{tab:distortion}. In the following lemmas in this section, we will assume that $\tau=2$. More general versions of these lemmas appear in Appendix \ref{sec:full_constants}. 

\subsection{Proofs of expansion bounds}

The bounds in Lemma~\ref{lem:expansion-weak} are summarized in Table \ref{tab:distortion}. We will prove each statement separately.

\input{maincontent/distortion_table}

%\begin{lem}   \label{lem:not-outlier-weak}
%Let $\alpha\from$ Algorithm %\ref{alg:extend}$((X,d),S,p,\alpha_S,\alpha_X,\randrangtwo)$ and $u,v\in S$. Then $\distp{\alpha(u)}{\alpha(v))}\leq \shrink\cdot c_S\cdot d(u,v)$.
%\end{lem}

Lemma \ref{lem:expansion-weak}~\ref{lem:not-outlier-weak} is automatically true by definition of $\alpha$. 
% Lemma \ref{lem:expansion-weak}~\eqref{lem:sameset} is also true because if $x,y\in K_i$ for a fixed $i$, then they are mapped to the same point and thus their distance is at  most $0\leq c_S\cdot d(x,y)$.

\begin{figure}
    \centering
    \begin{minipage}{.4\textwidth}
        \input{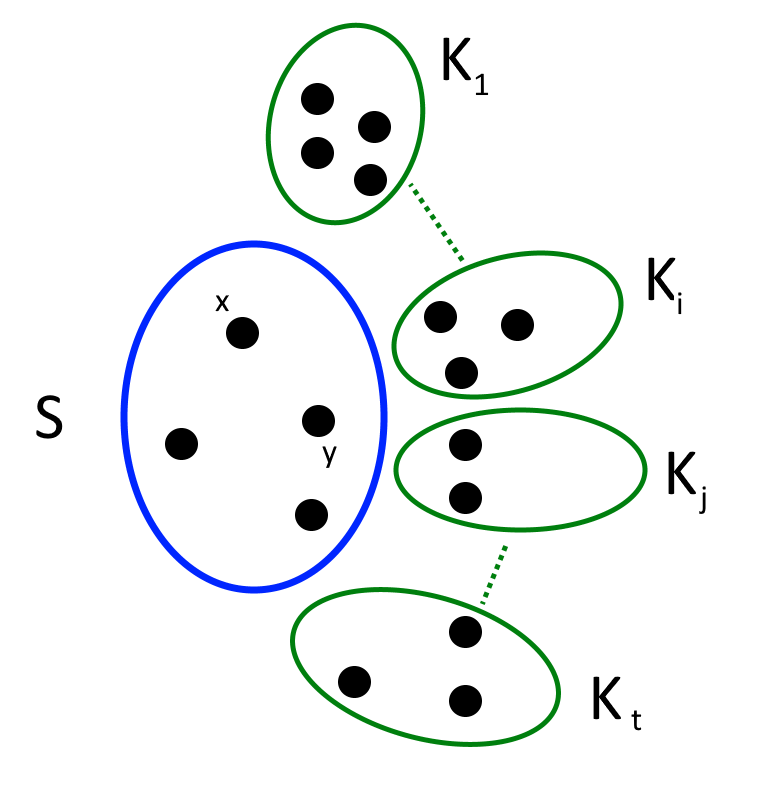} 
    \end{minipage}
    \begin{minipage}{.4\textwidth}
        \input{figures/both_in_ki}
    \end{minipage}
    
    \caption{Left: Visualization of nodes referenced in Lemma~\ref{lem:expansion-weak}~\eqref{lem:not-outlier-weak} and Lemma~\ref{lem:expansion};~\eqref{lem:not-outlier-weak} Right: Visualization of nodes referenced in Lemma~\ref{lem:expansion}~\eqref{lem:sameset}  }
    \label{fig:both_in_s}
\end{figure}

%\begin{lem}\label{lem:sameset}
%    Let $\alpha\from$ Algorithm \ref{alg:extend}$((X,d),S,p,\alpha_S,\alpha_X,\randrangtwo)$. Consider $x,y\in K_i$  for some $K_i$ as defined in line 7 of the algorithm. Then $\distp{\alpha(x)}{\alpha(y))}\leq \shrink \cdot \origdist \cdot d(x,y)$. 
%\end{lem}

Now consider the distortion between outliers and non-outliers (i.e. nodes in $S$ compared to nodes in $K$).

\begin{proofof}{Lemma~\ref{lem:expansion-weak}~\eqref{lem:ui-weak}}
    Let $i$ be the index of $x$'s cluster, that is, $x\in K_i$. First, suppose that $x=u_i$, that is, 
    %First, consider the case that $\pi(x)=i$ and $x\in K_i$ (i.e. 
    $x$ is the center of the cluster it's in. Then we get that $\alpha_S(x)=\alpha_S(\gamma(x))$. Clearly this implies that if $y=\gamma(x)$, the distortion is at most $c_S$. Otherwise, we have $d(y,x)\geq d(\gamma(x),x)$, so we get 
    \begin{align*}
        d_Y(\alpha(x),\alpha(y)) &= d_Y(\alpha_S(\gamma(x)),\alpha_S(y)) \\
        &\leq c_S\cdot d(y,\gamma(x)) \\
        &\leq c_S\cdot d(y,x)+c_S\cdot d(x,\gamma(x)) \\
        &\leq 2c_S\cdot d(x,y),
    \end{align*}

    where the second line is by the fact that distortion of $\alpha_S$ is at most $c_S$, the second is by the triangle inequality, and the third is by the fact that $d(x,\gamma(x)) \leq d(x,y)$ by definition of $\gamma$. 

    Now consider an arbitrary $x\in K_i$ for some $i$, and let $u_i$ be the center of $K_i$ (i.e. $\pi(u_i)=i$). We get the following:
    \begin{align*}
        d_Y(\alpha(x),\alpha(y)) &= d_Y(\alpha(u_i),\alpha(y)) \\
        &\leq 2c_S\cdot d(u_i,y) \\
        &\leq 2c_S\cdot (d(u_i,x)+d(x,y)) \\
        &\leq 10c_S\cdot d(x,y),
    \end{align*}

    where the first line is because $x$ is also assigned to the same position as $u_i$'s closest neighbor, the second line is by the argument we just made for $u_i$, the third line is by the triangle inequality, and the last line is by the fact that $d(u_i,x)\leq 4d(x,\gamma(x))\leq 4d(x,y)$ since $b\leq 4$ and $x\in K_i$.
\end{proofof}

% Next we want to consider the distortion between more general members of a set $K_i$ and members of $X\setminus K_i$. For the coming proofs, to show that $x$ and $y$ are not too distorted, we will generally focus on showing that the distance between $x$ and $y$ is at least a constant factor larger than the distance from $x$ or $y$ to some other point whose distance to $x$ or $y$ is already known not to be too distorted.

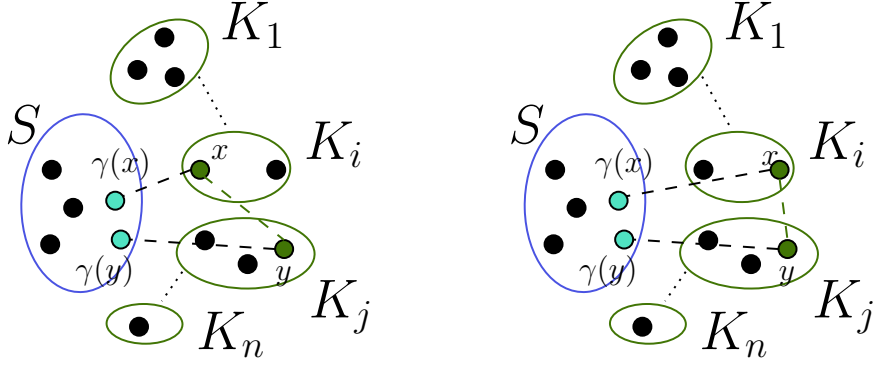
\begin{figure}
    \centering
    \begin{minipage}{.4\textwidth}
        \input{figures/far_xy}
    \end{minipage}
    \begin{minipage}{.4\textwidth}
        \input{figures/close_xy}
    \end{minipage}
    \caption{Left: Visualization of nodes referenced in Lemma~\ref{lem:expansion-weak}~\eqref{lem:farxy-weak} and Lemma~\ref{lem:expansion}~\eqref{lem:farxy}; Right: Visualization of nodes referenced in Lemma~\ref{lem:expansion-weak}~\eqref{lem:closexy-weak} and Lemma~\ref{lem:expansion}~\eqref{lem:closexy}  }
    \label{fig:far_xy}
\end{figure}

Next we consider comparing two the distance of two nodes $x,y\in X\setminus S$. First we consider the case that at least one of the nodes has a relatively short distance to $S$ compared to the distance to the other node.

%\begin{lem}\label{lem:farxy-weak}
%     Let $\alpha\from$ Algorithm \ref{alg:extend}$((X,d),S,p,\alpha_S,\alpha_X,\randrangtwo)$. Consider $x,y\in X-S$ with $d(x,\neigh(x))\leq \comparisontwo\cdot d(x,y)$. Then $\distp{\alpha(x)}{\alpha(y)}\leq \shrink \cdot [46\subdist + 3\comparison\cdot 66 \origdist ]\cdot d(x,y)$.
%\end{lem}

\begin{proofof}{Lemma~\ref{lem:expansion-weak}~\eqref{lem:farxy-weak}}
    Let $\neigh$ be as defined in line \eqref{def-gamma} of the algorithm.
    We get 
    \begin{align*}
        d_Y(\alpha(x),\alpha(y)) &\leq d_Y(\alpha(x),\alpha(\gamma(x))) + d_Y(\alpha(\gamma(x)),\alpha(y)) \\
        &\leq 10c_S\cdot d(x,\gamma(x)) + 10c_S\cdot d(\gamma(x),y) \\
        &\leq 20c_S\cdot d(x,\gamma(x)) + 10c_S \cdot d(x,y) \\
        &\leq 40c_S\cdot d(x,y)+10c_S\cdot d(x,y) \\
        &= 50c_S\cdot d(x,y),
    \end{align*}

    where the first line is by the triangle inequality on $Y$, the second is by Lemma~\ref{lem:expansion-weak}~\eqref{lem:ui-weak}, the third is by the triangle inequality on $X$, and the fourth is by the fact that $d(x,\gamma(x))\leq 2d(x,y)$, a condition of this lemma.
    % $d(x,u)\leq b\cdot d(x,\gamma(x))\leq 4d(x,\gamma(x))$ is a requirement for $x\in K_i$ since $u$ is the center.
\end{proofof}

Now we consider the final case, where we must consider expected distance.

\begin{proofof}{Lemma~\ref{lem:expansion-weak}~\eqref{lem:closexy-weak}}
    Let $\neigh$ be as defined in line ~\eqref{def-gamma} of the algorithm.
    We will say that $x$ and $y$ are ``split" if $x\in K_i,y\in K_j$ for $i\neq j$ and $K_i,K_j$ as defined in line \eqref{def-ki} of the algorithm. Let $u_i,u_j$ be as defined in line \eqref{ui-def} of the algorithm for the same choice of $i,j$ respectively. 

    \begin{itemize}
        \item First consider the worst-case distortion when $x$ and $y$ are \textit{not} split. Then $x,y\in K_i$ for some $i$, so $d_Y(\alpha(x),\alpha(y))=0$. 
        % We have $\distp{\alpha(x)}{\alpha(y)}\leq \origdist \cdot d(x,y)$ by Lemma~\ref{lem:expansion-weak}~\ref{lem:sameset}, no matter the choice of $\decider$. 

        \item Now consider the worst-case distortion when  $x$ and $y$ \textit{are} split.
        % such that $x$ is placed in $K_i$ and $y$ is placed in $K_j$ for $i<j$. We have
        % (We can assume this wlog by symmetry.) %Now consider the maximum distance $\distp{\alpha(x)}{\alpha(y)}$ in this case. We get 
            \begin{align*}
                d_Y(\alpha(x),\alpha(y)) &\leq d_Y(\alpha(x),\alpha(\gamma(x))) + d_Y(\alpha(\gamma(x)),\alpha(y)) \\
                &\leq 10c_S\cdot d(x,\gamma(x)) + 10c_S\cdot d(\gamma(x),y) \\
                &\leq 20 c_S\cdot d(x,\gamma(x)) + 10c_S\cdot d(x,y) \\
                &\leq 20 c_S\cdot d(x,\gamma(x)) + 5c_S\cdot d(x,\gamma(x)) \\
                &= 25c_S\cdot d(x,\gamma(x)),
            \end{align*}

            where the first inequality is by the triangle inequality on $Y$, the second is by Lemma~\ref{lem:expansion-weak}~\eqref{lem:ui-weak}, the third is by the triangle inequality on $X$, and the fourth is by the fact that $d(x,y)<\frac{1}{\comparisontwo}d(x,\gamma(x))$ as a condition of this part of the lemma. Note that an identical analysis shows that $d_Y(\alpha(x),\alpha(y))\leq 25c_S\cdot d(y,\gamma(y))$.

            % where the first inequality is by the triangle inequality, the second is by Lemma \ref{lem:ui-weak-separate}, the third inequality is by the triangle inequality on $d(\gamma(x),y)$, the fourth inequality is by the fact that $d(x,y)<\frac{1}{\comparisontwo}d(x,\gamma(x))$ by the condition of this lemma, and the final equality is by rearranging terms.

            %Note that $b$ is upper bounded in value by the upper bound of the range $b$ is chosen from.

 \end{itemize}

%Let $K_u$ be the cluster formed in line 8 such that $u$ is the center of that cluster. Consider $\Pr_{b,\pi}[x,y \text{ are split }| u \text{ decides }(x,y)]$. 

% In this case we have $d(x,\neigh(u_i))\leq b\cdot d(x,\neigh(x))$ \textit{and} $d(y,\neigh(u_i))>b\cdot d(y,\neigh(y))$, by how we defined $K_i$. 
Now consider the probability of $x$ and $y$ being split. First, let us fix some node $u$ chosen in some iteration of Step~\eqref{ui-def} of the algorithm and let $K_u$ be the cluster formed by this vertex. Suppose that the placement of $x$ and $y$ is undetermined prior to this point of time. We will bound the probability that $x$ and $y$ are split by $u$, that is, exactly one of these vertices ends up in the cluster $K_u$.
Without loss of generality, assume that $\frac{d(x,u)}{d(x,\neigh(x))}\leq \frac{d(y,u)}{d(y,\neigh(y))}$. This implies that if $x,y$ are split, then $x\in K_u,y\notin K_u$ and we have 
            \begin{align*}
                \frac{d(x,u)}{d(x,\neigh(x))} \leq b \leq \frac{d(y,u)}{d(y,\neigh(y))}.
            \end{align*}
This implies $b$ must fall in a range  of width 
            \begin{align*}
                W &\leq \frac{d(y,u)}{d(y,\neigh(y))}-\frac{d(x,u)}{d(x,\neigh(x))}\\
                &\leq \frac{d(x,u)+d(x,y)}{d(x,\neigh(y))-d(x,y)}-\frac{d(x,u)}{d(x,\neigh(x))} \\
                &\leq \frac{d(x,u)+d(x,y)}{d(x,\neigh(x))-d(x,y)}-\frac{d(x,u)}{d(x,\neigh(x))} \\
                %&= \frac{d(x,\neigh(u_i))\cdot d(x,\neigh(x)) + d(x,y)\cdot d(x,\neigh(x)) - d(x,\neigh(u_i)) \cdot d(x,\neigh(x)) +d(x,\neigh(u_i))\cdot d(x,y)}{d(x,\neigh(x))\cdot (d(x,\neigh(x))-d(x,y))}\\
                &= d(x,y)\cdot \frac{d(x,\neigh(x))+d(x,u)}{d(x,\neigh(x))\cdot (d(x,\neigh(x))-d(x,y))} \\
                %&\leq d(x,y)\cdot \frac{5\cdot d(x,\neigh(x))}{d(x,\neigh(x))\cdot (d(x,\neigh(x))-d(x,y))}\\
                &\le d(x,y)\cdot \frac{5}{ (d(x,\neigh(x))-d(x,y))} \\
                %&< d(x,y)\cdot \frac{5}{ \frac{1}{2}d(x,\neigh(x))} \\
                &\leq 10\cdot \frac{d(x,y)}{d(x,\neigh(x))},
            \end{align*}

            where the second inequality is by applying the triangle inequality twice, the third inequality is by the fact that $\neigh(x)$ is a closest node in $S$ to $x$, and the fourth is by cross-multiplying.
            %, and the fifth is by reducing. 
            The fifth line is by the fact that $x\in K_u$, which implies $d(x,u)\leq \text{max possible value of b}\cdot d(x,\neigh(x))$. Finally, the sixth follows from the fact that $d(x,y)<\frac{1}{\comparisontwo}\cdot d(x,\neigh(x))$, a condition of the lemma.

Next we will bound the overall probability that $x$ and $y$ are split by {\em some} node $u$. For a vertex $u$, define $\beta_u = \min\left\{\frac{d(x,u)}{d(x,\neigh(x))},\frac{d(y,u)}{d(y,\neigh(y))}\right\}$. This is the smallest value of $b$ at which the cluster $K_u$ formed by $u$ contains either $x$ or $y$. 
Consider ordering vertices $u$ in $K$ in increasing order of $\beta_u$, and let $\ind:K\rightarrow [k]$ denote this ordering. We 
say that a node $u$ ``decides" the pair $(x,y)$ if at least one of $x$ and $y$ is in $K_u$. $u$ can decide $(x,y)$ iff $b\ge\beta_u$ (i.e. at least one of $x,y$ meets the criteria to be in the cluster $u$ is the center of). This implies that if $u$ and $u'$ satisfy $\ind(u)<\ind(u')$ and $u$ appears before $u'$ in the ordering $\pi$, then at the time we consider $u'$ in Step~\eqref{ui-def}, either it is the case that $b\ge\beta_u$ and $(x,y)$ has already been decided by $u$, or it is the case that $b<\beta_u<\beta_{u'}$, in which case $u'$ cannot decide $(x,y)$. Therefore, in either case, $u'$ does not decide $(x,y)$, and consequently does not split them.

In other words, in order for a vertex $u$ to be able to split $(x,y)$, it must be the case that among the $\ind(u)$ vertices before $u$ in the $\ind$ ordering (and including $u$ itself), $u$ is the first vertex to appear in the ordering $\pi$. Let us call this latter event $E_u$, and observe that this event is independent of the choice of $b$ --- it only depends on the choice of the permutations $\pi$ and $\ind$. We also note that $\Pr_\pi[E_u]=1/\ind(u)$.

%of the nodes in $K$ such that $\min\left\{\frac{d(x,u)}{d(x,\neigh(x))},\frac{d(y,u)}{d(y,\neigh(y))}\right\}<\min\left\{\frac{d(x,v)}{d(x,\neigh(x))},\frac{d(y,v)}{d(y,\neigh(y))}\right\}$ implies $\ind(u)<\ind(v)$. Say that the pair $(x,y)$ is ``decided" by a node $u\in K$ if $u$ is the first node in the ordering given by $\pi$ (as chosen in line 5 of the algorithm) such that $u$ is the center of some $K_i$ and $x\in K_i$ or $y\in K_i$. We have that $\Pr_\pi[(x,y)\text{ is decided by } u]\leq \frac{1}{i}$, as $u$ decides the pair $x,y$ if and only if $\frac{d(x,u)}{d(x,\neigh(x))}\leq b$ or $\frac{d(y,u)}{d(y,\neigh(y))}\leq b$ and it is the first such node in the ordering to do this (i.e. if all preceding nodes $v$ in the ordering have $\frac{d(x,v)}{d(x,\neigh(x))}> b$ and $\frac{d(y,v)}{d(y,\neigh(y))}> b$). For any choice of $b$, if $u$ is such that $\frac{d(x,u)}{d(x,\neigh(x))}\leq b$ or $\frac{d(y,u)}{d(y,\neigh(y))}\leq b$, then  there are at least $\ind(u)-1$ other nodes that meet this criteria as well. Thus, there is at most a $1/\ind(u)$ chance that $u$ is the first node under the ordering in $\pi$ to meet this criterion, and there is at most a $1/i$ chance of $u$ deciding $(x,y)$. Say that $(x,y)$ is split by $u$ if $u$ decides $(x,y)$ and $(x,y)$ are split.

We can now write down the probability that $x$ and $y$ are split as follows, using the fact that from our discussion above, $\Pr_{b}[u\text{ splits }(x,y)| \neg E_u]=0$ for all $u$. 
%consider $\Pr_{\pi,b}[x,y$ are split $]$.

\begin{align*}
    \Pr_{\pi,b}[x,y\text{ are split}] &= \sum_{u\in K}\Pr_{b}[u\text{ splits }(x,y)| E_u] \cdot \Pr_\pi[E_u] + \Pr_{b}[u\text{ splits }(x,y)| \neg E_u] \cdot \Pr_\pi[\neg E_u]\\
    %& \ \ \ \ \ \ \ \ \ \ \ \ \ \ \ \ \ \ \ \Pr_\pi[\ind(u)<\ind(v)\forall v \text{ such that } \pi(v)<\pi(u)] \\
    &\leq \sum_{u\in K}\frac{W}{\tau}\cdot \frac{1}{\ind(u)} \\
    &= \sum_{u\in K}\frac{10}{2}\frac{d(x,y)}{d(x_u,\neigh(x_u))}\cdot \frac{1}{\ind(u)}.
\end{align*}
Here we used $x_u$ to denote the node in $\{x,y\}$ that is closer to $u$, and we substituted expressions from above for $W$, $\tau$, and $\Pr_\pi[E_u]$. 
%Note that if $(x,y)$ are split and $x_u$ is closer to $u$, then $x_u$ must be in the group constructed by $u$ and $y$ must be placed in a group after this point. Conditioning on the order in $\pi$ gives no information about $b$, and for $u$ to split $x,y$, we must have $b$ fall in the range $W$ as we have already discussed. 

Finally, let us consider the expected distance between $\alpha(x)$ and $\alpha(y)$. Note that by our earlier analysis, the distance between $x$ and $y$ when they are split by $u$ is at most $25c_S\cdot d(x_u,\neigh(x_u))$. Thus, we can compute the expected distance between $x$ and $y$ as follows:
\begin{align*}
    \mathop{E}_{\pi,b}[d_Y(\alpha(x),\alpha(y))] &\leq 0\cdot \Pr[(x,y)\text{ are not split}]  \\
    & \ \ \ \ \ \ + \sum_{u\in K}\Pr_{b}[x,y\text{ are split by } u|E_u]\cdot \Pr_{\pi}[E_u]\cdot25\subdist \cdot d(x_u,\neigh(x_u))\\
    &\leq  \sum_{u\in K|\beta_u\leq \tau+2}  \frac{5}{\ind(u)} \cdot\frac{d(x,y)}{d(x_u,\neigh(x_u))}\cdot 25\subdist \cdot d(x_u,\neigh(x_u)) \\
    &\leq 125\subdist \cdot d(x,y)\cdot \sum_{i=1}^k\frac{1}{i} \\
    &= 125\subdist \cdot H_k \cdot d(x,y),
\end{align*}
where $H_k$ is the $k$th Harmonic number.
\end{proofof}
%\kristin{old stuff below - unhide to see}

\hide{
Define $\eta(u):=\min\left\{\frac{d(x,u)}{d(x,\neigh(x))},\frac{d(y,u)}{d(y,\neigh(y))}\right\}$, and define $\eta'(u,\pi):=\min_{v|\pi(v)< \pi(u)}\left\{ \eta(v) \right\}$. (If the given set is empty, let $\eta'(u,\pi)$ be $\tau+2$.) If $\eta'(u,\pi)<\eta(u)$, reset this value to $\eta(u)$.  Define $w(u,\pi):=\eta'(u,\pi)-\eta(u)$. Note that $u$ decides $(x,y)$ if and only if $b$ falls in the range between $\eta'(u,\pi)$ and $\eta(u)$, as either $x$ or $y$ being in the set formed by $u$ implies $b\geq \eta(u)$ and $x$ and $y$ \textit{not} being in the set formed by some $v$ prior to $u$ in the ordering under $\pi$ implies that $b$ is smaller than $\eta(v)$. We consider only the smallest $\eta(v)$ for all $v$ preceding $u$ in the permutation and note that we need only consider the preceding $v$ with $\eta(v)>\eta(u)$, as otherwise it is impossible that $u$ decides $(x,y)$. Thus, we get that $b$ \textit{must} fall in a range of width $w(u,\pi)$ (which may be $0$ if there exists a  point with smaller $\eta$ value, as any choice of $b$ is impossible to obtain this result with the given $\pi$). 

Thus, the probability that $u$ decides $(x,y)$, given $\pi$, is at most $w(u,\pi)/\tau$. Also note that if $\pi$ is given, as well as the fact that $u$ decides $(x,y)$, then $b$ is uniformly distributed on a range of width at most $w(u,\pi)$, and if $x,y$ are also split under these conditions, then $b$ must \textit{also} be in a range of width $W$. Thus, this conditional probability happens with probability at most $W/w(u,\pi)$. 
% Define $t(x,y)$ to be the number of pairs $(u,\pi)$ for which $w(u,\pi)$ is non-zero. Note that if $u$ is fixed and $\pi$ is chosen uniformly at random, the probability that $w(u,\pi)\neq 0$ is the probability that $u$ is the first point under $\pi$ such that $\eta(u)\leq \tau+2$. (If $b$ is chosen to be larger than $\eta(v)$ for any $v$ and $v$ is the first such point that this holds for under $\pi$, then $v$ automatically decides $(x,y)$ even if it's impossible for $v$ to split $(x,y)$.) If there are $t'(x,y)$ values of $u$ meeting this criterion (ie $t'(x,y)$ points such that it is \textit{possible} that point can split $(x,y)$ for some choice of $(b,\pi)$), then for a uniformly random permutation, there is a $1/t'(x,y)$ chance that $u$ is the first such value. Thus, in total there are $k!/t'(x,y)$ permutations with this property (or $0$ if $\eta(u)>\tau+2$). Some of the values discussed here are represented in Figure \ref{fig:intervals}.
In total, there are at most $k!/\ind(u)$ permutations with this property (or $0$ if $\eta(u)>\tau+2$), as $w(u,\pi)$ was set to $0$ if there exists a previous vertex under $\pi$ with smaller value under $\eta$ or if $\eta(u)>\tau+2$. Some of the values discussed here are represented in Figure \ref{fig:intervals}.

\begin{figure}
    \centering
    \input{figures/intervals}
    \caption{This figure shows some of the values referenced in the proof of Lemma~\ref{lem:expansion-weak}~\eqref{lem:closexy-weak}. Here $u$ is a point that we are considering for splitting $(x,y)$. The dotted markers mark the $\eta(v)$ values for each $v\in K$ such that $\pi(v)>\pi(u)$. Here the line marked $\eta(v)$ marks $\eta(v)$ for some $v$ with $\pi(v)>\pi(u)$ and $\eta(v)<\eta(u)$. The light red dashes represent the $\eta(w)$ values for $w$ such that $\pi(w)<\pi(u)$. Here the marker labeled $\eta(v)$ is for a point $v$ with $\pi(v)>\pi(u)$ and $\eta(v)<\eta(u)$. The marker labeled $\eta(w)$ is for a point $w$ with $\pi(w)<\pi(u)$, and $\eta(w)>\eta(u)$. In particular, $w$ is $\eta(w)$ is $\eta'(u,\pi)$ for this choice of $\pi$.  \\    
    Note that if $\pi(v)<\pi(u)$ \textit{and} $\eta(v)\leq \eta(u)$ for any $v\neq u$, $u$ cannot decide $(x,y)$. If $u$ decides $(x,y)$, then $b$ must fall in such a range that it has value at least $\eta(u)$ and value no more than $\eta(w)$ (as otherwise $w$ would have decided $(x,y)$). The width of this range is defined to be exactly $w(u,\pi)$.  \\
    For a node $u$ to split a pair $(x,y)$ given that $u$ decides $(x,y)$, $b$ must fall in the range between $\frac{d(x,u)}{d(x,\neigh(x))}$ and $\frac{d(y,u)}{d(y,\neigh(y))}$, which has width exactly $W$ by definition of $W$. }
    \label{fig:intervals}
\end{figure}

Now consider $\Pr_{b,\pi}[(x,y)\text{ are split}]$:

\begin{align*}
     \Pr_{\pi,b}[x,y\text{ are split}] &= \sum_{u\in K}\Pr_{b,\pi}[x,y\text{ are split }|u\text{ decides }(x,y)]\cdot \Pr_{b,\pi}[u\text{ decides }(x,y)] \\
     &= \sum_{u\in K}\sum_\pi\Pr_{b}[x,y\text{ are split }|\pi \text{ and }u \text{ decides }(x,y)]\cdot \Pr_b[u \text{ decides }(x,y)|\pi] \cdot \Pr[\pi] \\
     &\leq \sum_{u\in K}\sum_{\pi|w(u,\pi)\neq 0} \frac{W}{w(u,\pi)}  \cdot \frac{w(u,\pi)}{\tau} \cdot \frac{1}{k!} \\
     &\leq \sum_{u\in K|\eta(u)\leq \tau+2} \frac{k!}{\ind(u)} \cdot \frac{W}{\tau}  \cdot \frac{1}{k!} \\
     &= \sum_{u\in K|\eta(u)\leq \tau+2}  \frac{1}{\ind(u)} \cdot \frac{10\cdot \frac{d(x,y)}{d(x_u,\neigh(x_u))}}{2} \\
     &= \sum_{u\in K|\eta(u)\leq \tau+2}  \frac{5}{\ind(u)} \cdot\frac{d(x,y)}{d(x_u,\neigh(x_u))},
\end{align*} 

where $x_u$ is $\arg\min_{x,y}\left\{\frac{d(x,u)}{d(x,\neigh(x))},\frac{d(y,u)}{d(y,\neigh(y))}\right\}$. More specifically, note that this analysis implies $\Pr_{b,\pi}[x,y\text{ are split }|u\text{ decides }(x,y)]\cdot \Pr_{b,\pi}[u\text{ decides }(x,y)]$  is at most $\frac{5}{\ind(u)} \cdot\frac{d(x,y)}{d(x,\neigh(x))}$ if $\eta(u)\leq \tau+2$ and $0$ otherwise.

Note that by our earlier analysis, the distance between $x$ and $y$ when they are split by $u$ is at most $[\frac{31}{2}c_S+\frac{45}{2}c_X]\cdot d(x_u,\neigh(x_u))$. Thus, we can compute the expected distance between $x$ and $y$ as follows:

\begin{align*}
    E_{\pi,b}[\distp{\alpha(x)}{\alpha(y)}] &\leq c_X\cdot d(x,y)\cdot \Pr[(x,y)\text{ are not split}]  \\
    & \ \ \ \ \ \ + \sum_{u\in K}\cdot \Pr_{b,\pi}[x,y\text{ are split }|u\text{ decides }(x,y)]\cdot \Pr_{b,\pi}[u\text{ decides }(x,y)]\cdot[\frac{31}{2}\subdist + \frac{45}{2}\origdist ] \cdot d(x_u,\neigh(x_u))\\
    &\leq \origdist\cdot d(x,y)+ \sum_{u\in K|\eta(u)\leq \tau+2}  \frac{5}{\ind(u)} \cdot\frac{d(x,y)}{d(x_u,\neigh(x_u))}\cdot [\frac{31}{2}\subdist + \frac{45}{2}\origdist ] \cdot d(x_u,\neigh(x_u)) \\
    &\leq \origdist\cdot d(x,y) + 5\cdot [\frac{31}{2}\subdist + \frac{45}{2}\origdist ]\cdot \sum_{i=1}^k\frac{1}{i}\cdot d(x,y) \\
    &= \left(\frac{155}{2}\cdot H_k\cdot \subdist + (\frac{225}{2}\cdot H_k+1)\cdot \origdist\right)\cdot d(x,y),
\end{align*}

where $H_k$ is the $k$th partial harmonic sum.

\hide{
\kristin{old stuff below here - didn't delete in case what I'm doing above is wrong}

\begin{align*}
    \Pr_{\pi,b}[x,y\text{ are split}] &= \sum_{u\in K}\Pr_{\pi,b}[x,y\text{ are split }| u \text{ decides }(x,y)]\cdot \Pr_{\pi,b}[u \text{ decides }(x,y)] \\
    &\leq \sum_{u\in K}10\cdot \frac{\frac{d(x,y)}{d(x_u,\neigh(x_u))}}{\tau+2-\min_{x,y}\left\{\frac{d(x,u)}{d(x,\neigh(x))},\frac{d(y,u)}{d(y,\neigh(y))}\right\}}\cdot \frac{1}{\ind(u)}\cdot \frac{\tau+2-\min_{x,y}\left\{\frac{d(x,u)}{d(x,\neigh(x))},\frac{d(y,u)}{d(y,\neigh(y))}\right\}}{\tau} \\
    &= \sum_{u\in K} 5\cdot \frac{d(x,y)}{d(x_u,\neigh(x_u))}\cdot \frac{1}{\ind(u)} 
\end{align*}

where $x_u$ is $\arg\min_{x,y}\left\{\frac{d(x,u)}{d(x,\neigh(x))},\frac{d(y,u)}{d(y,\neigh(y))}\right\}$. Here we have used the fact that in order for $u$ to decide $(x,y)$, we must have $b\geq \min_{x,y}\left\{\frac{d(x,u)}{d(x,\neigh(x))},\frac{d(y,u)}{d(y,\neigh(y))}\right\}$ since at least one of $x$ and $u$ must be in the group $u$ forms the center of. This happens with probability at most the width of the range from this minimum value to $\tau+2$, divided by the width of the total range, $\tau$. Additionally, $u$ must be the first node in the order which may split. At least $\ind(u)$  other points have index at most as large as $u$, so when we take a uniformly random permutation there is at most a $1/\ind(u)$ chance that $u$ is the first of these $\ind(u)$ points. Note that $b$ and $\pi$ are chosen independently, so we may multiply these probabilities. Additionally, the probability that $x$ and $y$ are split given that $u$ decides $(x,y)$ is the probability that $b$ falls in a range of width $W$ as defined earlier divided by the total range that $b$ may be drawn from. Knowing that $u$ splits $b$ tells us that $b$ must be drawn from the range of values between $\min_{x,y}\left\{\frac{d(x,u)}{d(x,\neigh(x))},\frac{d(y,u)}{d(y,\neigh(y))}\right\}$ and $\tau+2$. Thus, we are able to get our final result when we substitute in $2$ for $\tau$ for the last line. \kristin{I'm worried about how much knowing that $u$ splits $(x,y)$ may affect the range that $b$ could actually come from}

% nd we have used the fact that $b$ has a $W/\tau$ chance of falling within a width $W$ region. 
Note that by our earlier analysis, the distance between $x$ and $y$ when they are split by $u$ is at most $[\frac{31}{2}c_S+\frac{45}{2}c_X]\cdot d(x_u,\neigh(x_u))$. Thus, we can compute the expected distance between $x$ and $y$ as follows:

\begin{align*}
    E_{\pi,b}[\distp{\alpha(x)}{\alpha(y)}] &\leq c_X\cdot d(x,y)\cdot \Pr[(x,y)\text{ are not split}]  \\
    & \ \ \ \ \ \ + \sum_{u\in K}\Pr_{\pi,b}[x,y\text{ are split }| u \text{ decides }(x,y)]\cdot \Pr_{\pi,b}[u \text{ decides }(x,y)]\cdot [\frac{31}{2}c_S+\frac{45}{2}c_X]\cdot d(x_u,\neigh(x_u)) \\
    &\leq c_X\cdot d(x,y) + \sum_{u\in K}5\cdot[\frac{31}{2}c_S+\frac{45}{2}c_X]\cdot  \frac{d(x,y)}{d(x_u,\neigh(x_u))}\cdot \frac{1}{\ind(u)}\cdot d(x_u,\neigh(x_u)) \\
    &= c_X\cdot d(x,y)+5\cdot [\frac{31}{2}c_S+\frac{45}{2}c_X]\cdot d(x,y)\cdot \sum_{i=1}^k\frac{1}{i} \\
    &= (\frac{155}{2}\cdot H_k\cdot c_S+(\frac{225}{2}\cdot H_k+1)\cdot c_X)\cdot d(x,y),
\end{align*}

where $H_k$ is the $k$th partial harmonic sum.
}
}
%\end{proofof}

% \begin{figure}
%     \centering
%     \input{figures/close_xy}
%     \caption{Visualization of nodes referenced in Lemma~\ref{lem:expansion-weak}~\eqref{lem:closexy-weak} }
%     \label{fig:close_xy}
% \end{figure}

% Note that in many of the previous lemmas, we have upper bounded the distortion in terms of $b$. However, in all cases this distortion is linear in $b$, so we can plug in the upper bound $\rangwidth+2$ and obtain a new upper bound in terms of the input parameters for the algorithm.

\subsection{Deterministic extensions}
\label{sec:improvements}
%\kristin{maybe we should rename this section since we have used the word weak to mean unbounded contraction now?}

In this section we prove Theorem~\ref{thm:extension}.  Note that in the previous part, we were able to allow the target metric to be arbitrary, but now we require that it be some $\ell_p$ space so that we can ``average" the possible outputs of the embedding. Thus, in this section we assume that $Y$ is $\reals^n$ with the $\ell_p$ metric.
%and Corollary \ref{cor:exists-embedding}.

\begin{proof}[Proof of Theorem~\ref{thm:extension}]
 We have shown that for all $x,y\in X\setminus K$, $\distp{\alpha(x)}{\alpha(y)}=d(x,y)$ and for all $x,y\in X$, $E_{\pi,b}[\distl{2}{\alpha(x)}{\alpha(y)}] \leq 125\subdist\cdot d(x,y)$. We can now use the convexity of the $\ell_p$ norm to upper bound the distance between the {\em expected} points. In particular, note that the point that $x$ is mapped to is completely determined by the choice of subsets $K_i$. There are at most $2^{2^{|K|}}$ ways to partition $K$, so the set of points over which $\alpha(x)$ is chosen is finite, and we can assign finite probability $p_t$ to $\Pr[\alpha(x)=\alpha^t(x)]$ such that $\alpha^t$ is a function in the support of $\alpha$ and the sum over this probability for all points in the support is $1$. Let $E[\alpha(x)]:=E_{\pi,b}[\alpha(x)]$ be $\sum_{t}p_t\cdot \alpha^t(x)$ where the sum is over all functions $\alpha^t$ points in the support of $\alpha$. 
 % Assume we pad any extra indices with $0$s if the embeddings are different lengths due to different numbers of $K_i$ being formed.) 
 We claim that the embedding $\alpha\st:X\rightarrow \ell_p^n$ defined by $\alpha\st(x):=E[\alpha(x)]$ is a Lipschitz extension of $\alpha_S$. In particular, we will use the Banach space properties of the $\ell_p$ norm to obtain our desired bound.

 \begin{align*}
     \distl{p}{\alpha\st(x)}{\alpha\st(y)} &= \distl{p}{E[\alpha(x)]}{E[\alpha(y)]} \\
     &= \distl{p}{\sum_t p_t\cdot \alpha^t(x)}{\sum_t p_t\cdot \alpha^t(y)}\\
     &= ||\sum_t (p_t\alpha^t(x)-p_t\alpha^t(y))||_p \\
     &\leq \sum_{t} ||p_t\cdot(\alpha^t(x)-\alpha^t(y))||_p \\
     &= \sum_{t}p_t\distl{p}{\alpha^t(x)}{\alpha^t(y)} \\
     &= \mathop{E}_{\pi,b}[\distl{p}{\alpha(x)}{\alpha(y)}] \\
     &\leq  125\subdist \cdot H_k\cdot d(x,y),
\end{align*}

where the first line is by definition of $\alpha\st$, the second is by definition of $E[\alpha(x)],E[\alpha(y)]$, the third is by regrouping, the fourth is by the triangle inequality of $\ell_p$, the fifth is by scalar properties of $\ell_p$, the sixth is by the definition of expectation, and the seventh is by our bounds from the previous subsection. 

Note that for $x\in S$, $\alpha^t(x)=\alpha^{t'}(x)$ for all $\alpha^t,\alpha^{t'}$ in the support of $\alpha$. Thus, $\alpha\st(x)=\alpha^t(x)=\alpha_S(x)$.
\end{proof}

In fact, in the proof of the previous theorem, the properties of $\ell_p$ space that we used were properties of any Banach space, so Theorem \ref{thm:extension} applies when the destination metric is any Banach space.

%The proof of Corollary \ref{cor:exists-embedding-weak} is identical to that of Corollary \ref{cor:exists-embedding}.

%% file: maincontent/alg1.tex
\begin{algorithm}
\caption{Algorithm for finding a Liptschitz extension for finite metrics}
\label{alg:extend}
\textbf{Input:} Metric space $(X,d), |X|=n$, subset $S\subseteq X$, embedding $\alpha_S:X\setminus S\rightarrow Y$ for some target metric $(Y,d_Y)$, and %$\randrang$ for 
real number $\rangwidth>0$ \\
\textbf{Output:} A randomized embedding $\alpha:X\rightarrow Y$ such that for all $x,y\in X$, $E[d_Y(\alpha(x),\alpha(y))]\leq 125H_k\cdot c\cdot d(x,y)$ for $k=|X\setminus S|$, and for all $x\in S$, $\alpha(x)=\alpha_S(x)$
% \kristinEDIT{$d_{\ell_2}(\alpha(x),\alpha(y))\leq f(c_S)\cdot d(x,y)$ and for all $x,y\in X$, $d_{\ell_2}(\alpha(x),\alpha(y))\leq g(c_S, c_X)\cdot d(x,y)$}
% {$\distl{2}{\alpha(x)}{\alpha(y)}\leq f(c_S)\cdot d(x,y)$ and for all $x,y\in X$, $E[\distl{2}{\alpha(x)}{\alpha(y)}]\leq g(c_S, c_X)\cdot d(x,y)$}.
\begin{algorithmic}[1]
\State $K\from X\setminus S$.
\State \label{def-gamma} Define $\neigh:K\fto S$ such that 
$\gamma(u)\in\arg\min_{v\in S} d(u,v)$.
%$\gamma(u)\mapsto v$ where $v$ is an arbitrary vertex in $S$ such that $d(u,S)=d(u,v)$. 
\Comment{ie $\neigh(u)$ is one of $u$'s closest neighbors in $S$}
\State Select $\decider$ uniformly at random from the range $\randrang$
\State Select a uniformly random permutation $\perm:K\rightarrow [k]$ of the vertices in $K$
\State $K'\from K$
\For{$i=1$ to $k$}
    \State\label{ui-def} $u_i\from \pi^{-1}(i)$ 
    \State\label{def-ki} $K_i\from \{v\in K' \ | \ d(v,u_i)\leq b\cdot d(v,\neigh(v))\}$ \Comment{Let the ``center" of $K_i$ be $u_i$ } 
    \State $K'\from K'\setminus K_i$
\EndFor
\State Define an embedding $\alpha':X\fto \lspace{p}$ such that
    \begin{align*}
        \alpha'(v)=\begin{cases} 
        \alpha_S(v) & \text{if } v\in S \\
        \alpha_S(\neigh(u_i)) & \text{if } v\in K_i \text{ and with $u_i$ being the center of $K_i$}
        \end{cases}.
    \end{align*}
% \State Define an embedding $\alpha:X\fto \lspace{p}$ such that $\alpha(v)\mapsto (\alpha'(v)\concat\alpha_1(v)\concat\cdots \concat \alpha_t(v))$ \Comment{{here $\concat$ denotes concatenation }}
%$\alpha$ is the concatenation of all the embeddings we have defined and   $\cdot$  denotes element-wise multiplication  }
\State Output $\alpha'$ as $\alpha$
\end{algorithmic}
\end{algorithm}

%% file: maincontent/distortion_table.tex
\begin{table}[]
    \centering
    \begin{tabular}{c|c|cl}
        membership of $x$ and $y$ & restrictions on $d(x,y)$ & upper bound on expected distortion & \\
        \hline 
        $x,y\in S$ & none & $\subdist$ & \eqref{lem:not-outlier-weak} \\
        % $x,y\in K_i$ & none & $\subdist$ & \eqref{lem:sameset} \\
        $x\in S,y\in X\setminus S$ & none & $10\subdist$ & \eqref{lem:ui-weak} \\
        $x,y\in X\setminus S$ & $d(x,\neigh(x))\leq 2 \cdot d(x,y)$ & $50\subdist$ & \eqref{lem:farxy-weak} \\
        $x,y\in X\setminus S$ & $d(x,\neigh(x)),d(y,\neigh(y))> 2\cdot d(x,y)$ & $125\subdist\cdot H_k$ & \eqref{lem:closexy-weak}
    \end{tabular}
    \caption{Summary of the bounds in Lemma~\ref{lem:expansion-weak}.  %\ref{lem:not-outlier}, \ref{lem:sameset}, \ref{lem:ui}, \ref{lem:farxy}, and \ref{lem:closexy}. 
    Let $\alpha\from$ Algorithm \ref{alg:extend}$((X,d),S,\alpha_S,\rangwidth)$  where $\alpha_S:S\rightarrow \lspace{p}$ is an expanding embedding of distortion at most $c_S$. Then the third column of the table gives an upper bound on the the expected value of $d_Y(\alpha(x),\alpha(y))$ where $x$ and $y$ meet the criteria of the first two columns. {Here $\neigh$ and $K_i$ are as defined in lines \eqref{def-gamma} and \eqref{ui-def}-\eqref{def-ki} of the algorithm.} }
    \label{tab:distortion}
\end{table}

%% file: figures/both_in_s.tex
\tikzset{every picture/.style={line width=0.75pt}} %set default line width to 0.75pt        

\begin{tikzpicture}[x=0.75pt,y=0.75pt,yscale=-.4,xscale=.4]
%uncomment if require: \path (0,495); %set diagram left start at 0, and has height of 495

%Shape: Circle [id:dp7281444393949086] 
\draw  [fill={rgb, 255:red, 0; green, 0; blue, 0 }  ,fill opacity=1 ] (253.33,292.65) .. controls (253.33,286.39) and (258.41,281.32) .. (264.67,281.32) .. controls (270.93,281.32) and (276,286.39) .. (276,292.65) .. controls (276,298.91) and (270.93,303.98) .. (264.67,303.98) .. controls (258.41,303.98) and (253.33,298.91) .. (253.33,292.65) -- cycle ;
%Shape: Circle [id:dp011449811924921782] 
\draw  [fill={rgb, 255:red, 74; green, 144; blue, 226 }  ,fill opacity=1 ] (173.33,253.65) .. controls (173.33,247.39) and (178.41,242.32) .. (184.67,242.32) .. controls (190.93,242.32) and (196,247.39) .. (196,253.65) .. controls (196,259.91) and (190.93,264.98) .. (184.67,264.98) .. controls (178.41,264.98) and (173.33,259.91) .. (173.33,253.65) -- cycle ;
%Shape: Circle [id:dp2691096050802084] 
\draw  [fill={rgb, 255:red, 0; green, 0; blue, 0 }  ,fill opacity=1 ] (260.33,341.65) .. controls (260.33,335.39) and (265.41,330.32) .. (271.67,330.32) .. controls (277.93,330.32) and (283,335.39) .. (283,341.65) .. controls (283,347.91) and (277.93,352.98) .. (271.67,352.98) .. controls (265.41,352.98) and (260.33,347.91) .. (260.33,341.65) -- cycle ;
%Shape: Circle [id:dp13729815426604697] 
\draw  [fill={rgb, 255:red, 74; green, 144; blue, 226 }  ,fill opacity=1 ] (171.33,347.65) .. controls (171.33,341.39) and (176.41,336.32) .. (182.67,336.32) .. controls (188.93,336.32) and (194,341.39) .. (194,347.65) .. controls (194,353.91) and (188.93,358.98) .. (182.67,358.98) .. controls (176.41,358.98) and (171.33,353.91) .. (171.33,347.65) -- cycle ;
%Shape: Circle [id:dp37148216033690074] 
\draw  [fill={rgb, 255:red, 0; green, 0; blue, 0 }  ,fill opacity=1 ] (200.33,301.65) .. controls (200.33,295.39) and (205.41,290.32) .. (211.67,290.32) .. controls (217.93,290.32) and (223,295.39) .. (223,301.65) .. controls (223,307.91) and (217.93,312.98) .. (211.67,312.98) .. controls (205.41,312.98) and (200.33,307.91) .. (200.33,301.65) -- cycle ;
%Shape: Circle [id:dp5668609054436715] 
\draw  [fill={rgb, 255:red, 0; green, 0; blue, 0 }  ,fill opacity=1 ] (281.33,127.65) .. controls (281.33,121.39) and (286.41,116.32) .. (292.67,116.32) .. controls (298.93,116.32) and (304,121.39) .. (304,127.65) .. controls (304,133.91) and (298.93,138.98) .. (292.67,138.98) .. controls (286.41,138.98) and (281.33,133.91) .. (281.33,127.65) -- cycle ;
%Shape: Circle [id:dp4367461957922514] 
\draw  [fill={rgb, 255:red, 0; green, 0; blue, 0 }  ,fill opacity=1 ] (328.33,136.65) .. controls (328.33,130.39) and (333.41,125.32) .. (339.67,125.32) .. controls (345.93,125.32) and (351,130.39) .. (351,136.65) .. controls (351,142.91) and (345.93,147.98) .. (339.67,147.98) .. controls (333.41,147.98) and (328.33,142.91) .. (328.33,136.65) -- cycle ;
%Shape: Circle [id:dp7281530223464285] 
\draw  [fill={rgb, 255:red, 0; green, 0; blue, 0 }  ,fill opacity=1 ] (315.33,86.65) .. controls (315.33,80.39) and (320.41,75.32) .. (326.67,75.32) .. controls (332.93,75.32) and (338,80.39) .. (338,86.65) .. controls (338,92.91) and (332.93,97.98) .. (326.67,97.98) .. controls (320.41,97.98) and (315.33,92.91) .. (315.33,86.65) -- cycle ;
%Shape: Circle [id:dp19965111032544036] 
\draw  [fill={rgb, 255:red, 0; green, 0; blue, 0 }  ,fill opacity=1 ] (283.33,451.65) .. controls (283.33,445.39) and (288.41,440.32) .. (294.67,440.32) .. controls (300.93,440.32) and (306,445.39) .. (306,451.65) .. controls (306,457.91) and (300.93,462.98) .. (294.67,462.98) .. controls (288.41,462.98) and (283.33,457.91) .. (283.33,451.65) -- cycle ;
%Shape: Circle [id:dp7295351269258217] 
\draw  [fill={rgb, 255:red, 0; green, 0; blue, 0 }  ,fill opacity=1 ] (360.33,253.65) .. controls (360.33,247.39) and (365.41,242.32) .. (371.67,242.32) .. controls (377.93,242.32) and (383,247.39) .. (383,253.65) .. controls (383,259.91) and (377.93,264.98) .. (371.67,264.98) .. controls (365.41,264.98) and (360.33,259.91) .. (360.33,253.65) -- cycle ;
%Shape: Circle [id:dp9560417520731872] 
\draw  [fill={rgb, 255:red, 0; green, 0; blue, 0 }  ,fill opacity=1 ] (456.33,253.65) .. controls (456.33,247.39) and (461.41,242.32) .. (467.67,242.32) .. controls (473.93,242.32) and (479,247.39) .. (479,253.65) .. controls (479,259.91) and (473.93,264.98) .. (467.67,264.98) .. controls (461.41,264.98) and (456.33,259.91) .. (456.33,253.65) -- cycle ;
%Shape: Circle [id:dp13124137366839683] 
\draw  [fill={rgb, 255:red, 0; green, 0; blue, 0 }  ,fill opacity=1 ] (366.33,342.65) .. controls (366.33,336.39) and (371.41,331.32) .. (377.67,331.32) .. controls (383.93,331.32) and (389,336.39) .. (389,342.65) .. controls (389,348.91) and (383.93,353.98) .. (377.67,353.98) .. controls (371.41,353.98) and (366.33,348.91) .. (366.33,342.65) -- cycle ;
%Shape: Circle [id:dp6499665320196779] 
\draw  [fill={rgb, 255:red, 0; green, 0; blue, 0 }  ,fill opacity=1 ] (466.33,353.65) .. controls (466.33,347.39) and (471.41,342.32) .. (477.67,342.32) .. controls (483.93,342.32) and (489,347.39) .. (489,353.65) .. controls (489,359.91) and (483.93,364.98) .. (477.67,364.98) .. controls (471.41,364.98) and (466.33,359.91) .. (466.33,353.65) -- cycle ;
%Shape: Circle [id:dp4975726570075103] 
\draw  [fill={rgb, 255:red, 0; green, 0; blue, 0 }  ,fill opacity=1 ] (420.33,372.65) .. controls (420.33,366.39) and (425.41,361.32) .. (431.67,361.32) .. controls (437.93,361.32) and (443,366.39) .. (443,372.65) .. controls (443,378.91) and (437.93,383.98) .. (431.67,383.98) .. controls (425.41,383.98) and (420.33,378.91) .. (420.33,372.65) -- cycle ;
%Shape: Ellipse [id:dp7774033686279984] 
\draw  [color={rgb, 255:red, 74; green, 83; blue, 226 }  ,draw opacity=1 ] (219.81,183.72) .. controls (261.77,179.99) and (296.89,227.11) .. (298.26,288.97) .. controls (299.64,350.82) and (266.73,403.99) .. (224.77,407.72) .. controls (182.81,411.45) and (147.69,364.33) .. (146.31,302.47) .. controls (144.94,240.62) and (177.85,187.45) .. (219.81,183.72) -- cycle ;
%Shape: Ellipse [id:dp6364535216957987] 
\draw  [color={rgb, 255:red, 65; green, 117; blue, 5 }  ,draw opacity=1 ] (374.36,75.89) .. controls (390.49,94.36) and (379.37,127.79) .. (349.53,150.54) .. controls (319.68,173.29) and (282.4,176.76) .. (266.27,158.29) .. controls (250.13,139.81) and (261.25,106.39) .. (291.1,83.63) .. controls (320.94,60.88) and (358.22,57.41) .. (374.36,75.89) -- cycle ;
%Shape: Ellipse [id:dp1703715689659211] 
\draw  [color={rgb, 255:red, 65; green, 117; blue, 5 }  ,draw opacity=1 ] (349.36,250.73) .. controls (347.5,226.27) and (376.41,206.15) .. (413.94,205.8) .. controls (451.47,205.44) and (483.41,224.99) .. (485.27,249.45) .. controls (487.13,273.9) and (458.21,294.02) .. (420.68,294.37) .. controls (383.15,294.73) and (351.22,275.19) .. (349.36,250.73) -- cycle ;
%Shape: Ellipse [id:dp5535997174648917] 
\draw  [color={rgb, 255:red, 65; green, 117; blue, 5 }  ,draw opacity=1 ] (342.28,359.86) .. controls (340.41,335.32) and (377.78,315.06) .. (425.74,314.61) .. controls (473.7,314.16) and (514.09,333.68) .. (515.95,358.22) .. controls (517.82,382.76) and (480.46,403.02) .. (432.5,403.47) .. controls (384.54,403.92) and (344.15,384.4) .. (342.28,359.86) -- cycle ;
%Shape: Ellipse [id:dp6341825744536298] 
\draw  [color={rgb, 255:red, 65; green, 117; blue, 5 }  ,draw opacity=1 ] (253.81,448.49) .. controls (252.75,434.65) and (273.31,423.23) .. (299.71,422.98) .. controls (326.12,422.73) and (348.37,433.75) .. (349.43,447.59) .. controls (350.48,461.43) and (329.93,472.85) .. (303.52,473.1) .. controls (277.12,473.35) and (254.86,462.33) .. (253.81,448.49) -- cycle ;
%Straight Lines [id:da5852303179868603] 
\draw  [dash pattern={on 0.84pt off 2.51pt}]  (370,135.98) -- (406.33,201.32) ;
%Straight Lines [id:da20468509380277022] 
\draw  [dash pattern={on 0.84pt off 2.51pt}]  (349.33,382.32) -- (323.33,419.32) ;

% Text Node
\draw (122,168.98) node [anchor=north west][inner sep=0.75pt]  [font=\huge] [align=left] {$\displaystyle S$};
% Text Node
\draw (393,37.98) node [anchor=north west][inner sep=0.75pt]  [font=\huge] [align=left] {$\displaystyle K_{1}$};
% Text Node
\draw (499,188.98) node [anchor=north west][inner sep=0.75pt]  [font=\huge] [align=left] {$\displaystyle K_{i}$};
% Text Node
\draw (505,388.98) node [anchor=north west][inner sep=0.75pt]  [font=\huge] [align=left] {$\displaystyle K_{j}$};
% Text Node
\draw (363,428.98) node [anchor=north west][inner sep=0.75pt]  [font=\huge] [align=left] {$\displaystyle K_{n}$};
% Text Node
\draw (202,340.98) node [anchor=north west][inner sep=0.75pt]  [font=\large] [align=left] {$\displaystyle y$};
% Text Node
\draw (192,214.98) node [anchor=north west][inner sep=0.75pt]  [font=\large] [align=left] {$\displaystyle x$};
\end{tikzpicture}

%% file: figures/both_in_ki.tex
\tikzset{every picture/.style={line width=0.75pt}} %set default line width to 0.75pt        

\begin{tikzpicture}[x=0.75pt,y=0.75pt,yscale=-.4,xscale=.4]
%uncomment if require: \path (0,478); %set diagram left start at 0, and has height of 478

%Shape: Circle [id:dp9601191015117321] 
\draw  [fill={rgb, 255:red, 0; green, 0; blue, 0 }  ,fill opacity=1 ] (232.33,283.67) .. controls (232.33,277.41) and (237.41,272.33) .. (243.67,272.33) .. controls (249.93,272.33) and (255,277.41) .. (255,283.67) .. controls (255,289.93) and (249.93,295) .. (243.67,295) .. controls (237.41,295) and (232.33,289.93) .. (232.33,283.67) -- cycle ;
%Shape: Circle [id:dp1592786842961471] 
\draw  [fill={rgb, 255:red, 0; green, 0; blue, 0 }  ,fill opacity=1 ] (152.33,244.67) .. controls (152.33,238.41) and (157.41,233.33) .. (163.67,233.33) .. controls (169.93,233.33) and (175,238.41) .. (175,244.67) .. controls (175,250.93) and (169.93,256) .. (163.67,256) .. controls (157.41,256) and (152.33,250.93) .. (152.33,244.67) -- cycle ;
%Shape: Circle [id:dp11486051347907522] 
\draw  [fill={rgb, 255:red, 0; green, 0; blue, 0 }  ,fill opacity=1 ] (239.33,332.67) .. controls (239.33,326.41) and (244.41,321.33) .. (250.67,321.33) .. controls (256.93,321.33) and (262,326.41) .. (262,332.67) .. controls (262,338.93) and (256.93,344) .. (250.67,344) .. controls (244.41,344) and (239.33,338.93) .. (239.33,332.67) -- cycle ;
%Shape: Circle [id:dp6454395857960384] 
\draw  [fill={rgb, 255:red, 0; green, 0; blue, 0 }  ,fill opacity=1 ] (150.33,338.67) .. controls (150.33,332.41) and (155.41,327.33) .. (161.67,327.33) .. controls (167.93,327.33) and (173,332.41) .. (173,338.67) .. controls (173,344.93) and (167.93,350) .. (161.67,350) .. controls (155.41,350) and (150.33,344.93) .. (150.33,338.67) -- cycle ;
%Shape: Circle [id:dp8356839121896729] 
\draw  [fill={rgb, 255:red, 0; green, 0; blue, 0 }  ,fill opacity=1 ] (179.33,292.67) .. controls (179.33,286.41) and (184.41,281.33) .. (190.67,281.33) .. controls (196.93,281.33) and (202,286.41) .. (202,292.67) .. controls (202,298.93) and (196.93,304) .. (190.67,304) .. controls (184.41,304) and (179.33,298.93) .. (179.33,292.67) -- cycle ;
%Shape: Circle [id:dp40113233909503476] 
\draw  [fill={rgb, 255:red, 0; green, 0; blue, 0 }  ,fill opacity=1 ] (260.33,118.67) .. controls (260.33,112.41) and (265.41,107.33) .. (271.67,107.33) .. controls (277.93,107.33) and (283,112.41) .. (283,118.67) .. controls (283,124.93) and (277.93,130) .. (271.67,130) .. controls (265.41,130) and (260.33,124.93) .. (260.33,118.67) -- cycle ;
%Shape: Circle [id:dp9880352312475664] 
\draw  [fill={rgb, 255:red, 0; green, 0; blue, 0 }  ,fill opacity=1 ] (307.33,127.67) .. controls (307.33,121.41) and (312.41,116.33) .. (318.67,116.33) .. controls (324.93,116.33) and (330,121.41) .. (330,127.67) .. controls (330,133.93) and (324.93,139) .. (318.67,139) .. controls (312.41,139) and (307.33,133.93) .. (307.33,127.67) -- cycle ;
%Shape: Circle [id:dp36977527690528733] 
\draw  [fill={rgb, 255:red, 0; green, 0; blue, 0 }  ,fill opacity=1 ] (294.33,77.67) .. controls (294.33,71.41) and (299.41,66.33) .. (305.67,66.33) .. controls (311.93,66.33) and (317,71.41) .. (317,77.67) .. controls (317,83.93) and (311.93,89) .. (305.67,89) .. controls (299.41,89) and (294.33,83.93) .. (294.33,77.67) -- cycle ;
%Shape: Circle [id:dp3816812493080892] 
\draw  [fill={rgb, 255:red, 0; green, 0; blue, 0 }  ,fill opacity=1 ] (262.33,442.67) .. controls (262.33,436.41) and (267.41,431.33) .. (273.67,431.33) .. controls (279.93,431.33) and (285,436.41) .. (285,442.67) .. controls (285,448.93) and (279.93,454) .. (273.67,454) .. controls (267.41,454) and (262.33,448.93) .. (262.33,442.67) -- cycle ;
%Shape: Circle [id:dp7278612381374552] 
\draw  [fill={rgb, 255:red, 65; green, 117; blue, 5 }  ,fill opacity=1 ] (339.33,244.67) .. controls (339.33,238.41) and (344.41,233.33) .. (350.67,233.33) .. controls (356.93,233.33) and (362,238.41) .. (362,244.67) .. controls (362,250.93) and (356.93,256) .. (350.67,256) .. controls (344.41,256) and (339.33,250.93) .. (339.33,244.67) -- cycle ;
%Shape: Circle [id:dp9549940357379694] 
\draw  [fill={rgb, 255:red, 65; green, 117; blue, 5 }  ,fill opacity=1 ] (435.33,244.67) .. controls (435.33,238.41) and (440.41,233.33) .. (446.67,233.33) .. controls (452.93,233.33) and (458,238.41) .. (458,244.67) .. controls (458,250.93) and (452.93,256) .. (446.67,256) .. controls (440.41,256) and (435.33,250.93) .. (435.33,244.67) -- cycle ;
%Shape: Circle [id:dp8769805418842307] 
\draw  [fill={rgb, 255:red, 0; green, 0; blue, 0 }  ,fill opacity=1 ] (345.33,333.67) .. controls (345.33,327.41) and (350.41,322.33) .. (356.67,322.33) .. controls (362.93,322.33) and (368,327.41) .. (368,333.67) .. controls (368,339.93) and (362.93,345) .. (356.67,345) .. controls (350.41,345) and (345.33,339.93) .. (345.33,333.67) -- cycle ;
%Shape: Circle [id:dp4569910508515549] 
\draw  [fill={rgb, 255:red, 0; green, 0; blue, 0 }  ,fill opacity=1 ] (445.33,344.67) .. controls (445.33,338.41) and (450.41,333.33) .. (456.67,333.33) .. controls (462.93,333.33) and (468,338.41) .. (468,344.67) .. controls (468,350.93) and (462.93,356) .. (456.67,356) .. controls (450.41,356) and (445.33,350.93) .. (445.33,344.67) -- cycle ;
%Shape: Circle [id:dp796350512055223] 
\draw  [fill={rgb, 255:red, 0; green, 0; blue, 0 }  ,fill opacity=1 ] (399.33,363.67) .. controls (399.33,357.41) and (404.41,352.33) .. (410.67,352.33) .. controls (416.93,352.33) and (422,357.41) .. (422,363.67) .. controls (422,369.93) and (416.93,375) .. (410.67,375) .. controls (404.41,375) and (399.33,369.93) .. (399.33,363.67) -- cycle ;
%Shape: Ellipse [id:dp4614254829664912] 
\draw  [color={rgb, 255:red, 74; green, 83; blue, 226 }  ,draw opacity=1 ] (198.81,174.74) .. controls (240.77,171.01) and (275.89,218.13) .. (277.26,279.98) .. controls (278.64,341.84) and (245.73,395.01) .. (203.77,398.74) .. controls (161.81,402.47) and (126.69,355.35) .. (125.31,293.49) .. controls (123.94,231.63) and (156.85,178.47) .. (198.81,174.74) -- cycle ;
%Shape: Ellipse [id:dp32633612680982793] 
\draw  [color={rgb, 255:red, 65; green, 117; blue, 5 }  ,draw opacity=1 ] (353.36,66.9) .. controls (369.49,85.38) and (358.37,118.8) .. (328.53,141.56) .. controls (298.68,164.31) and (261.4,167.78) .. (245.27,149.3) .. controls (229.13,130.83) and (240.25,97.4) .. (270.1,74.65) .. controls (299.94,51.89) and (337.22,48.42) .. (353.36,66.9) -- cycle ;
%Shape: Ellipse [id:dp1350175418350148] 
\draw  [color={rgb, 255:red, 65; green, 117; blue, 5 }  ,draw opacity=1 ] (328.36,241.74) .. controls (326.5,217.28) and (355.41,197.17) .. (392.94,196.81) .. controls (430.47,196.46) and (462.41,216) .. (464.27,240.46) .. controls (466.13,264.92) and (437.21,285.04) .. (399.68,285.39) .. controls (362.15,285.74) and (330.22,266.2) .. (328.36,241.74) -- cycle ;
%Shape: Ellipse [id:dp512737086557087] 
\draw  [color={rgb, 255:red, 65; green, 117; blue, 5 }  ,draw opacity=1 ] (321.28,350.87) .. controls (319.41,326.34) and (356.78,306.08) .. (404.74,305.63) .. controls (452.7,305.17) and (493.09,324.7) .. (494.95,349.24) .. controls (496.82,373.77) and (459.46,394.03) .. (411.5,394.49) .. controls (363.54,394.94) and (323.15,375.41) .. (321.28,350.87) -- cycle ;
%Shape: Ellipse [id:dp37389704153063263] 
\draw  [color={rgb, 255:red, 65; green, 117; blue, 5 }  ,draw opacity=1 ] (232.81,439.51) .. controls (231.75,425.66) and (252.31,414.24) .. (278.71,413.99) .. controls (305.12,413.74) and (327.37,424.76) .. (328.43,438.6) .. controls (329.48,452.45) and (308.93,463.87) .. (282.52,464.12) .. controls (256.12,464.37) and (233.86,453.35) .. (232.81,439.51) -- cycle ;
%Straight Lines [id:da9691723243183035] 
\draw  [dash pattern={on 0.84pt off 2.51pt}]  (349,127) -- (385.33,192.33) ;
%Straight Lines [id:da5229659861125182] 
\draw  [dash pattern={on 0.84pt off 2.51pt}]  (328.33,373.33) -- (302.33,410.33) ;

% Text Node
\draw (101,160) node [anchor=north west][inner sep=0.75pt]  [font=\huge] [align=left] {$\displaystyle S$};
% Text Node
\draw (372,29) node [anchor=north west][inner sep=0.75pt]  [font=\huge] [align=left] {$\displaystyle K_{1}$};
% Text Node
\draw (478,180) node [anchor=north west][inner sep=0.75pt]  [font=\huge] [align=left] {$\displaystyle K_{i}$};
% Text Node
\draw (484,380) node [anchor=north west][inner sep=0.75pt]  [font=\huge] [align=left] {$\displaystyle K_{j}$};
% Text Node
\draw (342,420) node [anchor=north west][inner sep=0.75pt]  [font=\huge] [align=left] {$\displaystyle K_{n}$};
% Text Node
\draw (422,209) node [anchor=north west][inner sep=0.75pt]  [font=\large] [align=left] {$\displaystyle y$};
% Text Node
\draw (357,203) node [anchor=north west][inner sep=0.75pt]  [font=\large] [align=left] {$\displaystyle x$};

\end{tikzpicture}

%% file: figures/far_xy.tex
\tikzset{every picture/.style={line width=0.75pt}} %set default line width to 0.75pt        

\begin{tikzpicture}[x=0.75pt,y=0.75pt,yscale=-.4,xscale=.4]
%uncomment if require: \path (0,512); %set diagram left start at 0, and has height of 512

%Shape: Circle [id:dp6689870776686602] 
\draw  [fill={rgb, 255:red, 80; green, 227; blue, 194 }  ,fill opacity=1 ] (272.33,297.67) .. controls (272.33,291.41) and (277.41,286.33) .. (283.67,286.33) .. controls (289.93,286.33) and (295,291.41) .. (295,297.67) .. controls (295,303.93) and (289.93,309) .. (283.67,309) .. controls (277.41,309) and (272.33,303.93) .. (272.33,297.67) -- cycle ;
%Shape: Circle [id:dp3115714522210862] 
\draw  [fill={rgb, 255:red, 0; green, 0; blue, 0 }  ,fill opacity=1 ] (192.33,258.67) .. controls (192.33,252.41) and (197.41,247.33) .. (203.67,247.33) .. controls (209.93,247.33) and (215,252.41) .. (215,258.67) .. controls (215,264.93) and (209.93,270) .. (203.67,270) .. controls (197.41,270) and (192.33,264.93) .. (192.33,258.67) -- cycle ;
%Shape: Circle [id:dp2647522274767633] 
\draw  [fill={rgb, 255:red, 80; green, 227; blue, 194 }  ,fill opacity=1 ] (279.33,346.67) .. controls (279.33,340.41) and (284.41,335.33) .. (290.67,335.33) .. controls (296.93,335.33) and (302,340.41) .. (302,346.67) .. controls (302,352.93) and (296.93,358) .. (290.67,358) .. controls (284.41,358) and (279.33,352.93) .. (279.33,346.67) -- cycle ;
%Shape: Circle [id:dp4006913204618314] 
\draw  [fill={rgb, 255:red, 0; green, 0; blue, 0 }  ,fill opacity=1 ] (190.33,352.67) .. controls (190.33,346.41) and (195.41,341.33) .. (201.67,341.33) .. controls (207.93,341.33) and (213,346.41) .. (213,352.67) .. controls (213,358.93) and (207.93,364) .. (201.67,364) .. controls (195.41,364) and (190.33,358.93) .. (190.33,352.67) -- cycle ;
%Shape: Circle [id:dp9775035229554345] 
\draw  [fill={rgb, 255:red, 0; green, 0; blue, 0 }  ,fill opacity=1 ] (219.33,306.67) .. controls (219.33,300.41) and (224.41,295.33) .. (230.67,295.33) .. controls (236.93,295.33) and (242,300.41) .. (242,306.67) .. controls (242,312.93) and (236.93,318) .. (230.67,318) .. controls (224.41,318) and (219.33,312.93) .. (219.33,306.67) -- cycle ;
%Shape: Circle [id:dp6406706835490428] 
\draw  [fill={rgb, 255:red, 0; green, 0; blue, 0 }  ,fill opacity=1 ] (300.33,132.67) .. controls (300.33,126.41) and (305.41,121.33) .. (311.67,121.33) .. controls (317.93,121.33) and (323,126.41) .. (323,132.67) .. controls (323,138.93) and (317.93,144) .. (311.67,144) .. controls (305.41,144) and (300.33,138.93) .. (300.33,132.67) -- cycle ;
%Shape: Circle [id:dp06889656913027675] 
\draw  [fill={rgb, 255:red, 0; green, 0; blue, 0 }  ,fill opacity=1 ] (347.33,141.67) .. controls (347.33,135.41) and (352.41,130.33) .. (358.67,130.33) .. controls (364.93,130.33) and (370,135.41) .. (370,141.67) .. controls (370,147.93) and (364.93,153) .. (358.67,153) .. controls (352.41,153) and (347.33,147.93) .. (347.33,141.67) -- cycle ;
%Shape: Circle [id:dp20153576424906694] 
\draw  [fill={rgb, 255:red, 0; green, 0; blue, 0 }  ,fill opacity=1 ] (334.33,91.67) .. controls (334.33,85.41) and (339.41,80.33) .. (345.67,80.33) .. controls (351.93,80.33) and (357,85.41) .. (357,91.67) .. controls (357,97.93) and (351.93,103) .. (345.67,103) .. controls (339.41,103) and (334.33,97.93) .. (334.33,91.67) -- cycle ;
%Shape: Circle [id:dp8816319422446455] 
\draw  [fill={rgb, 255:red, 0; green, 0; blue, 0 }  ,fill opacity=1 ] (302.33,456.67) .. controls (302.33,450.41) and (307.41,445.33) .. (313.67,445.33) .. controls (319.93,445.33) and (325,450.41) .. (325,456.67) .. controls (325,462.93) and (319.93,468) .. (313.67,468) .. controls (307.41,468) and (302.33,462.93) .. (302.33,456.67) -- cycle ;
%Shape: Circle [id:dp18815194166319582] 
\draw  [fill={rgb, 255:red, 65; green, 117; blue, 5 }  ,fill opacity=1 ] (379.33,258.67) .. controls (379.33,252.41) and (384.41,247.33) .. (390.67,247.33) .. controls (396.93,247.33) and (402,252.41) .. (402,258.67) .. controls (402,264.93) and (396.93,270) .. (390.67,270) .. controls (384.41,270) and (379.33,264.93) .. (379.33,258.67) -- cycle ;
%Shape: Circle [id:dp8584853458992439] 
\draw  [fill={rgb, 255:red, 0; green, 0; blue, 0 }  ,fill opacity=1 ] (475.33,258.67) .. controls (475.33,252.41) and (480.41,247.33) .. (486.67,247.33) .. controls (492.93,247.33) and (498,252.41) .. (498,258.67) .. controls (498,264.93) and (492.93,270) .. (486.67,270) .. controls (480.41,270) and (475.33,264.93) .. (475.33,258.67) -- cycle ;
%Shape: Circle [id:dp5235504423764665] 
\draw  [fill={rgb, 255:red, 0; green, 0; blue, 0 }  ,fill opacity=1 ] (385.33,347.67) .. controls (385.33,341.41) and (390.41,336.33) .. (396.67,336.33) .. controls (402.93,336.33) and (408,341.41) .. (408,347.67) .. controls (408,353.93) and (402.93,359) .. (396.67,359) .. controls (390.41,359) and (385.33,353.93) .. (385.33,347.67) -- cycle ;
%Shape: Circle [id:dp9200601234551113] 
\draw  [fill={rgb, 255:red, 65; green, 117; blue, 5 }  ,fill opacity=1 ] (485.33,358.67) .. controls (485.33,352.41) and (490.41,347.33) .. (496.67,347.33) .. controls (502.93,347.33) and (508,352.41) .. (508,358.67) .. controls (508,364.93) and (502.93,370) .. (496.67,370) .. controls (490.41,370) and (485.33,364.93) .. (485.33,358.67) -- cycle ;
%Shape: Circle [id:dp7337723776415634] 
\draw  [fill={rgb, 255:red, 0; green, 0; blue, 0 }  ,fill opacity=1 ] (439.33,377.67) .. controls (439.33,371.41) and (444.41,366.33) .. (450.67,366.33) .. controls (456.93,366.33) and (462,371.41) .. (462,377.67) .. controls (462,383.93) and (456.93,389) .. (450.67,389) .. controls (444.41,389) and (439.33,383.93) .. (439.33,377.67) -- cycle ;
%Shape: Ellipse [id:dp1597583855987792] 
\draw  [color={rgb, 255:red, 74; green, 83; blue, 226 }  ,draw opacity=1 ] (238.81,188.74) .. controls (280.77,185.01) and (315.89,232.13) .. (317.26,293.98) .. controls (318.64,355.84) and (285.73,409.01) .. (243.77,412.74) .. controls (201.81,416.47) and (166.69,369.35) .. (165.31,307.49) .. controls (163.94,245.63) and (196.85,192.47) .. (238.81,188.74) -- cycle ;
%Shape: Ellipse [id:dp4919814983429309] 
\draw  [color={rgb, 255:red, 65; green, 117; blue, 5 }  ,draw opacity=1 ] (393.36,80.9) .. controls (409.49,99.38) and (398.37,132.8) .. (368.53,155.56) .. controls (338.68,178.31) and (301.4,181.78) .. (285.27,163.3) .. controls (269.13,144.83) and (280.25,111.4) .. (310.1,88.65) .. controls (339.94,65.89) and (377.22,62.42) .. (393.36,80.9) -- cycle ;
%Shape: Ellipse [id:dp9371763146435852] 
\draw  [color={rgb, 255:red, 65; green, 117; blue, 5 }  ,draw opacity=1 ] (368.36,255.74) .. controls (366.5,231.28) and (395.41,211.17) .. (432.94,210.81) .. controls (470.47,210.46) and (502.41,230) .. (504.27,254.46) .. controls (506.13,278.92) and (477.21,299.04) .. (439.68,299.39) .. controls (402.15,299.74) and (370.22,280.2) .. (368.36,255.74) -- cycle ;
%Shape: Ellipse [id:dp8524881937270623] 
\draw  [color={rgb, 255:red, 65; green, 117; blue, 5 }  ,draw opacity=1 ] (361.28,364.87) .. controls (359.41,340.34) and (396.78,320.08) .. (444.74,319.63) .. controls (492.7,319.17) and (533.09,338.7) .. (534.95,363.24) .. controls (536.82,387.77) and (499.46,408.03) .. (451.5,408.49) .. controls (403.54,408.94) and (363.15,389.41) .. (361.28,364.87) -- cycle ;
%Shape: Ellipse [id:dp8825535570385807] 
\draw  [color={rgb, 255:red, 65; green, 117; blue, 5 }  ,draw opacity=1 ] (272.81,453.51) .. controls (271.75,439.66) and (292.31,428.24) .. (318.71,427.99) .. controls (345.12,427.74) and (367.37,438.76) .. (368.43,452.6) .. controls (369.48,466.45) and (348.93,477.87) .. (322.52,478.12) .. controls (296.12,478.37) and (273.86,467.35) .. (272.81,453.51) -- cycle ;
%Straight Lines [id:da7862740692706967] 
\draw  [dash pattern={on 0.84pt off 2.51pt}]  (389,141) -- (425.33,206.33) ;
%Straight Lines [id:da41039648729551925] 
\draw  [dash pattern={on 0.84pt off 2.51pt}]  (368.33,387.33) -- (342.33,424.33) ;
%Straight Lines [id:da9474357114227299] 
\draw  [dash pattern={on 4.5pt off 4.5pt}]  (379.33,258.67) -- (293.33,292) ;
%Straight Lines [id:da4249112842956644] 
\draw  [dash pattern={on 4.5pt off 4.5pt}]  (485.33,358.67) -- (302,346.67) ;
%Straight Lines [id:da8310363854039857] 
\draw [color={rgb, 255:red, 65; green, 117; blue, 5 }  ,draw opacity=1 ] [dash pattern={on 4.5pt off 4.5pt}]  (394.33,266) -- (496.67,347.33) ;

% Text Node
\draw (141,174) node [anchor=north west][inner sep=0.75pt]  [font=\huge] [align=left] {$\displaystyle S$};
% Text Node
\draw (412,43) node [anchor=north west][inner sep=0.75pt]  [font=\huge] [align=left] {$\displaystyle K_{1}$};
% Text Node
\draw (518,194) node [anchor=north west][inner sep=0.75pt]  [font=\huge] [align=left] {$\displaystyle K_{i}$};
% Text Node
\draw (524,394) node [anchor=north west][inner sep=0.75pt]  [font=\huge] [align=left] {$\displaystyle K_{j}$};
% Text Node
\draw (382,434) node [anchor=north west][inner sep=0.75pt]  [font=\huge] [align=left] {$\displaystyle K_{n}$};
% Text Node
\draw (482,377) node [anchor=north west][inner sep=0.75pt]  [font=\large] [align=left] {$\displaystyle y$};
% Text Node
\draw (401,224) node [anchor=north west][inner sep=0.75pt]  [font=\large] [align=left] {$\displaystyle x$};
% Text Node
\draw (250,228) node [anchor=north west][inner sep=0.75pt]  [font=\large] [align=left] {$\displaystyle \gamma ( x)$};
% Text Node
\draw (230,363) node [anchor=north west][inner sep=0.75pt]  [font=\large] [align=left] {$\displaystyle \gamma ( y)$};

\end{tikzpicture}

%% file: figures/close_xy.tex
\tikzset{every picture/.style={line width=0.75pt}} %set default line width to 0.75pt        

\begin{tikzpicture}[x=0.75pt,y=0.75pt,yscale=-.4,xscale=.4]
%uncomment if require: \path (0,546); %set diagram left start at 0, and has height of 546

%Shape: Circle [id:dp3785837561169303] 
\draw  [fill={rgb, 255:red, 80; green, 227; blue, 194 }  ,fill opacity=1 ] (292.33,297.67) .. controls (292.33,291.41) and (297.41,286.33) .. (303.67,286.33) .. controls (309.93,286.33) and (315,291.41) .. (315,297.67) .. controls (315,303.93) and (309.93,309) .. (303.67,309) .. controls (297.41,309) and (292.33,303.93) .. (292.33,297.67) -- cycle ;
%Shape: Circle [id:dp3557081051069848] 
\draw  [fill={rgb, 255:red, 0; green, 0; blue, 0 }  ,fill opacity=1 ] (212.33,258.67) .. controls (212.33,252.41) and (217.41,247.33) .. (223.67,247.33) .. controls (229.93,247.33) and (235,252.41) .. (235,258.67) .. controls (235,264.93) and (229.93,270) .. (223.67,270) .. controls (217.41,270) and (212.33,264.93) .. (212.33,258.67) -- cycle ;
%Shape: Circle [id:dp1141148986422238] 
\draw  [fill={rgb, 255:red, 80; green, 227; blue, 194 }  ,fill opacity=1 ] (299.33,346.67) .. controls (299.33,340.41) and (304.41,335.33) .. (310.67,335.33) .. controls (316.93,335.33) and (322,340.41) .. (322,346.67) .. controls (322,352.93) and (316.93,358) .. (310.67,358) .. controls (304.41,358) and (299.33,352.93) .. (299.33,346.67) -- cycle ;
%Shape: Circle [id:dp6624323053620687] 
\draw  [fill={rgb, 255:red, 0; green, 0; blue, 0 }  ,fill opacity=1 ] (210.33,352.67) .. controls (210.33,346.41) and (215.41,341.33) .. (221.67,341.33) .. controls (227.93,341.33) and (233,346.41) .. (233,352.67) .. controls (233,358.93) and (227.93,364) .. (221.67,364) .. controls (215.41,364) and (210.33,358.93) .. (210.33,352.67) -- cycle ;
%Shape: Circle [id:dp8534167329095594] 
\draw  [fill={rgb, 255:red, 0; green, 0; blue, 0 }  ,fill opacity=1 ] (239.33,306.67) .. controls (239.33,300.41) and (244.41,295.33) .. (250.67,295.33) .. controls (256.93,295.33) and (262,300.41) .. (262,306.67) .. controls (262,312.93) and (256.93,318) .. (250.67,318) .. controls (244.41,318) and (239.33,312.93) .. (239.33,306.67) -- cycle ;
%Shape: Circle [id:dp11674921869384103] 
\draw  [fill={rgb, 255:red, 0; green, 0; blue, 0 }  ,fill opacity=1 ] (320.33,132.67) .. controls (320.33,126.41) and (325.41,121.33) .. (331.67,121.33) .. controls (337.93,121.33) and (343,126.41) .. (343,132.67) .. controls (343,138.93) and (337.93,144) .. (331.67,144) .. controls (325.41,144) and (320.33,138.93) .. (320.33,132.67) -- cycle ;
%Shape: Circle [id:dp6662214310342525] 
\draw  [fill={rgb, 255:red, 0; green, 0; blue, 0 }  ,fill opacity=1 ] (367.33,141.67) .. controls (367.33,135.41) and (372.41,130.33) .. (378.67,130.33) .. controls (384.93,130.33) and (390,135.41) .. (390,141.67) .. controls (390,147.93) and (384.93,153) .. (378.67,153) .. controls (372.41,153) and (367.33,147.93) .. (367.33,141.67) -- cycle ;
%Shape: Circle [id:dp2682061138462364] 
\draw  [fill={rgb, 255:red, 0; green, 0; blue, 0 }  ,fill opacity=1 ] (354.33,91.67) .. controls (354.33,85.41) and (359.41,80.33) .. (365.67,80.33) .. controls (371.93,80.33) and (377,85.41) .. (377,91.67) .. controls (377,97.93) and (371.93,103) .. (365.67,103) .. controls (359.41,103) and (354.33,97.93) .. (354.33,91.67) -- cycle ;
%Shape: Circle [id:dp8488116162934116] 
\draw  [fill={rgb, 255:red, 0; green, 0; blue, 0 }  ,fill opacity=1 ] (322.33,456.67) .. controls (322.33,450.41) and (327.41,445.33) .. (333.67,445.33) .. controls (339.93,445.33) and (345,450.41) .. (345,456.67) .. controls (345,462.93) and (339.93,468) .. (333.67,468) .. controls (327.41,468) and (322.33,462.93) .. (322.33,456.67) -- cycle ;
%Shape: Circle [id:dp6459569239697325] 
\draw  [fill={rgb, 255:red, 0; green, 0; blue, 0 }  ,fill opacity=1 ] (399.33,258.67) .. controls (399.33,252.41) and (404.41,247.33) .. (410.67,247.33) .. controls (416.93,247.33) and (422,252.41) .. (422,258.67) .. controls (422,264.93) and (416.93,270) .. (410.67,270) .. controls (404.41,270) and (399.33,264.93) .. (399.33,258.67) -- cycle ;
%Shape: Circle [id:dp3904177098967032] 
\draw  [fill={rgb, 255:red, 65; green, 117; blue, 5 }  ,fill opacity=1 ] (495.33,258.67) .. controls (495.33,252.41) and (500.41,247.33) .. (506.67,247.33) .. controls (512.93,247.33) and (518,252.41) .. (518,258.67) .. controls (518,264.93) and (512.93,270) .. (506.67,270) .. controls (500.41,270) and (495.33,264.93) .. (495.33,258.67) -- cycle ;
%Shape: Circle [id:dp09284398770972757] 
\draw  [fill={rgb, 255:red, 0; green, 0; blue, 0 }  ,fill opacity=1 ] (405.33,347.67) .. controls (405.33,341.41) and (410.41,336.33) .. (416.67,336.33) .. controls (422.93,336.33) and (428,341.41) .. (428,347.67) .. controls (428,353.93) and (422.93,359) .. (416.67,359) .. controls (410.41,359) and (405.33,353.93) .. (405.33,347.67) -- cycle ;
%Shape: Circle [id:dp70008525580856] 
\draw  [fill={rgb, 255:red, 65; green, 117; blue, 5 }  ,fill opacity=1 ] (505.33,358.67) .. controls (505.33,352.41) and (510.41,347.33) .. (516.67,347.33) .. controls (522.93,347.33) and (528,352.41) .. (528,358.67) .. controls (528,364.93) and (522.93,370) .. (516.67,370) .. controls (510.41,370) and (505.33,364.93) .. (505.33,358.67) -- cycle ;
%Shape: Circle [id:dp7794721589734246] 
\draw  [fill={rgb, 255:red, 0; green, 0; blue, 0 }  ,fill opacity=1 ] (459.33,377.67) .. controls (459.33,371.41) and (464.41,366.33) .. (470.67,366.33) .. controls (476.93,366.33) and (482,371.41) .. (482,377.67) .. controls (482,383.93) and (476.93,389) .. (470.67,389) .. controls (464.41,389) and (459.33,383.93) .. (459.33,377.67) -- cycle ;
%Shape: Ellipse [id:dp41013087160144646] 
\draw  [color={rgb, 255:red, 74; green, 83; blue, 226 }  ,draw opacity=1 ] (258.81,188.74) .. controls (300.77,185.01) and (335.89,232.13) .. (337.26,293.98) .. controls (338.64,355.84) and (305.73,409.01) .. (263.77,412.74) .. controls (221.81,416.47) and (186.69,369.35) .. (185.31,307.49) .. controls (183.94,245.63) and (216.85,192.47) .. (258.81,188.74) -- cycle ;
%Shape: Ellipse [id:dp4271737049243487] 
\draw  [color={rgb, 255:red, 65; green, 117; blue, 5 }  ,draw opacity=1 ] (413.36,80.9) .. controls (429.49,99.38) and (418.37,132.8) .. (388.53,155.56) .. controls (358.68,178.31) and (321.4,181.78) .. (305.27,163.3) .. controls (289.13,144.83) and (300.25,111.4) .. (330.1,88.65) .. controls (359.94,65.89) and (397.22,62.42) .. (413.36,80.9) -- cycle ;
%Shape: Ellipse [id:dp8961642835312253] 
\draw  [color={rgb, 255:red, 65; green, 117; blue, 5 }  ,draw opacity=1 ] (388.36,255.74) .. controls (386.5,231.28) and (415.41,211.17) .. (452.94,210.81) .. controls (490.47,210.46) and (522.41,230) .. (524.27,254.46) .. controls (526.13,278.92) and (497.21,299.04) .. (459.68,299.39) .. controls (422.15,299.74) and (390.22,280.2) .. (388.36,255.74) -- cycle ;
%Shape: Ellipse [id:dp2583936201810242] 
\draw  [color={rgb, 255:red, 65; green, 117; blue, 5 }  ,draw opacity=1 ] (381.28,364.87) .. controls (379.41,340.34) and (416.78,320.08) .. (464.74,319.63) .. controls (512.7,319.17) and (553.09,338.7) .. (554.95,363.24) .. controls (556.82,387.77) and (519.46,408.03) .. (471.5,408.49) .. controls (423.54,408.94) and (383.15,389.41) .. (381.28,364.87) -- cycle ;
%Shape: Ellipse [id:dp8704374402764121] 
\draw  [color={rgb, 255:red, 65; green, 117; blue, 5 }  ,draw opacity=1 ] (292.81,453.51) .. controls (291.75,439.66) and (312.31,428.24) .. (338.71,427.99) .. controls (365.12,427.74) and (387.37,438.76) .. (388.43,452.6) .. controls (389.48,466.45) and (368.93,477.87) .. (342.52,478.12) .. controls (316.12,478.37) and (293.86,467.35) .. (292.81,453.51) -- cycle ;
%Straight Lines [id:da5314383604545512] 
\draw  [dash pattern={on 0.84pt off 2.51pt}]  (409,141) -- (445.33,206.33) ;
%Straight Lines [id:da5278073334028348] 
\draw  [dash pattern={on 0.84pt off 2.51pt}]  (388.33,387.33) -- (362.33,424.33) ;
%Straight Lines [id:da7926356522222424] 
\draw  [dash pattern={on 4.5pt off 4.5pt}]  (495.33,258.67) -- (313.33,292) ;
%Straight Lines [id:da3290697456277969] 
\draw  [dash pattern={on 4.5pt off 4.5pt}]  (505.33,358.67) -- (322,346.67) ;
%Straight Lines [id:da21389110762924846] 
\draw [color={rgb, 255:red, 65; green, 117; blue, 5 }  ,draw opacity=1 ] [dash pattern={on 4.5pt off 4.5pt}]  (506.67,270) -- (516.67,347.33) ;

% Text Node
\draw (161,174) node [anchor=north west][inner sep=0.75pt]  [font=\huge] [align=left] {$\displaystyle S$};
% Text Node
\draw (432,43) node [anchor=north west][inner sep=0.75pt]  [font=\huge] [align=left] {$\displaystyle K_{1}$};
% Text Node
\draw (538,194) node [anchor=north west][inner sep=0.75pt]  [font=\huge] [align=left] {$\displaystyle K_{i}$};
% Text Node
\draw (544,394) node [anchor=north west][inner sep=0.75pt]  [font=\huge] [align=left] {$\displaystyle K_{j}$};
% Text Node
\draw (402,434) node [anchor=north west][inner sep=0.75pt]  [font=\huge] [align=left] {$\displaystyle K_{n}$};
% Text Node
\draw (502,377) node [anchor=north west][inner sep=0.75pt]  [font=\large] [align=left] {$\displaystyle y$};
% Text Node
\draw (480,231) node [anchor=north west][inner sep=0.75pt]  [font=\large] [align=left] {$\displaystyle x$};
% Text Node
\draw (270,228) node [anchor=north west][inner sep=0.75pt]  [font=\large] [align=left] {$\displaystyle \gamma ( x)$};
% Text Node
\draw (250,363) node [anchor=north west][inner sep=0.75pt]  [font=\large] [align=left] {$\displaystyle \gamma ( y)$};

\end{tikzpicture}

%% file: figures/intervals.tex
\tikzset{every picture/.style={line width=0.75pt}} %set default line width to 0.75pt        

\begin{tikzpicture}[x=0.75pt,y=0.75pt,yscale=-1,xscale=1]
%uncomment if require: \path (0,202); %set diagram left start at 0, and has height of 202

%Straight Lines [id:da07007727893048021] 
\draw    (41.33,100) -- (633.33,100) ;
\draw [shift={(635.33,100)}, rotate = 180] [color={rgb, 255:red, 0; green, 0; blue, 0 }  ][line width=0.75]    (10.93,-3.29) .. controls (6.95,-1.4) and (3.31,-0.3) .. (0,0) .. controls (3.31,0.3) and (6.95,1.4) .. (10.93,3.29)   ;
\draw [shift={(39.33,100)}, rotate = 0] [color={rgb, 255:red, 0; green, 0; blue, 0 }  ][line width=0.75]    (10.93,-3.29) .. controls (6.95,-1.4) and (3.31,-0.3) .. (0,0) .. controls (3.31,0.3) and (6.95,1.4) .. (10.93,3.29)   ;
%Straight Lines [id:da8050305541857101] 
\draw [color={rgb, 255:red, 155; green, 155; blue, 155 }  ,draw opacity=1 ] [dash pattern={on 4.5pt off 4.5pt}]  (139.33,90.33) -- (139.33,114.33) ;
%Straight Lines [id:da6525216157077252] 
\draw [color={rgb, 255:red, 208; green, 2; blue, 27 }  ,draw opacity=1 ]   (250.33,89.33) -- (250.33,110.33) ;
\draw [shift={(250.33,113.33)}, rotate = 270] [fill={rgb, 255:red, 208; green, 2; blue, 27 }  ,fill opacity=1 ][line width=0.08]  [draw opacity=0] (8.93,-4.29) -- (0,0) -- (8.93,4.29) -- cycle    ;
%Straight Lines [id:da12349643407931499] 
\draw [color={rgb, 255:red, 65; green, 117; blue, 5 }  ,draw opacity=1 ]   (330.33,90.33) -- (330.33,111.33) ;
\draw [shift={(330.33,114.33)}, rotate = 270] [fill={rgb, 255:red, 65; green, 117; blue, 5 }  ,fill opacity=1 ][line width=0.08]  [draw opacity=0] (8.93,-4.29) -- (0,0) -- (8.93,4.29) -- cycle    ;
%Straight Lines [id:da48687956709832414] 
\draw [color={rgb, 255:red, 155; green, 155; blue, 155 }  ,draw opacity=1 ] [dash pattern={on 4.5pt off 4.5pt}]  (290.33,89.33) -- (290.33,113.33) ;
%Straight Lines [id:da9744971998799346] 
\draw [color={rgb, 255:red, 155; green, 155; blue, 155 }  ,draw opacity=1 ] [dash pattern={on 4.5pt off 4.5pt}]  (160.33,93.33) -- (160.33,114.33) ;
\draw [shift={(160.33,90.33)}, rotate = 90] [fill={rgb, 255:red, 155; green, 155; blue, 155 }  ,fill opacity=1 ][line width=0.08]  [draw opacity=0] (8.93,-4.29) -- (0,0) -- (8.93,4.29) -- cycle    ;
%Straight Lines [id:da8797342998970543] 
\draw [color={rgb, 255:red, 155; green, 155; blue, 155 }  ,draw opacity=1 ] [dash pattern={on 4.5pt off 4.5pt}]  (420.33,90.33) -- (420.33,114.33) ;
%Straight Lines [id:da26518757107996604] 
\draw [color={rgb, 255:red, 155; green, 155; blue, 155 }  ,draw opacity=1 ] [dash pattern={on 4.5pt off 4.5pt}]  (489.33,90.33) -- (489.33,114.33) ;
%Straight Lines [id:da21682743882841127] 
\draw [color={rgb, 255:red, 226; green, 161; blue, 161 }  ,draw opacity=1 ]   (460.33,90.33) -- (460.33,111.33) ;
\draw [shift={(460.33,114.33)}, rotate = 270] [fill={rgb, 255:red, 226; green, 161; blue, 161 }  ,fill opacity=1 ][line width=0.08]  [draw opacity=0] (8.93,-4.29) -- (0,0) -- (8.93,4.29) -- cycle    ;
%Straight Lines [id:da13021655185583425] 
\draw [color={rgb, 255:red, 226; green, 161; blue, 161 }  ,draw opacity=1 ]   (480.33,90.33) -- (480.33,114.33) ;
%Straight Lines [id:da7099853047621081] 
\draw [color={rgb, 255:red, 226; green, 161; blue, 161 }  ,draw opacity=1 ]   (510.33,90.33) -- (510.33,114.33) ;
%Straight Lines [id:da3882109197181791] 
\draw [color={rgb, 255:red, 226; green, 161; blue, 161 }  ,draw opacity=1 ]   (530.33,90.33) -- (530.33,114.33) ;
%Straight Lines [id:da32076870058011475] 
\draw [color={rgb, 255:red, 226; green, 161; blue, 161 }  ,draw opacity=1 ]   (569.33,90.33) -- (569.33,114.33) ;
%Shape: Rectangle [id:dp16171384066231043] 
\draw  [color={rgb, 255:red, 255; green, 255; blue, 255 }  ,draw opacity=1 ][fill={rgb, 255:red, 184; green, 233; blue, 134 }  ,fill opacity=1 ] (249,27) -- (460.33,27) -- (460.33,41) -- (249,41) -- cycle ;
%Shape: Rectangle [id:dp09327976716750852] 
\draw  [color={rgb, 255:red, 255; green, 255; blue, 255 }  ,draw opacity=1 ][fill={rgb, 255:red, 225; green, 91; blue, 104 }  ,fill opacity=1 ] (250,62.33) -- (331.33,62.33) -- (331.33,76.67) -- (250,76.67) -- cycle ;
%Straight Lines [id:da07609027231860277] 
\draw    (99.33,40.33) -- (100.33,157.33) ;
%Straight Lines [id:da6117217692038901] 
\draw    (599.33,39.33) -- (599.33,154.33) ;

% Text Node
\draw (122,141) node [anchor=north west][inner sep=0.75pt]   [align=left] {$ $};
% Text Node
\draw (201.33,127.33) node [anchor=north west][inner sep=0.75pt]  [font=\footnotesize] [align=left] {$\displaystyle \textcolor[rgb]{0.82,0.01,0.11}{\eta ( u) =\frac{d( x,u)}{d( x,\gamma ( x))}} \ $};
% Text Node
\draw (313.33,126.33) node [anchor=north west][inner sep=0.75pt]  [font=\footnotesize] [align=left] {$\displaystyle \textcolor[rgb]{0.25,0.46,0.02}{\frac{d( y,u)}{d( y,\gamma ( y))} \ }$};
% Text Node
\draw (144,64) node [anchor=north west][inner sep=0.75pt]   [align=left] {$\displaystyle \textcolor[rgb]{0.61,0.61,0.61}{\eta ( v)}$};
% Text Node
\draw (418,123) node [anchor=north west][inner sep=0.75pt]   [align=left] {$\displaystyle \textcolor[rgb]{0.89,0.36,0.43}{\eta ( w) =\eta '( u,\pi )}$};
% Text Node
\draw (307,6) node [anchor=north west][inner sep=0.75pt]   [align=left] {$ $$\displaystyle w( u,\pi )$};
% Text Node
\draw (278,48) node [anchor=north west][inner sep=0.75pt]   [align=left] {$ $$\displaystyle W$};
% Text Node
\draw (95,167) node [anchor=north west][inner sep=0.75pt]   [align=left] {$\displaystyle 2$};
% Text Node
\draw (583,165.67) node [anchor=north west][inner sep=0.75pt]   [align=left] {$\displaystyle 2+\tau $};

\end{tikzpicture}

%% file: maincontent/full_nested.tex
In this section, we will give a modification of Algorithm \ref{alg:extend} when $(Y,d_Y)$ is $\ell_1$ space that will allow us to avoid excessive contraction. We will then use this to argue that compositions of nested embeddings exist for embeddings into $\ell_1$. This will allow us to prove Theorems \ref{thm:nested-comp} and \ref{thm:nested-comp-strong}.

As in the previous case, we first give Algorithm \ref{alg:nest}, a randomized algorithm that will have good expected distortion for every pair, but now we will work to ensure the distances are not too contracted in addition to not being too expanded.

\input{maincontent/alg2}

Note that Algorithm \ref{alg:nest} appends together a bunch of embeddings, one for $S$ and one for each $K_i$. The idea is that the main source of our contraction in the previous algorithm comes from the fact that many points may be mapped to the same place. To avoid this, we will use our embedding $\alpha_X$ to add some extra indices that will distinguish points that end up in the same set from each other and from their center's neighbor.

Let  $c_S$ and $c_X$ denote the distortion of $\alpha_S$ and $\alpha_X$ respectively. We now show that if $\alpha$ is the output of Algorithm \ref{alg:nest} on input $((X,d),S,\alpha_S,\alpha_X,\rangwidth)$ with $\rangwidth=2$, the distortion between elements in $S$ is at most $c_S$ and all other distortion is at most $ \closexytwo$ in expectation. Our argument is broken into two parts: Lemma~\ref{lem:expanding} shows that the embedding $\alpha$ has low contraction; Lemma~\ref{lem:expansion} bounds the amount by which every distance expands. 

We state the lemmas first and then prove them in the following subsections. Throughout these arguments we assume that $\rangwidth=2$, although it is possible to obtain slightly better distortion bounds by choosing a value for $\rangwidth$ carefully. We present general versions of the lemmas, exhibiting the dependence of the bounds on $\rangwidth$ in Appendix \ref{sec:full_constants}. 
% Theorem~\ref{thm:nested-comp} follows in a straightforward manner from these lemmas; Section~\ref{sec:improvements} contains a proof of this theorem as well as a proof of the strengthened composition result -- Theorem~\ref{thm:nested-comp-strong}.

\begin{lem}\label{lem:expanding}
Let $\alpha\from$ Algorithm \ref{alg:nest}$((X,d),S,\alpha_S,\alpha_X,\rangwidth)$ with $\rangwidth=2$. Then for all $x,y\in X$, we have $\distl{1}{\alpha(x)}{\alpha(y)}\geq d(x,y)$. 
% Furthermore, for $x,y\in S$, we have $\distl{1}{\alpha(x)}{\alpha(y)}\geq d(x,y)$. 
\end{lem}

\begin{lem}\label{lem:expansion}
Let $\alpha\from$ Algorithm \ref{alg:nest}$((X,d),S,\alpha_S,\alpha_X,\rangwidth)$ with $\rangwidth=2$. Then we have the following bounds on the expansion for each pair $x,y\in X$:
\begin{enumerate}[(a)]
    \item \label{lem:not-outlier} If $x,y\in S$, then $\distl{1}{\alpha(x)}{\alpha(y))}\leq  c_S\cdot d(x,y)$.
    \item \label{lem:sameset} If $x,y\in K_i$, then $\distl{1}{\alpha(x)}{\alpha(y))}\leq \bothinkitwo\cdot d(x,y)$.
    \item \label{lem:ui} If $x\in S,y\in X\setminus S$, then $\distl{1}{\alpha(x)}{\alpha(y))}\leq \oneineachtwo\cdot d(x,y)$.
    \item \label{lem:farxy} If $x,y\in X\setminus S$
    { and $d(x,\gamma(x))\leq 2\cdot d(x,y)$ for $\gamma$ as defined in line 2 of the algorithm}, then $\distl{1}{\alpha(x)}{\alpha(y))}\leq \farxytwo\cdot d(x,y)$.
    % \kristin{don't we need a condition on their relative distance for this one? otherwise it looks odd to have d and e. Also should we say expected expansion at the top?}
    \item \label{lem:closexy} If $x,y\in X\setminus S$ { and $d(x,\neigh(x)),d(y,\neigh(y))> 2\cdot d(x,y)$ for $\neigh$ as defined in line 2 of the algorithm}, then 
    % \kristinEDIT{$\distl{1}{\alpha(x)}{\alpha(y))}$}
    {$E_\alpha[\distl{1}{\alpha(x)}{\alpha(y))}]$} $\leq \closexytwo\cdot d(x,y)$.
\end{enumerate}
\end{lem}

\subsection{Proof of {contraction bounds}}

\begin{proofof}{Lemma~\ref{lem:expanding}}
    Let $\neigh$ be as defined in line \eqref{def-gamma-full} of the algorithm and let the $u_i$ and $K_i$ be as defined in lines \eqref{ui-def} and \eqref{def-ki} of the algorithm. We divide into cases.
    \begin{enumerate}
        \item If $x,y\in S$, then $\distl{1}{\alpha(x)}{\alpha(y)}= \distl{1}{\alpha'(x)}{\alpha'(y)}\geq d(x,y)$ because $\alpha_S$ is expanding 
        \item If $x,y\in K_i$ for some $K_i$ defined in line \eqref{def-ki-full} of the algorithm or if $x\in K_i,y=\neigh(u_i)$, their embeddings differ only on the coordinates associated with $\alpha_i$, so we have 
        \begin{align*}
            \distl{1}{\alpha(x)}{\alpha(y)} &= \distl{1}{\alpha_i(x)}{\alpha_i(y)} \\
            &=  \distl{1}{\alpha_X(x)}{\alpha_X(y)} \\
            &\geq  d(x,y),
        \end{align*}
        where the second line is by definition of $\alpha_i$ and the last line is because $\alpha_X$ is expanding.
        \item If $x\in S,y\notin S$, then let $y\in K_i$ for some $i$ as defined in line \eqref{def-ki-full} of the algorithm and let $u_i$ be the center for this $i$ defined in line \eqref{ui-def-full}. This implies that $\alpha(x)$ and $\alpha(y)$ differ only on coordinates associated with $\alpha'$ and $\alpha_i$.
        \begin{align*}
            \distl{1}{\alpha(x)}{\alpha(y)} &= \distl{1}{\alpha'(x)}{\alpha'(y)}+\distl{1}{\alpha_i(x)}{\alpha_i(y)} \\
            &\geq \distl{1}{\alpha_S(x)}{\alpha_S(\neigh(u_i))}+\distl{1}{\alpha_X(\neigh(u_i))}{\alpha_X(y)} \\
            &\geq d(x,\neigh(u_i))+d(\neigh(u_i),y) \\
            &\geq d(x,y),
        \end{align*}
        where the second line is by definition of $\alpha'$ and $\alpha_i$, the third is by cases 1 and 2 of this lemma,  and the fifth is by the triangle inequality.
        \item If $x\in K_i,y\in K_j$ for some $K_i,K_j$ defined in line \eqref{def-ki-full} of the algorithm. Let $u_i,u_j$ be the centers of these sets as defined in line \eqref{ui-def-full} of the algorithm. Note that $\alpha(x)$ and $\alpha(y)$ differ only on the indices associated with $\alpha',\alpha_i$, and $\alpha_j$. Thus we get
        \begin{align*}
            \distl{1}{\alpha(x)}{\alpha(y)}^p &= \distl{1}{\alpha'(x)}{\alpha'(y)}+\distl{1}{\alpha_i(x)}{\alpha_i(y)}+\distl{1}{\alpha_j(x)}{\alpha_j(y)} \\
            &\geq \distl{1}{\alpha_S(\neigh(u_i))}{\alpha_S(\neigh(u_j))}+\distl{1}{\alpha_X(\neigh(u_i))}{\alpha_X(x)}+\distl{1}{\alpha_X(\neigh(u_j))}{\alpha_X(y)} \\
            &\geq d(\neigh(u_i),\neigh(u_j))+d(\neigh(u_i),x)+d(\neigh(u_j),y)\\
            &\geq  d(x,y),
        \end{align*}
        where the second line is by definition of the three embeddings, the third line is by cases 1 and 2 of this lemma, and the fifth is by the triangle inequality. 
    \end{enumerate}
\end{proofof}

{Note that if we were to use Algorithm \ref{alg:nest} to nest embeddings into more general $\lspace{p}$ spaces, we can use the same analysis as above, but rather than adding the distance between the concatenated vectors, we will need to raise the distances to the power of $p$, add them, and take the sum to the power $1/p$. Then by applying the power mean inequality we will get that contraction is at most $\shrink$ on all pairs of points and at most $1$ on pairs in $S$.} 
% \kristin{should I leave this here or move to a footnote?}

\subsection{Proofs of expansion bounds}

The bounds in Lemma~\ref{lem:expansion} are summarized in Table \ref{tab:distortion}. We will prove each statement separately.

\input{maincontent/distortion_table_full}

%\begin{lem}   \label{lem:not-outlier}
%Let $\alpha\from$ Algorithm %\ref{alg:nest}$((X,d),S,p,\alpha_S,\alpha_X,\randrangtwo)$ and $u,v\in S$. Then $\distl{1}{\alpha(u)}{\alpha(v))}\leq \shrink\cdot c_S\cdot d(u,v)$.
%\end{lem}

\begin{proofof}{Lemma~\ref{lem:expansion}~\eqref{lem:not-outlier}} 
% \kristinEDIT{Note that $\alpha_j(u)=\alpha_j(v)$ for all $j>t'$\kristin{what's $t'$?}, as}{} 
$u,v\notin K_i$ for all $i$, so $\alpha_i(u)=\alpha_i(v)$ for all $i$. Thus,
\begin{align*}
    \distl{1}{\alpha(u)}{\alpha(v))} &=\distl{1}{\alpha'(u)}{\alpha'(v)} \\
    &=\distl{1}{\alpha_S(u)}{\alpha_S(v)}\\
    &\leq  c_S\cdot d(u,v),
\end{align*} 
where the last equality is by definition of $\alpha_S$. 
\end{proofof}

%\begin{lem}\label{lem:sameset}
%    Let $\alpha\from$ Algorithm \ref{alg:nest}$((X,d),S,p,\alpha_S,\alpha_X,\randrangtwo)$. Consider $x,y\in K_i$  for some $K_i$ as defined in line 7 of the algorithm. Then $\distl{1}{\alpha(x)}{\alpha(y))}\leq \shrink \cdot \origdist \cdot d(x,y)$. 
%\end{lem}

\begin{proofof}{Lemma~\ref{lem:expansion}~\eqref{lem:sameset}}
    Since $x,y\in K_i$ for a fixed $i$, we have that $\alpha(x),\alpha(y)$ differ only on coordinates associated with $\alpha_i$. Thus, we get 
    \begin{align*}
        \distl{1}{\alpha(x)}{\alpha(y))} &= \distl{1}{\alpha_i(x)}{\alpha_i(y))} \\
        &= \distl{1}{\alpha_X(x)}{\alpha_X(y))}\\
        &\leq \origdist\cdot  d(x,y),
    \end{align*}
    where we have used the fact that $\alpha_X(u)=\alpha_i(u)$ for $x\in K_i$.
\end{proofof}

% \begin{figure}
%     \centering
%     \input{figures/both_in_ki}
%     \caption{Visualization of nodes referenced in Lemma~\ref{lem:expansion}~\eqref{lem:sameset} }
%     \label{fig:both_in_ki}
% \end{figure}

Now we consider the distortion between outliers and non-outliers. To prove these bounds, we will need to define extra points in the $\ell_p$ space that $X$ is mapped to. Consider the points $u_i$ defined in Step~\eqref{ui-def-full} of the algorithm. We will now define a point $\beta(u_i)$ for each such $u_i\in K$. If $u_i\in K_i$ we set $\beta(u_i)=\alpha(u_i)$. Otherwise if $u_i\in K_j$ for $j<i$, we set all of the coordinates of $\beta(u_i)$ to be the same as $\alpha(u_i)$, except for the coordinates associated with $\alpha_j(u_i)$, that are replaced with those of $\alpha_X(\neigh(u_j))$; and the coordinates associated with $\alpha_i(u_i)$, that are replaced with those for $\alpha_X(u_i)$. In other words, $\beta(u_i)$ is the point that $u_i$ would be mapped to by $\alpha$ if it had so happened that $u_i$ belonged to $K_i$.

Now we are ready to prove Lemma~\ref{lem:expansion}~\eqref{lem:ui}. 
We break this proof up into two parts: Lemmas~\ref{lem:SvsKi} and \ref{lem:ui-separate}.

\begin{lem}\label{lem:SvsKi}
    Let $\alpha\from$ Algorithm \ref{alg:nest}$((X,d),S,\alpha_S,\alpha_X,\rangwidth)$ with $\rangwidth=2$. Consider $x\in S$; $u_i$ as defined in Step~\eqref{ui-def} of the algorithm for some $i$; and $\beta(u_i)$ as defined above.
    %and consider a new point $\beta(u_i)$. If $u_i\in K_i$, let $\beta(u_i)=\alpha(u_i)$. Otherwise if $u_i\in K_j$ for $j<i$, let $\beta(u_i)$ be the same as $\alpha(u_i)$ where the values on coordinates associated with $\alpha_j(u_i)$ are replaced with those of $\alpha_X(\neigh(u_j))$ and those for $\alpha_i(u_i)$ are replaced with  those for $\alpha_X(u_i)$.    
    If $x=\neigh(u_i)$ where $\neigh:X\setminus S\fto S$ is as defined in Step~\eqref{def-gamma} of the algorithm, then $\distl{1}{\alpha(x)}{\beta(u_i)}\leq \origdist \cdot d(x,u_i)$. For all other $x$, $\distl{1}{\alpha(x)}{\beta(u_i)}\leq (\subdist+\origdist) \cdot d(x,u_i)$. 
\end{lem}
\begin{proof}
    First, note that the point $\beta(u_i)$ is exactly the point that would have been assigned to $\alpha(u_i)$ if it was not placed in some $K_j$ for $j<i$. Additionally, this means that $\beta(u_i)$ and $\alpha(x)$ are the same on the indices associated with $\alpha_j$ for all $j\neq i$ (namely they are assigned $\alpha_X(\neigh(u_j))$).

    Now we have $\alpha_j(x)=\alpha_j(u_i)$ for all $j\neq i$. Thus, we have \begin{align*}
        \distl{1}{\alpha(x)}{\beta(u_i)}&=\distl{1}{\alpha'(x)\concat\alpha_i(x)}{\alpha'(u_i)\concat\alpha_X(u_i)}\\ 
        &\leq \distl{1}{\alpha'(x)}{\alpha'(u_i)}+ \distl{1}{\alpha_i(x)}{\alpha_X(u_i)}.
    \end{align*} 
    
   We have two cases to consider.

    \begin{enumerate}
        \item If $x=\neigh(u_i)$, then we have $\alpha'(u_i)=\alpha'(x)$ by definition, so we have \begin{align*}
            \distl{1}{\alpha(x)}{\beta(u_i)} &\leq \distl{1}{\alpha_i(x)}{\alpha_X(u_i)} \\
            &=\distl{1}{\alpha_X(x)}{\alpha_X(u_i)} \\
            &\leq \origdist \cdot d(x,u_i),
        \end{align*}
        where the last line is by the fact that $\alpha_X$ is an embedding with at most $\origdist$ distortion.
        \item If $x\neq \neigh(u_i)$, we have $d(\neigh(u_i),u_i)\leq d(x,u_i)$ by definition of $\neigh(u_i)$. Then we use the triangle inequality as follows.
        \begin{align*}
            \distl{1}{\alpha(x)}{\beta(u_i))} &\leq  \distl{1}{\alpha(x)}{\alpha(\neigh(u_i))}+\distl{1}{\alpha(\neigh(u_i))}{\beta(u_i)} \\
            % &= \distl{1}{\alpha(x)}{\alpha(\neigh(u_i))}+\distl{1}{\alpha_X(\neigh(u_i))}{\alpha_X(u_i)} \\
            &\leq  \subdist \cdot d(x,u_i) +  \origdist\cdot d(\neigh(u_i),u_i) \\
            &\leq  \subdist \cdot d(x,u_i) + \origdist\cdot d(x,u_i)\\
            &= (\subdist+\origdist)d(x,u_i)
        \end{align*}
        where the  second line comes from Lemma~\ref{lem:expansion}~\eqref{lem:not-outlier} and the first case of this lemma. 
    \end{enumerate}
\end{proof}

\begin{figure}
    \centering
    \input{figures/one_in_each}
    \caption{Visualization of nodes referenced in Lemmas  \ref{lem:SvsKi} and \ref{lem:ui-separate} when $u_i\in K_i$. }
    \label{fig:one_in_each}
\end{figure}
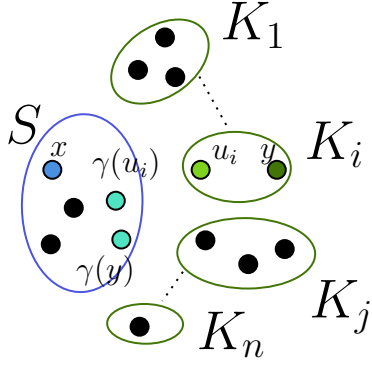

Next we want to consider the distortion between more general members of a set $K_i$ and members of $X\setminus K_i$. For the coming proofs, to show that $x$ and $y$ are not too distorted, we will generally focus on showing that the distance between $x$ and $y$ is at least a constant factor larger than the distance from $x$ or $y$ to some other point whose distance to $x$ or $y$ is already known not to be too distorted.

\begin{lem}\label{lem:ui-separate}
         Let $\alpha\from$ Algorithm \ref{alg:nest}$((X,d),S,\alpha_S,\alpha_X,\rangwidth)$ with $\rangwidth=2$. Let $K_i$ be as defined in Step~\eqref{def-ki} of the algorithm for some $i$. Consider $x\in S$ and $y\in K_i$ for some $i$. If $x=\neigh(y)$ where $\neigh$ is as defined in line \eqref{def-gamma} of the algorithm, then we have $\distl{1}{\alpha(x)}{{\alpha(y)}}\leq  [5\cdot \subdist+9\cdot \origdist] \cdot d(x,y)$. For all other $x$, $\distl{1}{\alpha(x)}{\alpha(y)}\leq   [7\cdot \subdist+9\cdot \origdist] \cdot d(x,y)$. 
\end{lem}

\begin{proof}
    % Note that $\alpha_j(x)=\alpha_j(y)$ for all $j\neq i$ 
    Let $\neigh:X\setminus S\fto S$ be as defined in Step~\eqref{def-gamma} of the algorithm, $u_i$ be as defined in Step~\eqref{ui-def} for $i$ consistent with the lemma statement, and $\beta(u_i)$ be as defined above. We divide analysis into two cases. 
    %Define a new point $\beta(u_i)$. If $u_i\in K_i$, let $\beta(u_i)=\alpha(u_i)$. Otherwise if $u_i\in K_j$ for $j<i$, let $\beta(u_i)$ be the same as $\alpha(u_i)$ where the input associated with $\alpha_j(u_i)$ is replaced with $\alpha_S(\neigh(u_j))$ and the indices for $\alpha_i(u_i)$ are replaced with  $\alpha_X(u_i)$. (This is the same as it is defined in the previous lemma statement.)
    \begin{enumerate}   
    \item If $x=\neigh(y)$, the lemma essentially comes from the triangle inequality and $y$'s relative closeness to $u_i$ that it must have if it is assigned to $K_i$.
    \begin{align*}
        \distl{1}{\alpha(x)}{\alpha(y)} & =\\
        \distl{1}{\alpha(\neigh(y))}{\alpha(y)} &\leq \distl{1}{\alpha(x)}{\beta(u_i)} + \distl{1}{\alpha(y)}{\beta(u_i)} \\
        &\leq [(\subdist+\origdist)\cdot d(x=\gamma(y),u_i) + \origdist d(y,u_i)] \\
        &\leq (\subdist+\origdist)\cdot [d(\neigh(y),y)+d(y,u_i)] +\origdist\cdot d(y,u_i) \\
        &= (\subdist+\origdist)\cdot d(\gamma(y),y) + (\subdist+2\origdist)\cdot d(y,u_i) \\
        &\leq  (\subdist+\origdist)\cdot d(\gamma(y),y) + 4(\subdist+2\origdist)\cdot d(y,\neigh(y))  \\
        &= [5\subdist+9\origdist]\cdot d(y,\neigh(y)=x).
    \end{align*}

    Here the first inequality is by the fact that $\alpha(x)$ and $\alpha(y)$ may only differ on the indices associated with $\alpha'$ and the indices associated with $\alpha_i$. The second inequality is by Lemma \ref{lem:SvsKi} and Lemma~\ref{lem:expansion}~\eqref{lem:sameset}, the third is by the triangle inequality, and the fourth line is by rearranging terms. The fifth line is by the fact that $d(y,u_i)\leq \decider \cdot d(y,\neigh(y))$ since $y$ is in $K_i$ and $u_i$ is the center, and the fact that $\decider\leq 4$ no matter the result of the random choice in the algorithm. 
    
    \item If $x\neq \neigh(y)$, we have $d(\neigh(y),y)\leq d(x,y)$ by definition of $\neigh(y)$. Thus, we get 
    \begin{align*}
        \distl{1}{\alpha(x)}{\alpha(y)} &\leq \distl{1}{\alpha(x)}{\alpha(\neigh(y))} +\distl{1}{\alpha(\neigh(y))}{\alpha(y)} \\
        &\leq  [ \subdist \cdot d(x,\neigh(y)) +[5\subdist+9\origdist]\cdot d(\neigh(y),y) ] \\
        &\leq [\subdist \cdot [d(x,y)+d(y,\neigh(y))] +[5\subdist+9\origdist]\cdot d(\neigh(y),y)] \\
        &\leq  [7\subdist+9\origdist]\cdot d(x,y),
    \end{align*}
    where the second inequality is by Lemma~\ref{lem:expansion}~\ref{lem:not-outlier} and by the previous case of this lemma, the third inequality is by the triangle inequality, and the last inequality is because $\neigh(y)$ is a closest node in $S$ to $y$.
    \end{enumerate}
\end{proof}

Finally, we consider comparing two the distance of two nodes $x,y\in X\setminus S$. First we consider the case that at least one of the nodes has a relatively short distance to $S$ compared to the distance to the other node.

%\begin{lem}\label{lem:farxy}
%     Let $\alpha\from$ Algorithm \ref{alg:nest}$((X,d),S,p,\alpha_S,\alpha_X,\randrangtwo)$. Consider $x,y\in X-S$ with $d(x,\neigh(x))\leq \comparisontwo\cdot d(x,y)$. Then $\distl{1}{\alpha(x)}{\alpha(y)}\leq \shrink \cdot [46\subdist + 3\comparison\cdot 66 \origdist ]\cdot d(x,y)$.
%\end{lem}

\begin{proofof}{Lemma~\ref{lem:expansion}~\eqref{lem:farxy}}
    Let $\neigh$ be as defined in line \eqref{def-gamma} of the algorithm.
    We get 
    \begin{align*}
        \distl{1}{\alpha(x)}{\alpha(y)} &\leq \distl{1}{\alpha(x)}{\alpha(\neigh(x))} + \distl{1}{\alpha(\neigh(x))}{\alpha(y)}\\
        &\leq  [5\subdist+9\origdist] \cdot d(x,\neigh(x)) +  [7\subdist+9\origdist] \cdot d(\neigh(x),y) \\
        &\leq [12\subdist+18\origdist] \cdot d(x,\neigh(x)) +  [7\subdist+9\origdist] \cdot d(x,y) \\
        &\leq [31\subdist + 45 c_X]\cdot  d(x,y),
    \end{align*}
    where the first inequality is by the triangle inequality, the second is by Lemma \ref{lem:SvsKi}, the third is by the triangle inequality applied to $d(\neigh(x),y)$ and rearranging terms, and the fourth is by the fact that $d(x,\neigh(x))\leq \comparisontwo \cdot d(x,y)$ and rearranging terms.
\end{proofof}

Now we consider the final case, where we must consider expected distance.

\begin{proofof}{Lemma~\ref{lem:expansion}~\eqref{lem:closexy}}
    Let $\neigh$ be as defined in line ~\eqref{def-gamma} of the algorithm.
    We will say that $x$ and $y$ are ``split" if $x\in K_i,y\in K_j$ for $i\neq j$ and $K_i,K_j$ as defined in line \eqref{def-ki} of the algorithm. Let $u_i,u_j$ be as defined in line \eqref{ui-def} of the algorithm for the same choice of $i,j$ respectively. 

    \begin{itemize}
        \item First consider the worst-case distortion when $x$ and $y$ are \textit{not} split. Then $x,y\in K_i$ for some $i$. We have $\distl{1}{\alpha(x)}{\alpha(y)}\leq \origdist \cdot d(x,y)$ by Lemma~\ref{lem:expansion}~\ref{lem:sameset}, no matter the choice of $\decider$. 

        \item Now consider the worst-case distortion when  $x$ and $y$ \textit{are} split such that $x$ is placed in $K_i$ and $y$ is placed in $K_j$ for $i<j$. We have
        % (We can assume this wlog by symmetry.) %Now consider the maximum distance $\distl{1}{\alpha(x)}{\alpha(y)}$ in this case. We get 
            \begin{align*}
                \distl{1}{\alpha(x)}{\alpha(y)} &\leq \distl{1}{\alpha(x)}{\alpha(\neigh(x))} + \distl{1}{\alpha(\neigh(x))}{\alpha(y)}  \\
                &\leq  [5\subdist+9\origdist]\cdot d(x,\neigh(x)) + [7\subdist+9\origdist]\cdot d(\neigh(x),y)  \\
                &\leq  [5\subdist+9\origdist]\cdot d(x,\neigh(x)) + [7\subdist+9\origdist]\cdot [d(x,y)+d(x,\neigh(x)) ] \\
                &\leq  [5\subdist+9\origdist]\cdot d(x,\neigh(x))  + [7\subdist+9\origdist]\cdot \frac{3}{2} d(x,\neigh(x))  \\
                &=  [\frac{31}{2}\subdist + \frac{45}{2}\origdist ] \cdot d(x,\neigh(x)),
            \end{align*}

            where the first inequality is by the triangle inequality, the second is by Lemma \ref{lem:ui-separate}, the third inequality is by the triangle inequality on $d(\gamma(x),y)$, the fourth inequality is by the fact that $d(x,y)<\frac{1}{\comparisontwo}d(x,\gamma(x))$ by the condition of this lemma, and the final equality is by rearranging terms.

            %Note that $b$ is upper bounded in value by the upper bound of the range $b$ is chosen from.

 \end{itemize}

%Let $K_u$ be the cluster formed in line 8 such that $u$ is the center of that cluster. Consider $\Pr_{b,\pi}[x,y \text{ are split }| u \text{ decides }(x,y)]$. 

% In this case we have $d(x,\neigh(u_i))\leq b\cdot d(x,\neigh(x))$ \textit{and} $d(y,\neigh(u_i))>b\cdot d(y,\neigh(y))$, by how we defined $K_i$. 
Note that the probability of $x$ and $y$ being split is no different in this case than in the proof of Lemma~\ref{lem:expansion-weak}~\eqref{lem:closexy}. The only changes are to the amount of distortion incurred when they are split versus when they are not split. Using the same definitions as in the proof of Lemma~\ref{lem:expansion-weak}~\eqref{lem:closexy} and our earlier analysis that the distance between $x$ and $y$ when they are split by $u$ is at most $[\frac{31}{2}c_S+\frac{45}{2}c_X]\cdot d(x_u,\neigh(x_u))$, we can compute the expected distance between $x$ and $y$ as follows:
\begin{align*}
    \mathop{E}_{\pi,b}[\distl{1}{\alpha(x)}{\alpha(y)}] &\leq c_X\cdot d(x,y)\cdot \Pr[(x,y)\text{ are not split}]  \\
    & \ \ \ \ \ \ + \sum_{u\in K}\Pr_{b}[x,y\text{ are split by } u|E_u]\cdot \Pr_{\pi}[E_u]\cdot\left(\frac{31}{2}\subdist + \frac{45}{2}\origdist \right) \cdot d(x_u,\neigh(x_u))\\
    &\leq \origdist\cdot d(x,y)+ \sum_{u\in K|\beta_u\leq \tau+2}  \frac{5}{\ind(u)} \cdot\frac{d(x,y)}{d(x_u,\neigh(x_u))}\cdot \left(\frac{31}{2}\subdist + \frac{45}{2}\origdist \right) \cdot d(x_u,\neigh(x_u)) \\
    &\leq \origdist\cdot d(x,y) + 5\cdot \left(\frac{31}{2}\subdist + \frac{45}{2}\origdist \right)\cdot d(x,y)\cdot \sum_{i=1}^k\frac{1}{i} \\
    &= \left(\frac{155}{2}\cdot H_k\cdot \subdist + \left(\frac{225}{2}\cdot H_k+1\right)\cdot \origdist\right)\cdot d(x,y),
\end{align*}
where $H_k$ is the $k$th Harmonic number.
\end{proofof}

\subsection{Deterministic nested embeddings}

In this section, we show that we can obtain $O(H_k)(c_S+c_X)$-nested embeddings when the target metric is $\ell_1$. As in the proof of Lipschitz extensions for general $\ell_p$ spaces, this method is not efficient and only serves as a proof of existence.

% \kristin{to do: add in the parts about proving 2.6 and 2.7 and combining things in general/how it applies to $\ell_1$ specifically in this case due to the good contraction bounds as well}

\begin{proof}[Proof of Theorem~\ref{thm:nested-comp}]
% The idea behind the extension to $\ell_1$ distances comes from the proof of Theorem \ref{thm:extension}, where we note that when $p=1$, the third line of the equation should actually be an equality. However, we may still obtain contraction 
Consider the embedding $\alpha^{**}$ that is $\alpha^{**}(x):=||_tp_t\alpha^t(x)$, where $||_t$ implies concatenation over all choices of $t$ and $p_t,\alpha^t$ are as defined in the proof of Theorem \ref{thm:extension}. Now we show that $\alpha^{**}(x)=E_{\pi,b}[x]$ for all $x\in X$. 

\begin{align*}
    \distl{1}{\alpha^{**}(x)}{\alpha^{**}(y)} &= \sum_t \distl{1}{p_t\alpha^t(x)}{p_t\alpha^t(y)} \\
    &= \sum_t p_t\distl{1}{\alpha^t(x)}{\alpha^t(y)} \\
    &= E_{\pi,b}[\distl{1}{\alpha(x)}{\alpha(y)}],
\end{align*}
where the first line is by definition of $\alpha^{**}$ and the fact that for $\ell_1$, the norm of two concatenated vectors equals the sum of their individual norms. The second line is by the scalar properties of $\ell_p$ norms, and the third line is by how we defined the $p_t$ and $\alpha^t$. Thus, we get that our bounds on distortion for individual pairs in our randomized embedding holds for all pairs simultaneously in this choice of embedding and we get existence as desired. Note that the embedding presented here may have a large (but finite) number of coordinates, but any $\ell_p^d$ metric on $n$ points is isometrically embeddable into $\ell_p^n$, so we obtain existence of an embedding with low distortion and at most $n$ dimensions.
\end{proof}

Note that trying to expand the above proof to work for other $\ell_p$ bounds we have the problem that the norm of vectors $u$ concatenated with $v$ is $(||u||_p^p + ||v||_p^p)^{1/p}$, which is smaller than $||u||_p+||v||_p$ for $p>1$. However, the randomized analysis goes through for any $p\geq 1$.

\begin{proof}[Proof of Theorem \ref{thm:nested-comp-strong}]
Notice that Algorithm \ref{alg:nest} can replace its use of $\alpha_X$ in line 8 with any embedding of $K_i\cup\{\neigh(u_i)\}$, and if we have an upper bound on such an embedding's distortion, we can replace $c_X$ in all of the theorems and lemmas in this section with that bound. 
\end{proof}

%% file: maincontent/alg2.tex
\begin{algorithm}
\caption{Algorithm for finding a nested embedding}
\label{alg:nest}
\textbf{Input:} Metric space $(X,d)$, subset $S\subseteq X$, expanding embedding $\alpha_S:X\setminus S\rightarrow \lspace{1}$, expanding embedding $\alpha_X:X\rightarrow\lspace{1}$, and %$\randrang$ for 
real number $\rangwidth>0$ \\
\textbf{Output:} A randomized expanding embedding $\alpha:X\rightarrow \lspace{1}$ such that for all $x,y\in S$, 
% \kristinEDIT{$d_{\ell_2}(\alpha(x),\alpha(y))\leq f(c_S)\cdot d(x,y)$ and for all $x,y\in X$, $d_{\ell_2}(\alpha(x),\alpha(y))\leq g(c_S, c_X)\cdot d(x,y)$}
{$d(x,y)\leq \distl{1}{\alpha(x)}{\alpha(y)}\leq c_S\cdot d(x,y)$ and for all $x,y\in X$, $d(x,y)\leq E[\distl{1}{\alpha(x)}{\alpha(y)}]\leq g(c_S, c_X)\cdot d(x,y)$}.
\begin{algorithmic}[1]
\State $K\from X\setminus S$.
\State \label{def-gamma-full}Define a function $\neigh:K\fto S$ such that 
$\gamma(u)\in\arg\min_{v\in S} d(u,v)$.
%$\gamma(u)\mapsto v$ where $v$ is an arbitrary vertex in $S$ such that $d(u,S)=d(u,v)$. 
\Comment{ie $\neigh(u)$ is one of $u$'s closest neighbors in $S$}
\State Select $\decider$ uniformly at random from the range $\randrang$
\State Select a uniformly random permutation $\perm:K\rightarrow [k]$ of the vertices in $K$
\State $K'\from K,i\from 1$
\For{$i=1$ to $k$}
    \State\label{ui-def-full} $u_i\from \pi^{-1}(i)$ 
    \State\label{def-ki-full} $K_i\from \{v\in K' \ | \ d(v,u_i)\leq b\cdot d(v,\neigh(v))\}$ \Comment{Let  the ``center" of $K_i$ be $u_i$ } 
    \State \change{Define an embedding $\alpha_i:X\fto \lspace{1}$ such that  \begin{align*}\alpha_i(v)=\begin{cases} \alpha_X(v) & v\in K_i\\ \alpha_X(\neigh(u_i)) & v\notin K_i \end{cases}\end{align*} }
    \State $K'\from K'\setminus K_i$
\EndFor
\State Define an embedding $\alpha':X\fto \lspace{1}$ such that
    \begin{align*}
        \alpha'(v)=\begin{cases} 
        \alpha_S(v) & \text{if } v\in S \\
        \alpha_S(\neigh(u_i)) & \text{if } v\in K_i \text{ and with $u_i$ being the center of $K_i$}
        \end{cases}.
    \end{align*}
\State \change{Define an embedding $\alpha:X\fto \lspace{1}$ such that $\alpha(v)\mapsto (\alpha'(v)\concat\alpha_1(v)\concat\cdots \concat \alpha_t(v))$} \Comment{{here $\concat$ denotes concatenation }}
%$\alpha$ is the concatenation of all the embeddings we have defined and   $\cdot$  denotes element-wise multiplication  }
\State Output $\alpha$
\end{algorithmic}
\end{algorithm}

%% file: maincontent/distortion_table_full.tex
\begin{table}[]
    \centering
    \begin{tabular}{c|c|cl}
        membership of $x$ and $y$ & restrictions on $d(x,y)$ & upper bound on expected distortion & \\
        \hline 
        $x,y\in S$ & none & $\bothinstwo$ & \eqref{lem:not-outlier} \\
        $x,y\in K_i$ & none & $\bothinkitwo$ & \eqref{lem:sameset} \\
        $x\in S,y\in X\setminus S$ & none & $\oneineachtwo$ & \eqref{lem:ui} \\
        $x,y\in X\setminus S$ & $d(x,\neigh(x))\leq 2 \cdot d(x,y)$ & $\farxytwo$ & \eqref{lem:farxy} \\
        $x,y\in X\setminus S$ & $d(x,\neigh(x)),d(y,\neigh(y))> 2\cdot d(x,y)$ & $\closexytwo$ & \eqref{lem:closexy}
    \end{tabular}
    \caption{Summary of the bounds in Lemma~\ref{lem:expansion}.  %\ref{lem:not-outlier}, \ref{lem:sameset}, \ref{lem:ui}, \ref{lem:farxy}, and \ref{lem:closexy}. 
    Let $\alpha\from$ Algorithm \ref{alg:nest}$((X,d),S,p,\alpha_S,\alpha_X,\rangwidth)$  where $\alpha_S:S\rightarrow \lspace{p}$ is an expanding embedding of distortion at most $c_S$ and $\alpha_X:X\rightarrow \lspace{p}$ is an expanding embedding of distortion at most $c_X$. Then the third column of the table gives an upper bound on the the expected value of $\distp{\alpha(x)}{\alpha(y)}$ where $x$ and $y$ meet the criteria of the first two columns. {Here $\neigh$ and $K_i$ are as defined in lines \eqref{def-gamma-full} and \eqref{ui-def-full}-\eqref{def-ki-full} of the algorithm.} }
    \label{tab:distortion}
\end{table}

%% file: figures/one_in_each.tex
\tikzset{every picture/.style={line width=0.75pt}} %set default line width to 0.75pt        

\begin{tikzpicture}[x=0.75pt,y=0.75pt,yscale=-.4,xscale=.4]
%uncomment if require: \path (0,556); %set diagram left start at 0, and has height of 556

%Shape: Circle [id:dp8982475433822312] 
\draw  [fill={rgb, 255:red, 80; green, 227; blue, 194 }  ,fill opacity=1 ] (252.33,297.67) .. controls (252.33,291.41) and (257.41,286.33) .. (263.67,286.33) .. controls (269.93,286.33) and (275,291.41) .. (275,297.67) .. controls (275,303.93) and (269.93,309) .. (263.67,309) .. controls (257.41,309) and (252.33,303.93) .. (252.33,297.67) -- cycle ;
%Shape: Circle [id:dp6557481011661439] 
\draw  [fill={rgb, 255:red, 74; green, 144; blue, 226 }  ,fill opacity=1 ] (172.33,258.67) .. controls (172.33,252.41) and (177.41,247.33) .. (183.67,247.33) .. controls (189.93,247.33) and (195,252.41) .. (195,258.67) .. controls (195,264.93) and (189.93,270) .. (183.67,270) .. controls (177.41,270) and (172.33,264.93) .. (172.33,258.67) -- cycle ;
%Shape: Circle [id:dp17590897490465363] 
\draw  [fill={rgb, 255:red, 80; green, 227; blue, 194 }  ,fill opacity=1 ] (259.33,346.67) .. controls (259.33,340.41) and (264.41,335.33) .. (270.67,335.33) .. controls (276.93,335.33) and (282,340.41) .. (282,346.67) .. controls (282,352.93) and (276.93,358) .. (270.67,358) .. controls (264.41,358) and (259.33,352.93) .. (259.33,346.67) -- cycle ;
%Shape: Circle [id:dp3163026395163051] 
\draw  [fill={rgb, 255:red, 0; green, 0; blue, 0 }  ,fill opacity=1 ] (170.33,352.67) .. controls (170.33,346.41) and (175.41,341.33) .. (181.67,341.33) .. controls (187.93,341.33) and (193,346.41) .. (193,352.67) .. controls (193,358.93) and (187.93,364) .. (181.67,364) .. controls (175.41,364) and (170.33,358.93) .. (170.33,352.67) -- cycle ;
%Shape: Circle [id:dp2775734937719956] 
\draw  [fill={rgb, 255:red, 0; green, 0; blue, 0 }  ,fill opacity=1 ] (199.33,306.67) .. controls (199.33,300.41) and (204.41,295.33) .. (210.67,295.33) .. controls (216.93,295.33) and (222,300.41) .. (222,306.67) .. controls (222,312.93) and (216.93,318) .. (210.67,318) .. controls (204.41,318) and (199.33,312.93) .. (199.33,306.67) -- cycle ;
%Shape: Circle [id:dp6985810571932616] 
\draw  [fill={rgb, 255:red, 0; green, 0; blue, 0 }  ,fill opacity=1 ] (280.33,132.67) .. controls (280.33,126.41) and (285.41,121.33) .. (291.67,121.33) .. controls (297.93,121.33) and (303,126.41) .. (303,132.67) .. controls (303,138.93) and (297.93,144) .. (291.67,144) .. controls (285.41,144) and (280.33,138.93) .. (280.33,132.67) -- cycle ;
%Shape: Circle [id:dp8855460991224491] 
\draw  [fill={rgb, 255:red, 0; green, 0; blue, 0 }  ,fill opacity=1 ] (327.33,141.67) .. controls (327.33,135.41) and (332.41,130.33) .. (338.67,130.33) .. controls (344.93,130.33) and (350,135.41) .. (350,141.67) .. controls (350,147.93) and (344.93,153) .. (338.67,153) .. controls (332.41,153) and (327.33,147.93) .. (327.33,141.67) -- cycle ;
%Shape: Circle [id:dp03636018613882075] 
\draw  [fill={rgb, 255:red, 0; green, 0; blue, 0 }  ,fill opacity=1 ] (314.33,91.67) .. controls (314.33,85.41) and (319.41,80.33) .. (325.67,80.33) .. controls (331.93,80.33) and (337,85.41) .. (337,91.67) .. controls (337,97.93) and (331.93,103) .. (325.67,103) .. controls (319.41,103) and (314.33,97.93) .. (314.33,91.67) -- cycle ;
%Shape: Circle [id:dp21615288996053184] 
\draw  [fill={rgb, 255:red, 0; green, 0; blue, 0 }  ,fill opacity=1 ] (282.33,456.67) .. controls (282.33,450.41) and (287.41,445.33) .. (293.67,445.33) .. controls (299.93,445.33) and (305,450.41) .. (305,456.67) .. controls (305,462.93) and (299.93,468) .. (293.67,468) .. controls (287.41,468) and (282.33,462.93) .. (282.33,456.67) -- cycle ;
%Shape: Circle [id:dp4203173320326572] 
\draw  [fill={rgb, 255:red, 126; green, 211; blue, 33 }  ,fill opacity=1 ] (359.33,258.67) .. controls (359.33,252.41) and (364.41,247.33) .. (370.67,247.33) .. controls (376.93,247.33) and (382,252.41) .. (382,258.67) .. controls (382,264.93) and (376.93,270) .. (370.67,270) .. controls (364.41,270) and (359.33,264.93) .. (359.33,258.67) -- cycle ;
%Shape: Circle [id:dp6688852272636758] 
\draw  [fill={rgb, 255:red, 65; green, 117; blue, 5 }  ,fill opacity=1 ] (455.33,258.67) .. controls (455.33,252.41) and (460.41,247.33) .. (466.67,247.33) .. controls (472.93,247.33) and (478,252.41) .. (478,258.67) .. controls (478,264.93) and (472.93,270) .. (466.67,270) .. controls (460.41,270) and (455.33,264.93) .. (455.33,258.67) -- cycle ;
%Shape: Circle [id:dp17833065102347856] 
\draw  [fill={rgb, 255:red, 0; green, 0; blue, 0 }  ,fill opacity=1 ] (365.33,347.67) .. controls (365.33,341.41) and (370.41,336.33) .. (376.67,336.33) .. controls (382.93,336.33) and (388,341.41) .. (388,347.67) .. controls (388,353.93) and (382.93,359) .. (376.67,359) .. controls (370.41,359) and (365.33,353.93) .. (365.33,347.67) -- cycle ;
%Shape: Circle [id:dp3324656279422187] 
\draw  [fill={rgb, 255:red, 0; green, 0; blue, 0 }  ,fill opacity=1 ] (465.33,358.67) .. controls (465.33,352.41) and (470.41,347.33) .. (476.67,347.33) .. controls (482.93,347.33) and (488,352.41) .. (488,358.67) .. controls (488,364.93) and (482.93,370) .. (476.67,370) .. controls (470.41,370) and (465.33,364.93) .. (465.33,358.67) -- cycle ;
%Shape: Circle [id:dp7395766703116189] 
\draw  [fill={rgb, 255:red, 0; green, 0; blue, 0 }  ,fill opacity=1 ] (419.33,377.67) .. controls (419.33,371.41) and (424.41,366.33) .. (430.67,366.33) .. controls (436.93,366.33) and (442,371.41) .. (442,377.67) .. controls (442,383.93) and (436.93,389) .. (430.67,389) .. controls (424.41,389) and (419.33,383.93) .. (419.33,377.67) -- cycle ;
%Shape: Ellipse [id:dp909784674963249] 
\draw  [color={rgb, 255:red, 74; green, 83; blue, 226 }  ,draw opacity=1 ] (218.81,188.74) .. controls (260.77,185.01) and (295.89,232.13) .. (297.26,293.98) .. controls (298.64,355.84) and (265.73,409.01) .. (223.77,412.74) .. controls (181.81,416.47) and (146.69,369.35) .. (145.31,307.49) .. controls (143.94,245.63) and (176.85,192.47) .. (218.81,188.74) -- cycle ;
%Shape: Ellipse [id:dp42577328965867234] 
\draw  [color={rgb, 255:red, 65; green, 117; blue, 5 }  ,draw opacity=1 ] (373.36,80.9) .. controls (389.49,99.38) and (378.37,132.8) .. (348.53,155.56) .. controls (318.68,178.31) and (281.4,181.78) .. (265.27,163.3) .. controls (249.13,144.83) and (260.25,111.4) .. (290.1,88.65) .. controls (319.94,65.89) and (357.22,62.42) .. (373.36,80.9) -- cycle ;
%Shape: Ellipse [id:dp09860201047085115] 
\draw  [color={rgb, 255:red, 65; green, 117; blue, 5 }  ,draw opacity=1 ] (348.36,255.74) .. controls (346.5,231.28) and (375.41,211.17) .. (412.94,210.81) .. controls (450.47,210.46) and (482.41,230) .. (484.27,254.46) .. controls (486.13,278.92) and (457.21,299.04) .. (419.68,299.39) .. controls (382.15,299.74) and (350.22,280.2) .. (348.36,255.74) -- cycle ;
%Shape: Ellipse [id:dp7938629085986322] 
\draw  [color={rgb, 255:red, 65; green, 117; blue, 5 }  ,draw opacity=1 ] (341.28,364.87) .. controls (339.41,340.34) and (376.78,320.08) .. (424.74,319.63) .. controls (472.7,319.17) and (513.09,338.7) .. (514.95,363.24) .. controls (516.82,387.77) and (479.46,408.03) .. (431.5,408.49) .. controls (383.54,408.94) and (343.15,389.41) .. (341.28,364.87) -- cycle ;
%Shape: Ellipse [id:dp9021722324901666] 
\draw  [color={rgb, 255:red, 65; green, 117; blue, 5 }  ,draw opacity=1 ] (252.81,453.51) .. controls (251.75,439.66) and (272.31,428.24) .. (298.71,427.99) .. controls (325.12,427.74) and (347.37,438.76) .. (348.43,452.6) .. controls (349.48,466.45) and (328.93,477.87) .. (302.52,478.12) .. controls (276.12,478.37) and (253.86,467.35) .. (252.81,453.51) -- cycle ;
%Straight Lines [id:da3477371282842876] 
\draw  [dash pattern={on 0.84pt off 2.51pt}]  (369,141) -- (405.33,206.33) ;
%Straight Lines [id:da0119044468701337] 
\draw  [dash pattern={on 0.84pt off 2.51pt}]  (348.33,387.33) -- (322.33,424.33) ;

% Text Node
\draw (121,174) node [anchor=north west][inner sep=0.75pt]  [font=\huge] [align=left] {$\displaystyle S$};
% Text Node
\draw (392,43) node [anchor=north west][inner sep=0.75pt]  [font=\huge] [align=left] {$\displaystyle K_{1}$};
% Text Node
\draw (498,194) node [anchor=north west][inner sep=0.75pt]  [font=\huge] [align=left] {$\displaystyle K_{i}$};
% Text Node
\draw (504,394) node [anchor=north west][inner sep=0.75pt]  [font=\huge] [align=left] {$\displaystyle K_{j}$};
% Text Node
\draw (362,434) node [anchor=north west][inner sep=0.75pt]  [font=\huge] [align=left] {$\displaystyle K_{n}$};
% Text Node
\draw (442,223) node [anchor=north west][inner sep=0.75pt]  [font=\large] [align=left] {$\displaystyle y$};
% Text Node
\draw (177,220) node [anchor=north west][inner sep=0.75pt]  [font=\large] [align=left] {$\displaystyle x$};
% Text Node
\draw (381,224) node [anchor=north west][inner sep=0.75pt]  [font=\large] [align=left] {$\displaystyle u_{i}$};
% Text Node
\draw (230,228) node [anchor=north west][inner sep=0.75pt]  [font=\large] [align=left] {$\displaystyle \gamma ( u_{i})$};
% Text Node
\draw (210,363) node [anchor=north west][inner sep=0.75pt]  [font=\large] [align=left] {$\displaystyle \gamma ( y)$};

\end{tikzpicture}

%% file: maincontent/conclusion.tex
In this paper, we give a bi-criteria approximation algorithm that given a constant $c$ and metric $(X,d)$ finds an $(O(k\log^2k), O(c))$-outlier embedding into $\lspace{2}$ if the metric has a $(k,c)$-outlier embedding into $\lspace{2}$. In doing so, we show that given a metric space $(X,d)$, a $c_S$-distortion embedding of a subset $S\subseteq X$ into $\lspace{1}$, there exists a Lipschitz extension with Lipschitz factor at most $O(H_k\cdot c_S)$ on every pair of points. Additionally, when the target metric is $\lspace{1}$ and we have an embedding of the entire space with distortion at most $c_X$, there exists a single (composition) embedding of $X$ into $\lspace{1}$ such that distortion between pairs of points in $S$ is at most $c_S$ and distortion between all pairs of points is at most $O(c_S+H_k\cdot c_X)$. We also leave several open questions on this topic. Among them, we ask:

\begin{itemize}
    \item Is there a polynomial time algorithm that given constant $c$, finds an $(O(k),O(c))$-embedding into $\lspace{2}$?
    \item What bicriteria approximations can be obtained for outlier embeddings into $\lpspace$ for other values of $p$?
    \item Do there exist nested compositions into $\ell_p$ for $p\ne 1$ that do not incur the contraction presented in this paper (i.e. ``strong" nested embeddings)?
    %\item Can the results for $\ell_1$ be expanded to obtain similar nested compositions for other values of $\ell_p$?
    \item Can the parameters for our nested embedding or Lipschitz extension algorithms be improved?
    \item Can our hardness of approximation result be extended to non-isometric outlier embeddings for large constants or slow-growing functions in $n$?
    \item Can we obtain any larger concrete lower bounds on approximation factors for an optimal outlier set?
\end{itemize}

%% file: maincontent/full_constants.tex
Lemma \ref{lem:extension-full} gives a version of Lemma~\ref{lem:expansion-weak}  more generally in terms of $\rangwidth$ and other flexible parameters. The proof would be identical to that given in Section \ref{sec:nested}.

\begin{lem}\label{lem:extension-full}
Let $\alpha\from$ Algorithm \ref{alg:extend}$((X,d),S,\alpha_S,\rangwidth)$. Then we have the following bounds on the expansion for each pair $x,y\in X$:
\begin{enumerate}[(a)]
    \item \label{lem:not-outlier-full} If $x,y\in S$, then $d_Y(\alpha(x),\alpha(y))\leq \subdist\cdot d(x,y)$.
    % \item \label{lem:sameset-full} If $x,y\in K_i$, then $d_Y(\alpha(x),\alpha(y))\leq \subdist\cdot d(x,y)$.
    \item \label{lem:ui-full} If $x\in S,y\in X\setminus S$, then $d_Y(\alpha(x),\alpha(y))\leq (2\tau+6)\subdist\cdot d(x,y)$.
    % , where $b$ is as chosen in line x of the algorithm.
    \item \label{lem:farxy-full} If $x,y\in X\setminus S$ and $d(x,\gamma(x))\leq \comparison\cdot d(x,y)$ for positive $\comparison$ and $\neigh$ as defined in line 2 of the algorithm, then $d_Y(\alpha(x),\alpha(y))\leq (2\comparison +1)(2\tau+6)\subdist \cdot d(x,y)$.  
    % where $b$ is as chosen in line x of the algorithm.
    % \kristin{don't we need a condition on their relative distance for this one? otherwise it looks odd to have d and e. Also should we say expected expansion at the top?}
    \item \label{lem:closexy-full} If $x,y\in X\setminus S$ and $d(x,\neigh(x)),d(y,\neigh(y))> \comparison\cdot d(x,y)$ for $\neigh$ as defined in line 2 of the algorithm and real  number $\comparison>1$, then 
    \begin{align*}
        \mathop{E}_{\alpha}[d_Y(\alpha(x),\alpha(y))] &\leq \left(\frac{(\tau+3)(2\tau+6)(2\comparison+1)}{\tau(\kappa-1)} \right) \cdot c_S\cdot d(x,y).
    \end{align*}
\end{enumerate}
\end{lem}

Lemma \ref{lem:expansion-full} gives a version of Lemma~\ref{lem:expansion}  more generally in terms of $\rangwidth$ and other flexible parameters. The proof would be identical to that given in Section \ref{sec:full_nested}.

\begin{lem}\label{lem:expansion-full}
Let $\alpha\from$ Algorithm \ref{alg:nest}$((X,d),S,\alpha_S,\alpha_X,\rangwidth)$. Then we have the following bounds on the expansion for each pair $x,y\in X$:
\begin{enumerate}[(a)]
    \item \label{lem:not-outlier-full} If $x,y\in S$, then $\distl{1}{\alpha(x)}{\alpha(y))}\leq \subdist\cdot d(x,y)$.
    \item \label{lem:sameset-full} If $x,y\in K_i$, then $\distl{1}{\alpha(x)}{\alpha(y))}\leq \bothinkitwo\cdot d(x,y)$.
    \item \label{lem:ui-full} If $x\in S,y\in X\setminus S$, then $\distl{1}{\alpha(x)}{\alpha(y))}\leq \oneineach d(x,y)$.
    % , where $b$ is as chosen in line x of the algorithm.
    \item \label{lem:farxy-full} If $x,y\in X\setminus S$ and $d(x,\gamma(x))\leq \comparison\cdot d(x,y)$ for positive $\comparison$ and $\neigh$ as defined in line 2 of the algorithm, then \begin{align*}
        \distl{1}{\alpha(x)}{\alpha(y)}\leq [((2\tau+8)\kappa+\tau+5)\cdot \subdist + ((4\tau+10)\kappa+2\tau+5)\cdot \origdist]\cdot d(x,y).
    \end{align*}
    % where $b$ is as chosen in line x of the algorithm.
    % \kristin{don't we need a condition on their relative distance for this one? otherwise it looks odd to have d and e. Also should we say expected expansion at the top?}
    \item \label{lem:closexy-full} If $x,y\in X\setminus S$ and $d(x,\neigh(x)),d(y,\neigh(y))> \comparison\cdot d(x,y)$ for $\neigh$ as defined in line 2 of the algorithm and real  number $\comparison>1$, then 
    \begin{align*}
        &\mathop{E}_\alpha[\distl{1}{\alpha(x)}{\alpha(y))}]\\
        & \ \ \ \ \ \ \ \ \leq \closexy\cdot d(x,y).
    \end{align*}
\end{enumerate}
\end{lem}

%% file: maincontent/hardness.tex
Sidiropoulos \textit{et al.} \cite{sidiropolous17} showed that for any $t\geq 2$, it is NP-hard to determine the size of the smallest outlier set for a finite metric $(X,d)$ such that the metric without the outlier set is isometrically embeddable into $\lspace{2}^t$.  Additionally, because they reduce from Vertex Cover, they show that under the Unique Games Conjecture it is NP-hard to approximate the size of such a set to a factor better than $2-\epsilon$. 
% This result naturally extends to $\ell_p$ for $1<p<\infty$.

% \kristin{I think their result extends to lp but I'm not totally sure - I need to think it through a bit to make sure that we get inconsistency in one instance but consistency in the other. It is not an unweighted graph metric though }

In this appendix, we will give an alternate proof of a similar conclusion, but we extend their result to show that it holds even if the input metric is an unweighted graph metric. We note that unlike the Sidiropolous \textit{et al.} proof, our proof does not apply for arbitrary choice of dimension $d$.
% , and in fact Chubarian and Sidiropoulos \cite{Chubarian20} showed that it takes only polynomial time to find the size of the smallest outlier set for embedding an unweighted graph metric into the real number line with distortion at most $c$, for any $c\geq 1$.

First, we claim the following Lemma \ref{lem:l1-hard} which we will prove later. Note that it is NP-hard to decide if a general metric is isometrically $\ell_1$-embeddable and thus it is hard to decide if the minimum outlier set for such an embedding has size $0$ or size larger than $0$, implying hardness of any approximation for this value. However, $\ell_1$-embeddability can be decided in polynomial time for unweighted graph metrics \cite{Shpectorov,dezaShpectorov,DezaLaurentText}, and we show that even with this restriction on the input, it is hard to determine minimum outlier set size.

\begin{lem}\label{lem:l1-hard}
    Let $(X,d)$ be the distance metric for an unweighted graph $G=(V,E)$. then, given $(X,d,k)$ it is NP-hard to decide if there exists a subset $K\subseteq X$ with $|K|=k$ such that $(X\setminus K,d|_{X\setminus K})$ is isometrically embeddable into $\lspace{1}$, even when the input metric is an unweighted graph metric. 
    
    Under the unique games conjecture, it is NP-hard to find a $2-\epsilon$ approximation for the minimum such $k$, for any $\epsilon>0$.    
\end{lem}

We can now use this result to prove Theorem \ref{thm:np-hard}.

\begin{figure}
    \centering
    \input{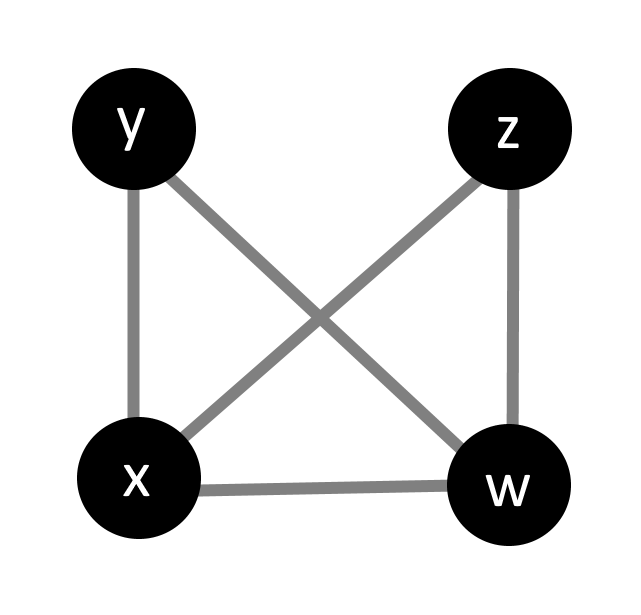}
    \caption{Example of a subgraph formed by nodes $\{u_1=x,u_2=y,v_1=w,v_2=z\}$ as defined in the proof of Theorem \ref{thm:np-hard}. }
    \label{fig:subgraph}
\end{figure}

\begin{proof}[Proof of Theorem \ref{thm:np-hard}] 
We appeal to Lemma \ref{lem:l1-hard} to show that the theorem holds for $p=1$ and here we show that it holds for $1<p<\infty$.

Consider a graph $G=(V,E)$. Given $G$, we give a polynomial time construction of an unweighted graph metric $(V',d_{G'})$ such that the size of the minimum outlier set for embedding the metric into $\lspace{p}$ is the same as the size of the minimum vertex cover on $G$. In this proof, minimum outlier set refers to the minimum outlier set for distortion $1$ and we consider embeddings into $\lspace{p}$ for finite integer $p> 1$.

\textbf{Construction:} We will construct an unweighted graph $G'=(V',E')$ and let $d_{G'}$ be the distance metric on this graph. In particular, let $V'=\{u_1|u\in V\}\cup \{u_2|u\in V\}$. Add edges between every pair of nodes in the graph, but omit edges $u_2v_2$ for $uv\in E$. 

\textbf{Correctness:} Let $K$ be a minimum outlier set on $(V',d_{G'})$ and let $\hat V$ be a minimum vertex cover on $G$. We claim $|K|=|\hat V|$.
\begin{description}
\item[$|K|\leq |\hat V|$]: We construct an outlier set of size at most $|\hat V|$. Define $K':=\{u_2|u\in \hat V\}$.  We claim that $(V'\setminus K',d|_{V'\setminus K'})$ is the equidistant metric with distance $1$ between all points, which is always embeddable in $\lspace{p}$ for any $p$. Assume this is not the case. Note that all distances in $G'$ are $1$ or $2$, so there exists a pair of nodes with distance $2$ between them. The only such pairs are of the form $\{u_2,v_2\}$ for $uv\in E$. However, $u_2,v_2\in V'\setminus K'$ implies that $\hat V$ does not cover edge $uv$ so it is not a vertex cover and we reach a contradiction. Thus since $K$ is minimum, we get $|K|\leq |K'|\leq |\hat V|$.
\item[$|\hat V|\leq |K|$]:  We construct a vertex cover of size at most $|K|$. Define $\hat V':=\{u|u_1\in K\text{ or }u_2\in K\}$. We see $|\hat V'|\leq |K|$ and we claim it is a vertex cover. Assume otherwise. Then there exists an edge $uv$ that is not covered by $\hat V'$ and there is a subgraph of the form in Figure \ref{fig:subgraph} in the induced subgraph of $G'$ on $V'\setminus K'$, $G'[V'\setminus K']$. Note that distances in this subgraph are exactly distances in the entire graph by our construction. Thus we need only show that this subgraph is not isometrically embeddable in $\lspace{p}$. 

Note that $\lspace{p}$ is a strictly convex space for $1<p<\infty$ \cite{Clarkson1936}. Thus, for $a,b\in \reals^t$ for any fixed $t$, if $||a||_p=||b||_p=1$ and $a\neq b$, then $||\frac{a+b}{2}||_p<1$, which implies $||a+b||_p<2^{1/p}$. Let $w,x,y,z$ be four points in $\lspace{p}$ such that the distance between all pairs of points is $1$, except between $y$ and $z$ which are a distance $2$ apart. Then let $a=y-x$ and $b=x-z$. We have $||y-x||_p=||x-z||_p=1$, so $||y-z||_p< 2^{1/p}$ unless $y-x=x-z$. Since $2^{1/p}<2$ for $p>1$, we get that $y-x=x-z\implies x=\frac{y+z}{2}$. However, equivalent analysis on $w$ implies $w=\frac{y+z}{2}$. This means that $x$ and $w$ are the same point and their distance is $0$, not $1$. Thus, this set of four points cannot exist in $\lspace{p}$ for $1<p<\infty$ and the subgraph in Figure \ref{fig:subgraph} is not isometrically embeddable into this space.
% \kristin{It took me a while to find a source that was showing the $\ell_p$ had this property and I tried to simplify it to match the properties we needed, but lmk if it's confusing as it is. It's also a pretty old result so I'm not sure if it's the kind of thing you're supposed to cite or not }
% Let $w,x,y,z$ be four points  in $\ell_p^t$ for some $t$ such that the distance is $1$ between all points except $y$ and $z$. We have that in a strictly convex
\end{description}
\end{proof}

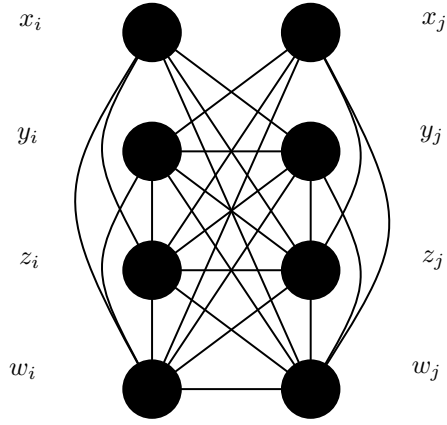
\begin{figure}
    \centering
    \input{figures/unembeddable_graph}
    \caption{Example of a subgraph formed by nodes $\{x_i,y_i,z_i,w_i,x_j,y_i,z_j,w_j\}$ as defined in the proof of Lemma \ref{lem:l1-hard}, where $ij\in E$. }
    \label{fig:unembeddablegraph}
\end{figure}

\begin{proof}[Proof of Lemma \ref{lem:l1-hard}]
As in the previous lemma, we reduce the vertex cover problem to this problem. Consider a graph $G=(V,E)$. We give a polynomial time construction of an unweighted graph metric $(V',d_{G'})$ such that that size of the minimum outlier set for embedding the metric into $\lspace{1}$ is the same as the size of the minimum vertex cover on $G$. In this proof, minimum outlier set refers to the minimum outlier set for distortion $1$ and we consider embeddings into $\lspace{1}$.

\textbf{Construction:} We will construct an unweighted graph $G'=(V',E')$ and let $d_{G'}$ be the distance metric on this graph. In particular, let $V'=\{x_i,y_i,z_i,w_i | i\in V\}$. Add edges between every pair of nodes in the graph, but omit edges $x_iy_i$ for all $i$. Additionally, omit edges $x_ix_j$ if  $ij\in E$.

\textbf{Correctness:} Let $K$ be a minimum outlier set on $(V',d_{G'})$ and let $\hat V$ be a minimum vertex cover on $G$. We claim $|K|=|\hat V|$.

\begin{description}
    \item[$|K|\leq |\hat V|$]: Define $K':=\{x_i|i\in \hat V\}$. We claim that $(V'\setminus K',d_{G'}|_{V'\setminus K'})$ is $\ell_1$ embeddable. In particular, \cite{Shpectorov,dezaShpectorov}, and \cite{DezaLaurentText} show that if an unweighted graph is such that each node has at most one other node it does not have an edge to (i.e. if the graph is a subgraph of a cocktail party graph), then the graph is $\ell_1$-embeddable. Note that $d_{G'}$ restricted to the nodes in $V'\setminus K'$ is in fact the distance metric on the induced subgraph of $G'$ on those same nodes, $G'[V'\setminus K']$. This is because we only removed some of the $x_i$, which cannot affect distances between any remaining pairs. 
    
    Thus, we are left with showing that $G'[V'\setminus K']$ is such that each node has at most one other node to which it does not have an edge. Assume otherwise. Then there exists $a,b,c\in V'\setminus K'$ such that $a$ does not have an edge to $b$ or to $c$. The only nodes in $G'$ that are missing an edge to more than one other node in $G'$ are some of the $x_i$, so $a$ must be $x_i$ for some $i$. Additionally, the nodes that $x_i$ does not have an edge to are $y_i$ and all $x_j$ such that $ij\in E$. Thus, at least one of $b$ and $c$ must be $x_j$ for some $j$ such that $ij\in E$. However, this implies $x_i,x_j\notin K'\implies i,j\notin \hat V$, so $\hat V$ is not a vertex cover and we reach a contradiction. Thus $K'$ is an outlier set and we get $|K|\leq |K'|=|\hat V|$
    
    \item[$|\hat V|\leq K$]: Define $\hat V':=\{i|x_i,y_i,z_i,\text{ or }w_i\in K\}$. We have $|\hat V'|\leq |K|$, so if $\hat V'$ is a valid vertex cover then we obtain the desired bound. Assume $\hat V'$ is not a vertex cover. Then there exists $u_iu_j\in E$ such that $\{x_i,y_i,z_i,w_i,x_j,y_j,z_j,w_j\}\subseteq V'\setminus K$. These nodes form the subgraph pictured in Figure \ref{fig:unembeddablegraph}, and $d_{G'}$ on this subset of nodes has the same value as the graph metric on this subgraph. Thus, we need only show that the subgraph in Figure \ref{fig:unembeddablegraph}, which we call $G''$, is not $\ell_1$-embeddable.
    
    By \cite{DezaLaurentText}, an unweighted graph is $\ell_1$-embeddable if and only if there exists an integer $t\in \mathbb{Z}^+$ such that the same graph with all edge weights set to $t$ is hypercube embeddable. Deza and Shpectorov \cite{Shpectorov,dezaShpectorov} show that if an unweighted graph is $\ell_1$-embeddable and it is not ``reducible" as defined by \cite{graham1985}, then it must be an isometric subgraph of a cocktail party graph or a half-cube (a type of graph that is hypercube embeddable at scale $2$). In Lemma \ref{lem:not-reducible} we use Graham and Winkler's \cite{graham1985} techniques to show that $G''$ is not reducible. 

    Additionally, $x_i$ lacks neighbors $y_i$ and $x_j$, so this is not a subgraph of a cocktail party graph, which is a graph in which each node is a neighbor of all but one node. This leaves us with showing that $G''$ is not a subgraph of a half-cube, which we prove by showing that it is not hypercube embeddable at scale $2$ in Lemma \ref{lem:not-halfcube}. Thus, we conclude $G''$ cannot be a subgraph of $G'[V'\setminus K]$ and $\hat V'$ must be a vertex cover.
\end{description}
\end{proof}

\begin{lem}\label{lem:not-reducible}
    The graph appearing in Figure \ref{fig:unembeddablegraph} is not reducible, where reducible is as defined by \cite{graham1985} 
\end{lem}

\begin{proof}
    Graham and Winkler show that a graph is not reducible if all of its edges are in the same equivalence class of the equivalence relation $\hat\theta$. $\hat \theta$ is defined to be the transitive closure of $\theta$, which is defined on the edges of a graph as follows:

    For a graph $G''=(V'',E'')$, edges $ab,cd\in E''$ are related by $\theta$ if and only if 
    \begin{align*}
        [d_{G''}(a,c)-d_{G''}(a,d)]-[d_{G''}(b,c)-d_{G''}(b,d)]\neq 0.
    \end{align*}

    We will show that all edges of $G''$ are in the same equivalence class of $\hat \theta$.

    First, notice that in the big/main clique of $y_i,y_j,z_i,z_j,w_i,w_j$ in $G''$, all edges must be related by $\hat\theta$. Take two adjacent edges $ab,bc$. Since all distances in the clique are $1$, we get $[d_{G''}(a,b)-d_{G''}(a,c)]-[d_{G''}(b,b)-d_{G''}(b,c)]=1\neq 0$. Thus all adjacent edges in the clique are related by $\theta$ and thus all edges in the clique are related by $\hat\theta$. This just leaves us to consider the edges to $x_i$ and $x_j$. We see that $z_ix_i\ \theta \ z_iw_i$ because again we have adjacent edges with distances between all three vertices being $1$. The same analysis goes for all the other edges to $x_i$ and to $x_j$ since they are part of forming a smaller clique with $x_i$ or $x_j$ and two members of the main clique. This means they are all in the same equivalence class as the edges in the main clique and thus all edges are in the same equivalence class, meaning it is irreducible.
\end{proof}

\begin{lem}\label{lem:not-halfcube}
    The graph appearing in Figure \ref{fig:unembeddablegraph} is not hypercube embeddable, and the same graph with all edge weights scaled to $2$ is also not hypercube embeddable.
\end{lem}

\begin{proof}
    The graph is not bipartite, so it is not hypercube embeddable at scale $1$ \cite{djokovic1973distance}.

    For the second part of this proof, we need only verify that the graph is not an isometric subgraph of a halved cube. To do this, we could run the algorithm of Deza and Shpectorov \cite{dezaShpectorov} for doing so. However, we will present the proof in a different way because their algorithm is more generalized than what is needed for our specific purposes.

    To begin the second part of this part of the proof, we first consider an alternative view of a hypercube embedding. In particular, if we have a hypercube embedding in dimension $t$ for a metric space on $n$ nodes, we can write an $n\times t$ matrix in which each row of the matrix is the binary string associated with a particular node. The number of of columns where two rows differ is then the distance between the corresponding nodes in the graph. Notice that this matrix defines a hypercube embedding and a hypercube embedding defines this matrix. Additionally, we can remove any columns in which all rows have the same value without affecting the ``distance" between rows. Additionally, if we swap the order of the columns or if we pick a column of this matrix and flip all bits in that column, it has no effect on the distance between the rows. Thus, we can always assume that the top row of the matrix for a hypercube embedding is made up of all $0$s, as such a choice of embedding must exist if some hypercube embedding does. 

Let's begin constructing a hypercube embedding for the graph $G''$. In particular, we define a matrix in which the top row corresponds to the node $x_i\in V(G'')$ and the row consists of all $0$s (which we've already argued is a fine assumption). Then we know that all rows except those corresponding to $y_i$ and $x_j$ must have exactly two $1$s in their rows, and $y_i$ and $x_j$ have exactly four $1$s in their rows (in order to make their distance from $x_i$ correct). We will assume that all the columns in the matrix we construct have at least two distinct values in each column, as we have argued such an embedding must exist if any embedding exists since we can delete columns where all values are equal. Thus, we can assume that $M$ has at most $18$ columns since there are at most eighteen $1$s in the entire matrix. This leads us to the following partial embedding where most values are unassigned so far.
\begin{center}
    \begin{tabular}{c|cccccccccccccccccc}
         & 1 & 2 & 3 & 4 & 5 & 6 & 7 & 8 & 9 & 10 & 11 & 12 & 13 & 14 & 15 & 16 & 17 & 18  \\
         \hline 
         $x_i$ & 0 & 0 & 0 & 0 & 0 & 0 & 0 & 0 & 0 & 0 & 0 & 0 & 0 & 0 & 0 & 0 & 0 & 0 \\
         $z_i$ & & & & & & & & & & & & & & & & & & \\
         $y_i$ & & & & & & & & & & & & & & & & & & \\
         $x_j$ & & & & & & & & & & & & & & & & & & \\
         $w_i$ & & & & & & & & & & & & & & & & & & \\
         $w_j$ & & & & & & & & & & & & & & & & & & \\
         $z_j$ & & & & & & & & & & & & & & & & & & \\
         $y_j$ & & & & & & & & & & & & & & & & & & 
    \end{tabular}
\end{center}

We will next assume that the two $1$s in $z_i$'s row are in the first two columns. (As we mentioned, we can always rearrange the columns of a hypercube embedding matrix to make one that looks this way since those columns at this point are indistinguishable.) This gives us the following matrix:

\begin{center}
    \begin{tabular}{c|cccccccccccccccccc}
         & 1 & 2 & 3 & 4 & 5 & 6 & 7 & 8 & 9 & 10 & 11 & 12 & 13 & 14 & 15 & 16 & 17 & 18  \\
         \hline 
         $x_i$ & 0 & 0 & 0 & 0 & 0 & 0 & 0 & 0 & 0 & 0 & 0 & 0 & 0 & 0 & 0 & 0 & 0 & 0 \\
         $z_i$ & 1 & 1 & 0 & 0 & 0 & 0 & 0 & 0 & 0 & 0 & 0 & 0 & 0 & 0 & 0 & 0 & 0 & 0  \\
         $y_i$ & & & & & & & & & & & & & & & & & & \\
         $x_j$ & & & & & & & & & & & & & & & & & & \\
         $w_i$ & & & & & & & & & & & & & & & & & & \\
         $w_j$ & & & & & & & & & & & & & & & & & & \\
         $z_j$ & & & & & & & & & & & & & & & & & & \\
         $y_j$ & & & & & & & & & & & & & & & & & & 
    \end{tabular}
\end{center}

Next, consider node $y_i$ that is a distance $2$ closer to $z_i$ than to $x_i$ (in the scaled graph, as it is a distance $1$ closer in the original graph). Notice that in all columns after the first two columns, if $y_i$ doesn't match $x_i$, then it also doesn't match $z_i$. Thus, the only place where it can have these two differences with $x_i$ that it does not have those differences with $z_i$ is the first two columns. Thus, $y_i$ must have $1$s in the first two columns. The same goes for $x_j$ because it is also a distance $2$ closer to $z_i$ than to $x_i$. For all the other nodes, they are equidistant from $x_i$ and $z_i$. Thus, because they must have the same number of differences with $x_i$'s and $z_i$'s rows after the first two columns, they must also have the same number of differences in the first two columns. This means they either have $10$ or $01$ in the first two columns. Because we don't know which of these to put in those columns, for now we don't fill those values in yet. Because $y_i$ and $x_j$ are a distance $4$ from $x_i$, they must have four $1$s total. We will assume that the first four $1$s for $y_i$ are in the first four columns, because we can rearrange any matrix in which the other two $1$s are in later columns such that they are in the second two columns. This leaves us with the following matrix constructed so far.

\begin{center}
    \begin{tabular}{c|cccccccccccccccccc}
         & 1 & 2 & 3 & 4 & 5 & 6 & 7 & 8 & 9 & 10 & 11 & 12 & 13 & 14 & 15 & 16 & 17 & 18  \\
         \hline 
         $x_i$ & 0 & 0 & 0 & 0 & 0 & 0 & 0 & 0 & 0 & 0 & 0 & 0 & 0 & 0 & 0 & 0 & 0 & 0 \\
         $z_i$ & 1 & 1 & 0 & 0 & 0 & 0 & 0 & 0 & 0 & 0 & 0 & 0 & 0 & 0 & 0 & 0 & 0 & 0  \\
         $y_i$ & 1 & 1 & 1 & 1 & 0 & 0 & 0 & 0 & 0 & 0 & 0 & 0 & 0 & 0 & 0 & 0 & 0 & 0  \\
         $x_j$ & 1 & 1 & & & & & & & & & & & & & & & & \\
         $w_i$ & & & & & & & & & & & & & & & & & & \\
         $w_j$ & & & & & & & & & & & & & & & & & & \\
         $z_j$ & & & & & & & & & & & & & & & & & & \\
         $y_j$ & & & & & & & & & & & & & & & & & & 
    \end{tabular}
\end{center}

Now we can consider the distance between $x_j$ and $y_i$. They are a distance $2$ apart in the scaled graph, and we only have two more $1$s that we can place in $x_j$'s row. If we put both of these $1$s in positions $3$ and $4$, then there are no differences between $x_j$ and $y_i$, and if we put none of these $1$s in positions $3$ or $4$, the two have a distance $4$ apart. Thus, we must have exactly a single $1$ in columns $3$ and $4$. We will assume that this is in column $3$, as at the moment these columns are indistinguishable so if a good embedding has the opposite assignment, we can swap the columns to get something consistent with this embedding. We will also assume that the last $1$ is in column $5$ since the columns after $4$ are indistinguishable at this point.

\begin{center}
    \begin{tabular}{c|cccccccccccccccccc}
         & 1 & 2 & 3 & 4 & 5 & 6 & 7 & 8 & 9 & 10 & 11 & 12 & 13 & 14 & 15 & 16 & 17 & 18  \\
         \hline 
         $x_i$ & 0 & 0 & 0 & 0 & 0 & 0 & 0 & 0 & 0 & 0 & 0 & 0 & 0 & 0 & 0 & 0 & 0 & 0 \\
         $z_i$ & 1 & 1 & 0 & 0 & 0 & 0 & 0 & 0 & 0 & 0 & 0 & 0 & 0 & 0 & 0 & 0 & 0 & 0  \\
         $y_i$ & 1 & 1 & 1 & 1 & 0 & 0 & 0 & 0 & 0 & 0 & 0 & 0 & 0 & 0 & 0 & 0 & 0 & 0  \\
         $x_j$ & 1 & 1 & 1 & 0 & 1 & 0 & 0 & 0 & 0 & 0 & 0 & 0 & 0 & 0 & 0 & 0 & 0 & 0  \\
         $w_i$ & & & & & & & & & & & & & & & & & & \\
         $w_j$ & & & & & & & & & & & & & & & & & & \\
         $z_j$ & & & & & & & & & & & & & & & & & & \\
         $y_j$ & & & & & & & & & & & & & & & & & & 
    \end{tabular}
\end{center}

Now we can consider labeling $w_i,w_j,z_j$. We already established that due to the fact they are equidistant from $x_i$ and $z_i$, they must all have either $10$ or $01$ in the first two columns. Then they have a single $1$ left for the rest of the columns since they are a distance $2$ from $x_i$. We notice that if we put this second $1$ in column $4$, then the distance between the node and $x_j$ is $4$ (one difference in the first two columns and one each in each of columns $3$ through $5$), when it should be $2$. Analogously, if the $1$ is in column $5$, the distance to $y_i$ is $4$ instead of $2$. If the $1$ is in column $6$ or larger, the distance to $y_i$ \textit{and} $x_j$ is $4$ instead of $2$. Thus, this leaves us with putting the $1$ in column $3$ for all three of these nodes. Everything after this must be $0$s since we used up our only other $1$ in the first two columns. This gives us the following partial embedding.

\begin{center}
    \begin{tabular}{c|cccccccccccccccccc}
         & 1 & 2 & 3 & 4 & 5 & 6 & 7 & 8 & 9 & 10 & 11 & 12 & 13 & 14 & 15 & 16 & 17 & 18  \\
         \hline 
         $x_i$ & 0 & 0 & 0 & 0 & 0 & 0 & 0 & 0 & 0 & 0 & 0 & 0 & 0 & 0 & 0 & 0 & 0 & 0 \\
         $z_i$ & 1 & 1 & 0 & 0 & 0 & 0 & 0 & 0 & 0 & 0 & 0 & 0 & 0 & 0 & 0 & 0 & 0 & 0  \\
         $y_i$ & 1 & 1 & 1 & 1 & 0 & 0 & 0 & 0 & 0 & 0 & 0 & 0 & 0 & 0 & 0 & 0 & 0 & 0  \\
         $x_j$ & 1 & 1 & 1 & 0 & 1 & 0 & 0 & 0 & 0 & 0 & 0 & 0 & 0 & 0 & 0 & 0 & 0 & 0  \\
         $w_i$ & & & 1 & 0 & 0 & 0 & 0 & 0 & 0 & 0 & 0 & 0 & 0 & 0 & 0 & 0 & 0 & 0 \\
         $w_j$ & & & 1 & 0 & 0 & 0 & 0 & 0 & 0 & 0 & 0 & 0 & 0 & 0 & 0 & 0 & 0 & 0 \\
         $z_j$ & & & 1 & 0 & 0 & 0 & 0 & 0 & 0 & 0 & 0 & 0 & 0 & 0 & 0 & 0 & 0 & 0 \\
         $y_j$ & & & & & & & & & & & & & & & & & & 
    \end{tabular}
\end{center}

Now we notice that $w_i,w_j,z_j$ have no differences after column $3$. 
Thus, all of their differences must be in the first two columns. This means that we must come up with three length $2$ binary strings that are all Hamming distance $2$ from each other, which is impossible. Because all of the decisions we made in constructing this embedding were necessary, this partial construction is required which means it's impossible to construct a hypercube embedding for this graph at scale $2$.
\end{proof}

%% file: figures/forbidden_subgraph.tex
\tikzset{every picture/.style={line width=0.75pt}} %set default line width to 0.75pt        

\begin{tikzpicture}[x=0.75pt,y=0.75pt,yscale=-1,xscale=1]
%uncomment if require: \path (0,214); %set diagram left start at 0, and has height of 214

%Shape: Circle [id:dp21074073193559095] 
\draw  [fill={rgb, 255:red, 0; green, 0; blue, 0 }  ,fill opacity=1 ] (43,51) .. controls (43,42.16) and (50.16,35) .. (59,35) .. controls (67.84,35) and (75,42.16) .. (75,51) .. controls (75,59.84) and (67.84,67) .. (59,67) .. controls (50.16,67) and (43,59.84) .. (43,51) -- cycle ;
%Shape: Circle [id:dp18919472664826764] 
\draw  [fill={rgb, 255:red, 0; green, 0; blue, 0 }  ,fill opacity=1 ] (43,162) .. controls (43,153.16) and (50.16,146) .. (59,146) .. controls (67.84,146) and (75,153.16) .. (75,162) .. controls (75,170.84) and (67.84,178) .. (59,178) .. controls (50.16,178) and (43,170.84) .. (43,162) -- cycle ;
%Shape: Circle [id:dp10681777634724576] 
\draw  [fill={rgb, 255:red, 0; green, 0; blue, 0 }  ,fill opacity=1 ] (163,52) .. controls (163,43.16) and (170.16,36) .. (179,36) .. controls (187.84,36) and (195,43.16) .. (195,52) .. controls (195,60.84) and (187.84,68) .. (179,68) .. controls (170.16,68) and (163,60.84) .. (163,52) -- cycle ;
%Shape: Circle [id:dp03299421713480255] 
\draw  [fill={rgb, 255:red, 0; green, 0; blue, 0 }  ,fill opacity=1 ] (163,163) .. controls (163,154.16) and (170.16,147) .. (179,147) .. controls (187.84,147) and (195,154.16) .. (195,163) .. controls (195,171.84) and (187.84,179) .. (179,179) .. controls (170.16,179) and (163,171.84) .. (163,163) -- cycle ;
%Straight Lines [id:da99065555723437] 
\draw    (59,51) -- (59,162) ;
%Straight Lines [id:da054411742558318155] 
\draw    (179,52) -- (179,163) ;
%Straight Lines [id:da38137592828962696] 
\draw    (59,51) -- (179,163) ;
%Straight Lines [id:da357961640740621] 
\draw    (179,52) -- (59,162) ;
%Straight Lines [id:da6604367699026488] 
\draw    (59,162) -- (179,163) ;

% Text Node
\draw (23,163) node [anchor=north west][inner sep=0.75pt]  [font=\Large] [align=left] {$\displaystyle x$};
% Text Node
\draw (204,164) node [anchor=north west][inner sep=0.75pt]  [font=\Large] [align=left] {$\displaystyle w$};
% Text Node
\draw (24,17) node [anchor=north west][inner sep=0.75pt]  [font=\Large] [align=left] {$\displaystyle y$};
% Text Node
\draw (204,24) node [anchor=north west][inner sep=0.75pt]  [font=\Large] [align=left] {$\displaystyle z$};

\end{tikzpicture}

%% file: figures/unembeddable_graph.tex
\tikzset{every picture/.style={line width=0.75pt}} %set default line width to 0.75pt        

\begin{tikzpicture}[x=0.75pt,y=0.75pt,yscale=-1,xscale=1]

%Shape: Circle [id:dp3975261670808363] 
\draw  [fill={rgb, 255:red, 0; green, 0; blue, 0 }  ,fill opacity=1 ] (136,39.5) .. controls (136,31.49) and (142.49,25) .. (150.5,25) .. controls (158.51,25) and (165,31.49) .. (165,39.5) .. controls (165,47.51) and (158.51,54) .. (150.5,54) .. controls (142.49,54) and (136,47.51) .. (136,39.5) -- cycle ;
%Shape: Circle [id:dp28056570538558634] 
\draw  [fill={rgb, 255:red, 0; green, 0; blue, 0 }  ,fill opacity=1 ] (136,99.5) .. controls (136,91.49) and (142.49,85) .. (150.5,85) .. controls (158.51,85) and (165,91.49) .. (165,99.5) .. controls (165,107.51) and (158.51,114) .. (150.5,114) .. controls (142.49,114) and (136,107.51) .. (136,99.5) -- cycle ;
%Shape: Circle [id:dp15865794558192303] 
\draw  [fill={rgb, 255:red, 0; green, 0; blue, 0 }  ,fill opacity=1 ] (136,159.5) .. controls (136,151.49) and (142.49,145) .. (150.5,145) .. controls (158.51,145) and (165,151.49) .. (165,159.5) .. controls (165,167.51) and (158.51,174) .. (150.5,174) .. controls (142.49,174) and (136,167.51) .. (136,159.5) -- cycle ;
%Shape: Circle [id:dp5453978951905287] 
\draw  [fill={rgb, 255:red, 0; green, 0; blue, 0 }  ,fill opacity=1 ] (136,219.5) .. controls (136,211.49) and (142.49,205) .. (150.5,205) .. controls (158.51,205) and (165,211.49) .. (165,219.5) .. controls (165,227.51) and (158.51,234) .. (150.5,234) .. controls (142.49,234) and (136,227.51) .. (136,219.5) -- cycle ;
%Shape: Circle [id:dp1877118572427139] 
\draw  [fill={rgb, 255:red, 0; green, 0; blue, 0 }  ,fill opacity=1 ] (216,39.5) .. controls (216,31.49) and (222.49,25) .. (230.5,25) .. controls (238.51,25) and (245,31.49) .. (245,39.5) .. controls (245,47.51) and (238.51,54) .. (230.5,54) .. controls (222.49,54) and (216,47.51) .. (216,39.5) -- cycle ;
%Shape: Circle [id:dp8444087315860302] 
\draw  [fill={rgb, 255:red, 0; green, 0; blue, 0 }  ,fill opacity=1 ] (216,99.5) .. controls (216,91.49) and (222.49,85) .. (230.5,85) .. controls (238.51,85) and (245,91.49) .. (245,99.5) .. controls (245,107.51) and (238.51,114) .. (230.5,114) .. controls (222.49,114) and (216,107.51) .. (216,99.5) -- cycle ;
%Shape: Circle [id:dp5961074161982878] 
\draw  [fill={rgb, 255:red, 0; green, 0; blue, 0 }  ,fill opacity=1 ] (216,159.5) .. controls (216,151.49) and (222.49,145) .. (230.5,145) .. controls (238.51,145) and (245,151.49) .. (245,159.5) .. controls (245,167.51) and (238.51,174) .. (230.5,174) .. controls (222.49,174) and (216,167.51) .. (216,159.5) -- cycle ;
%Shape: Circle [id:dp6151239712666448] 
\draw  [fill={rgb, 255:red, 0; green, 0; blue, 0 }  ,fill opacity=1 ] (216,219.5) .. controls (216,211.49) and (222.49,205) .. (230.5,205) .. controls (238.51,205) and (245,211.49) .. (245,219.5) .. controls (245,227.51) and (238.51,234) .. (230.5,234) .. controls (222.49,234) and (216,227.51) .. (216,219.5) -- cycle ;
%Straight Lines [id:da07543964063463893] 
\draw    (150.5,39.5) -- (230.5,99.5) ;
%Straight Lines [id:da10507653594047839] 
\draw    (150.5,99.5) -- (230.5,39.5) ;
%Straight Lines [id:da7025946117563342] 
\draw    (150.5,159.5) -- (230.5,99.5) ;
%Straight Lines [id:da12462598966379912] 
\draw    (150.5,219.5) -- (230.5,159.5) ;
%Straight Lines [id:da46891147201153616] 
\draw    (150.5,99.5) -- (230.5,159.5) ;
%Straight Lines [id:da7480665422768003] 
\draw    (150.5,159.5) -- (230.5,219.5) ;
%Straight Lines [id:da05275070525469605] 
\draw    (150.5,99.5) -- (230.5,99.5) ;
%Straight Lines [id:da45304486557923873] 
\draw    (150.5,159.5) -- (230.5,159.5) ;
%Straight Lines [id:da5877070789914338] 
\draw    (150.5,219.5) -- (230.5,219.5) ;
%Straight Lines [id:da6475575514665155] 
\draw    (150.5,99.5) -- (150.5,159.5) ;
%Straight Lines [id:da1844590728342863] 
\draw    (230.5,99.5) -- (230.5,159.5) ;
%Straight Lines [id:da38914533752679503] 
\draw    (230.5,159.5) -- (230.5,219.5) ;
%Straight Lines [id:da31918515566583516] 
\draw    (150.5,159.5) -- (150.5,219.5) ;
%Straight Lines [id:da41230921243097374] 
\draw    (150.5,39.5) -- (230.5,159.5) ;
%Straight Lines [id:da5848462571222055] 
\draw    (150.5,99.5) -- (230.5,219.5) ;
%Straight Lines [id:da7662660580445166] 
\draw    (230.5,39.5) -- (150.5,159.5) ;
%Straight Lines [id:da794783113040044] 
\draw    (230.5,99.5) -- (150.5,219.5) ;
%Straight Lines [id:da07802693873436106] 
\draw    (150.5,39.5) -- (230.5,219.5) ;
%Straight Lines [id:da7762324482515128] 
\draw    (230.5,39.5) -- (150.5,219.5) ;
%Curve Lines [id:da4364737957603171] 
\draw    (150.5,219.5) .. controls (116.33,156.67) and (117.33,157.67) .. (150.5,99.5) ;
%Curve Lines [id:da54581460657506] 
\draw    (150.5,159.5) .. controls (116.33,96.67) and (117.33,97.67) .. (150.5,39.5) ;
%Curve Lines [id:da48218465609365135] 
\draw    (150.5,219.5) .. controls (96.33,123) and (101.33,118) .. (150.5,39.5) ;
%Curve Lines [id:da9655575484565975] 
\draw    (230.5,39.5) .. controls (263.84,102.78) and (264.43,101.77) .. (230.5,159.5) ;
%Curve Lines [id:da028345567613481837] 
\draw    (230.5,99.5) .. controls (263.84,162.78) and (264.43,161.77) .. (230.5,219.5) ;
%Curve Lines [id:da16359220145061015] 
\draw    (230.5,39.5) .. controls (284.33,138) and (282.33,139) .. (230.5,219.5) ;

% Text Node
\draw (82,29) node [anchor=north west][inner sep=0.75pt]   [align=left] {$\displaystyle x_{i}$};
% Text Node
\draw (81,86) node [anchor=north west][inner sep=0.75pt]   [align=left] {$\displaystyle y_{i}$};
% Text Node
\draw (82,149) node [anchor=north west][inner sep=0.75pt]   [align=left] {$\displaystyle z_{i}$};
% Text Node
\draw (77,205) node [anchor=north west][inner sep=0.75pt]   [align=left] {$\displaystyle w_{i}$};
% Text Node
\draw (285,28) node [anchor=north west][inner sep=0.75pt]   [align=left] {$\displaystyle x_{j}$};
% Text Node
\draw (284,85) node [anchor=north west][inner sep=0.75pt]   [align=left] {$\displaystyle y_{j}$};
% Text Node
\draw (285,148) node [anchor=north west][inner sep=0.75pt]   [align=left] {$\displaystyle z_{j}$};
% Text Node
\draw (280,204) node [anchor=north west][inner sep=0.75pt]   [align=left] {$\displaystyle w_{j}$};

\end{tikzpicture}